\documentclass[journal]{IEEEtran}
\IEEEoverridecommandlockouts

\usepackage{amsmath,amssymb,amsfonts}
\usepackage{algorithm,algorithmic}
\usepackage{graphicx}
\usepackage{textcomp}
\usepackage[table]{xcolor}
\usepackage{booktabs}
\usepackage[square,sort,compress,comma,numbers]{natbib}
\usepackage{todonotes}
\usepackage{glossaries}
\usepackage{balance}
\usepackage{subfig}
\usepackage{caption}
\usepackage{subcaption}
\usepackage{xurl}
\usepackage{comment}
\usepackage{setspace}
\usepackage{makecell} 
\usepackage{xspace}
\usepackage{mathrsfs}
\usepackage{wasysym}
\usepackage{multirow}
\usepackage{tablefootnote}
\usepackage[flushleft]{threeparttable}
\usepackage[inline]{enumitem}
\usepackage{bm}
\usepackage{diagbox}
\usepackage{pifont}
\usepackage{anyfontsize}
\usepackage[bookmarks=false]{hyperref}
\usepackage[font=small]{caption}
\usepackage[most]{tcolorbox}

\usepackage{array}
\usepackage{siunitx}
\usepackage{pgf}        

\newlength{\extraarrayvspace}

\definecolor{redborder}{HTML}{d62728}
\definecolor{blueborder}{HTML}{1f77b4}
\definecolor{mybackground}{HTML}{f1f1f1}
\newtcolorbox{_objbox}[1][]{%
    enhanced,
    frame hidden,
    colback=mybackground,
    boxsep=0pt,
    before skip=2pt,
    after skip=2pt,
    breakable,
    overlay={%
    \fill[redborder]
      ([xshift=-1pt]interior.west|-interior.north)
      rectangle
      ([xshift=1pt]interior.west|-interior.south);
    },
    fonttitle={\normalsize\bfseries},
    coltitle=black,
    title={\textcolor{redborder}{Objective.}},
    attach title to upper,
    #1
}

\specialcomment{draft}{\begingroup\color{red}}{\endgroup}
\specialcomment{refine}{\begingroup\color{blue}}{\endgroup}

\newcommand{\fsl}{$\mathtt{FSL}$\xspace}
\newcommand{\fscil}{$\mathtt{FSCIL}$\xspace}
\newcommand{\cil}{$\mathtt{CIL}$\xspace}
\newcommand{\rfs}{$\mathtt{RFS}$\xspace}
\newcommand{\memp}{$\mathtt{Mem}\text{+}$\xspace}
\newcommand{\nafn}{$\mathtt{NAFN}$\xspace}
\newcommand{\quotes}[1]{``#1''}

\newcommand{\Cbase}{\mathcal{C}_{\mathrm{base}}}
\newcommand{\Cnewt}{\mathcal{C}^{(t)}_{\mathrm{new}}}
\newcommand{\Coldt}{\mathcal{C}^{(t)}_{\mathrm{old}}}
\newcommand{\Callt}{\mathcal{C}^{(t)}_{\mathrm{all}}}

\newcommand{\x}{\bm{x}}                      
\newcommand{\e}{\mathbf{e}}                      
\newcommand{\Ec}{\mathbf{E}_c}                   
\newcommand{\Echat}{\widehat{\mathbf{E}}_{c}}    

\newcommand{\LCE}{\mathcal{L}_{\mathrm{CE}}}
\newcommand{\LKL}{\mathcal{L}_{\mathrm{KL}}}
\newcommand{\Lrecon}{\mathcal{L}_{\text{Recon}}}
\newcommand{\Lcontrast}{\mathcal{L}_{\text{Contrast}}}
\newcommand{\Ltotal}{\mathcal{L}_{\text{Total}}}
\DeclareMathOperator{\Cov}{Cov}
\DeclareMathOperator{\mean}{mean}
\DeclareMathOperator{\Topr}{Top\text{-}r}
\DeclareMathOperator{\simop}{sim}

\newcommand{\Normal}{\mathcal{N}}

\newcommand{\LOF}{$\mathtt{LOF}$\xspace}
\newcommand{\IF}{$\mathtt{IF}$\xspace}
\newcommand{\PM}{$\mathtt{ProtoMargin}$\xspace}
\newcommand{\ECL}{$\mathtt{E\text{-}CL}$\xspace}
\newcommand{\wodenoise}{$\mathtt{w/o~Denoise}$\xspace}
\newcommand{\CSIDN}{$\mathtt{CSIDN}$\xspace}
\newcommand{\NCR}{$\mathtt{NCR}$\xspace}

\newcommand{\cmark}{\textcolor{green!60!black}{\ding{51}}}
\newcommand{\xmark}{\textcolor{red!70!black}{\ding{55}}}

\newcommand{\NN}{$\mathtt{NN}$\xspace}
\newcommand{\LR}{$\mathtt{LR}$\xspace}

\newcommand{\ie}{i.e.~}
\newcommand{\viz}{viz.~}
\newcommand{\wrt}{w.r.t.~}

\newcommand{\EDGEIIOT}{$\mathtt{Edge\text{-}IIoT}$\xspace}
\newcommand{\CIC}{$\mathtt{CSE\text{-}CIC\text{-}IDS2018}$\xspace}
\newcommand{\IOTNID}{$\mathtt{IoT\text{-}NID}$\xspace}

\newcommand{\qone}{\raisebox{-0.8ex}{\includegraphics[height=1.3em]{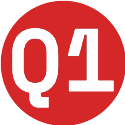}}}
\newcommand{\qtwo}{\raisebox{-0.8ex}{\includegraphics[height=1.3em]{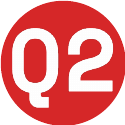}}}
\newcommand{\qthree}{\raisebox{-0.8ex}{\includegraphics[height=1.3em]{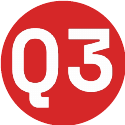}}}
\newcommand{\lone}{\raisebox{-0.8ex}{\includegraphics[height=1.3em]{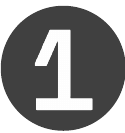}}}
\newcommand{\ltwo}{\raisebox{-0.8ex}{\includegraphics[height=1.3em]{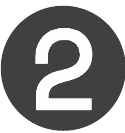}}}
\newcommand{\lthree}{\raisebox{-0.8ex}{\includegraphics[height=1.3em]{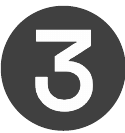}}}
\newcommand{\fix}{\raisebox{-0.8ex}{\includegraphics[height=1.2em]{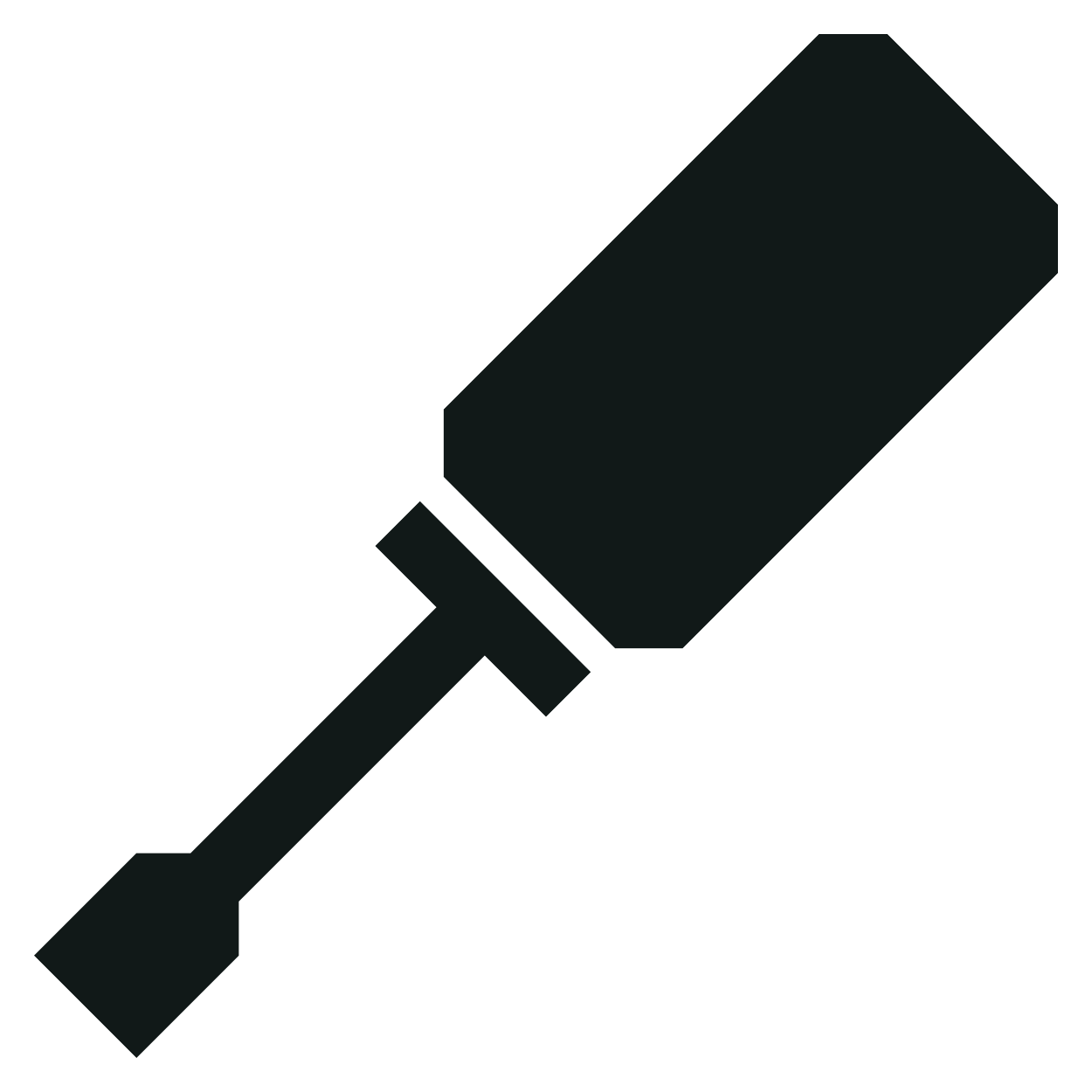}}}
\newcommand{\rob}{\raisebox{-0.8ex}{\includegraphics[height=1.2em]{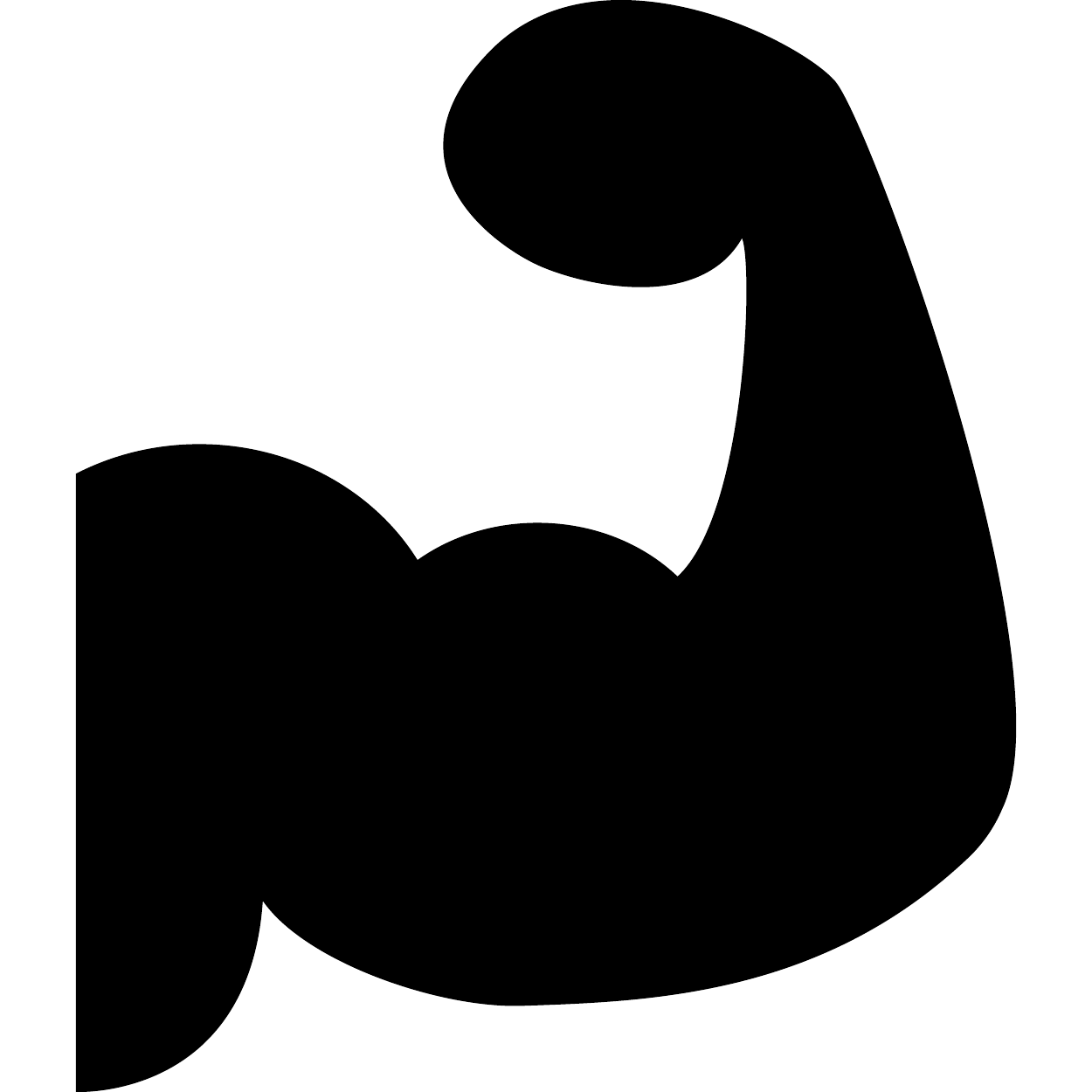}}}

\newcommand{\std}[1]{{\scriptsize$\,\pm$#1}}

\newcommand{\obj}[1]{\begin{_objbox} #1 \end{_objbox}}

\newcommand{\takeaway}[1]{%
\noindent {\textcolor{blueborder}{\bfseries Takeaway.}}\enspace 
{\emph{#1}}%
}

\newcommand{\para}[1]
{
\noindent\textbf{#1}\xspace}

\definecolor{lblue}{HTML}{1f77b4}

\definecolor{lb}{HTML}{81aad1}
\definecolor{lb2}{HTML}{90b6d4}
\definecolor{lr}{HTML}{db7575}
\definecolor{lg}{HTML}{C9F299}
\definecolor{dg}{HTML}{86C73B}
\newcommand{\bc}{\cellcolor{lb}}
\newcommand{\rc}{\cellcolor{lr}}

\def\BibTeX{{\rm B\kern-.05em{\sc i\kern-.025em b}\kern-.08em
    T\kern-.1667em\lower.7ex\hbox{E}\kern-.125emX}}
\begin{document}


\title{Network Intrusion Detection Adaptability under Spurious Labels: A Noise-Aware Few-Shot\\Class-Incremental Learning Framework}


\author{
Zijian Zhao,
Davide Di Monda,~\IEEEmembership{Student Member,~IEEE,}
Domenico Ciuonzo,~\IEEEmembership{Senior Member,~IEEE,}\\
Yipeng Wang,~\IEEEmembership{Senior Member,~IEEE,} and
Antonio Pescapè,~\IEEEmembership{Senior Member,~IEEE}
    \thanks{
    Z. Zhao and Y. Wang are with College of Computer Science, Beijing University of Technology, China (e-mail: b202374012@emails.bjut.edu.cn; wangyipeng@bjut.edu.cn).
    D. Ciuonzo and A. Pescapè are with the Department of Electrical Engineering and Information Technologies (DIETI), University of Naples Federico II, Italy (e-mail: domenico.ciuonzo@unina.it; pescape@unina.it).
    D. Di Monda is both with DIETI and IMT School for Advanced Studies, Lucca, Italy (e-mail: davide.dimonda@unina.it).\\
    This work was supported by the National Natural Science Foundation of China under Grant 62472011.
    }
}

\markboth{IEEE Transactions on Network and Service Management, ~Vol.~*, No.~*, MONTH~YYYY}%
{Surname \MakeLowercase{\textit{et al.}}: Paper Title}

\maketitle

\begin{abstract}
Network Intrusion Detection Systems (NIDSs) must incrementally adapt to emerging attacks, even when only a limited amount of labeled network traffic is available.
%
Consequently, Few-Shot Class-Incremental Learning (FSCIL)-based NIDSs have
gained growing attention.    
%
However, existing FSCIL-based NIDSs are highly vulnerable to label noise
in newly-observed attack classes.
To address this limitation, we propose NAFN, a \emph{noise-aware} FSCIL-based NIDS framework that reduces the impact of mislabeled data during incremental adaptation.
%
NAFN consists of two components: (\emph{i}) Noise Detection, which identifies and removes mislabeled biflows using noise-detection algorithms, and (\emph{ii}) Feature Sampling, which estimates class-conditional embeddings from clean biflows, calibrates them with base-class statistics, and resamples synthetic features to recover the original biflow count.
The framework follows the Rethink Few-Shot (RFS) teacher–student paradigm with lightweight nearest-neighbor or logistic regression classifiers, and various pretraining \& memory selection strategies.
%
%
Experiments on CSE-CIC-IDS2018, Edge-IIoT, and IoT-NID datasets show that NAFN effectively preserves prior knowledge and maintains strong incremental performance even under 50\% label noise, consistently improving new-class F1-scores over non-denoised baselines, 
up to $\bm{+52\%}$, 
while being time-efficient and robust in cross-network and high-noise incremental settings.
The implementation is publicly available to foster reproducibility.
%
\end{abstract}


\section{Introduction}
\label{sec:intro}

\IEEEPARstart{N}{etwork} Intrusion Detection Systems (NIDSs) play a critical role in safeguarding network security by monitoring traffic and detecting malicious activities. 
As intrusion techniques become increasingly sophisticated, NIDSs have evolved from early approaches based on port analysis and deep packet inspection to modern detection methods based on Machine Learning (ML) and Deep Learning (DL)~\cite{Rakine2025Access}.
%
However, traditional ML \& DL-based NIDSs are typically trained on closed-world datasets, containing only known benign and attack traffic.
%
In real-world network environments, however, new attacks continually emerge. Recent threat-intelligence reports confirm this trend, driven by ransomware and AI-enabled malicious campaigns.
For instance, in Oct.~$2025$, organizations worldwide faced an average of $1,938$ cyberattacks per week, an increase of $2\%$ over the previous month and $5\%$ compared to Oct.~$2024$~\cite{CheckPoint2025Oct}. 
Consequently, 
NIDSs struggle to adapt to this ever-changing threat landscape, resulting in limited effectiveness in real deployments.
%
To address this limitation, recent works have explored \emph{Class-Incremental Learning} (\cil)-based NIDSs~\cite{song20232,CIL-NIDS2} to continually learn new attacks while retaining previously acquired knowledge.
%
Nonetheless, \cil suffers from the scarcity of labeled traffic samples for new attacks, particularly in the case of zero-day attacks.
The integration of \emph{Few-Shot Learning} (\fsl)~\cite{zhao2025MUFFIN,Fewshot2} with \cil---\ie \emph{Few-Shot Class-Incremental Learning} (\fscil)---has emerged as a promising research direction for attack-traffic detection. 
\fscil-based NIDSs aim to incrementally classify new attacks with only a few labeled traffic samples (\viz few shots) without losing performance on previously learned classes.

Despite promising progress~\cite{FSCIL-NIDS2, FSCIL-NIDS1, cerasuolo2025attack}, existing \fscil-based NIDS \emph{remain fragile in real-world network environments}.
%
Most studies assume that the few available samples from new attacks are accurately labeled, which rarely holds true. 
Unlike base-class datasets that are carefully collected and labeled in controlled settings, new-class attack samples are often gathered from uncontrolled network environments where labels are generated through heuristic rules, automated scripts, or weak supervision, making them prone to errors. 
%
Hence, \emph{label noise} can severely degrade the performance of \fscil-based NIDS.
%
Since each new attack class contains only a few labeled traffic samples, 
every mislabeled instance can severely distort the feature distribution, degrading detection performance and introducing classification bias~\cite{noisylabel}.
In contrast, large-sample learning can tolerate noise as its influence is diluted~\cite{rolnick2017arXiv}.
%
Although manual relabeling can mitigate this problem, it is costly and undermines the real-time responsiveness required for intrusion detection. 
Therefore, label noise remains a major challenge that limits the robustness and practical deployment of \fscil-based NIDS.

In this paper, we propose \nafn, a \underline{\textbf{N}}oise-\underline{\textbf{A}}ware \underline{\textbf{F}}SCIL-based \underline{\textbf{N}}IDS framework designed to mitigate label noise
and enhance robustness in real-world scenarios.
%
\nafn introduces a \textbf{Noise-Aware Module}, consisting of two components: (\emph{i}) noise detection and (\emph{ii}) feature sampling.
Specifically, the noise detector removes mislabeled traffic samples from new attacks, while the feature sampler generates calibrated embeddings from the remaining clean data to mitigate the impact of sample loss.
%
%
%
Since this component is designed as a modular part of \nafn, it can accommodate different noise detection methods. To determine the most suitable choice, we evaluate several candidates, including Local Outlier Factor, Isolation Forest, Example-Level Curriculum Learning, and Prototype-Margin strategy. Ultimately, we adopt Prototype-Margin as the default option due to its superior stability and accuracy.
%
For feature sampling, the module estimates the feature distribution of clean new-class samples (after noise detection) and generates new embeddings based on this distribution.
To alleviate distribution bias under few-shot conditions, 
inspired by the literature on \fsl~\cite{yang2021free},
we further introduce a distribution calibration mechanism that leverages statistics from the base classes to align feature distributions of new and existing classes, thereby improving stability and generalization.

\nafn is trained in two stages.
First, a base training phase, and then multiple incremental learning phases.
During the base training phase, the model learns from the initial classes. When new attacks appear, \nafn performs incremental learning to quickly adapt.
We adopt the \textit{Rethinking Few-Shot} (\rfs) approach~\cite{RFS}, tailored to incremental learning~\cite{FSCIL-NIDS2}, as the core mechanism enabling \nafn to adapt to new attacks.
\rfs employs a teacher–student distillation strategy: at each incremental phase, the model from the previous phase acts as the teacher to guide the student, while a small set of samples from old classes (\viz memory) is retained to prevent forgetting. 
This design enables rapid adaptation to new attacks while maintaining knowledge stability. 
%
To further enhance the 
representation capability 
of the base model, 
we explore several pretraining strategies, including reconstruction-based~\cite{zhao2025MUFFIN, Ivandro2022DAE}, contrastive~\cite{he2020momentum, Ma2024contra}, and hybrid methods~\cite{WANG2025MTRC}.
%
In addition, we propose \memp, which integrates various exemplar selection strategies to populate the memory.
We consider the following strategies: Herding~\cite{iCaRL}, Rainbow~\cite{Rainbow}, and Random Sampling.
%
To summarize, our main \textbf{contributions} are as follows.
\begin{itemize}[leftmargin=*]
    \item We propose \nafn, a Noise-Aware Few-Shot Class-Incremental Learning framework for Network Intrusion Detection Systems (NIDS).
    It enables continual adaptation to novel attacks with few samples while effectively mitigating \emph{catastrophic forgetting}, thus fulfilling the core goal of \fscil-based NIDS.
    Experiments on three benchmark datasets (\CIC, \EDGEIIOT, and \IOTNID) demonstrate that \nafn achieves superior performance in incremental adaptation.
    
    \item We design a pluggable Noise-Aware Module to mitigate the impact of label noise. This module can identify and remove label noise in new attack traffic and generate new feature embeddings to restore the original data size.
    Under label noise conditions (noise levels ranging from $10\%$ to $50\%$), experimental results show a significant performance improvement in F1-score for new attacks, 
    up to $+52\%$ \wrt the baseline without denoising strategies.

    \item We perform a comprehensive evaluation of \nafn under both individual and combined challenges of cross-network adaptation and label noise. The results show that \nafn achieves consistent and generalizable performance across heterogeneous network datasets and remains robust under severe noise, with the largest observed pAUC improvements reaching $13\%$ for new attacks and $11\%$ for previously unseen attacks in high-noise settings.
\end{itemize}


%
The rest of this paper is organized as follows.
Sec.~\ref{sec:related} reviews recent works relevant to our study and positions our contribution within the existing literature.
%
Sec.~\ref{sec:problem} formalizes the problem of label noise on \fscil-based NIDSs and defines our research questions and objectives.  
%
Sec.~\ref{sec:methodology} provides
an overview 
of \nafn and details each of its components.  
Sec.~\ref{sec:exp_setup} describes the experimental setup, while Sec.~\ref{sec:exp_results} presents a comprehensive evaluation of \nafn on three recent and
public datasets.
%
Finally, Sec.~\ref{sec:conclusion} 
ends
the paper.


\section{Related Works}
\label{sec:related}
\setlength{\extraarrayvspace}{-1.0ex}
\definecolor{Gray}{gray}{0.92}

\begin{table*}[t]
\renewcommand{\arraystretch}{1.25}
\caption{Related studies focusing on network intrusion detection with \fscil and noisy labels.
Papers are listed in chronological order by publication year.
The final row provides an overview of the current work. 
The meaning of acronyms is provided at the bottom of the table.
}
\label{tab:sota}
\centering
\resizebox{\textwidth}{!}{
\begin{threeparttable}
\begin{tabular}{rcclcclcclccl}
    \specialrule{.1em}{.05em}{.05em}
    \multirow{2}{*}{\textbf{Reference}} &
    \multirow{2}{*}{\textbf{Year}} &
    \multicolumn{3}{c}{\textbf{FSCIL}} &
    \multicolumn{4}{c}{\textbf{Noise-Aware Capability}} &
    \multicolumn{4}{c}{\textbf{Traffic Data}} 
    \\
    \cmidrule(lr){3-5}
    \cmidrule(lr){6-9}
    \cmidrule(lr){10-13}
    
    & & \textbf{Used} & \textbf{Approach} & \textbf{CN} & \textbf{Present} & \begin{tabular}[c]{@{}l@{}}\textbf{Denoising}\\[\extraarrayvspace]\textbf{Approach}\end{tabular} & \textbf{Strat.} & \textbf{CN} & \textbf{Dataset} & \textbf{Raw} & \textbf{Early} & \textbf{Input} \\
    \hline

    \rowcolor{Gray}
    \citet{wang2021fscil} &
    $2021$ &
    \cmark &
    $\mathtt{MAML} + \mathtt{MM}$ &
    \xmark &
    \xmark &
    -- &
    -- &
    \xmark &
    $\mathtt{NSL\text{-}KDD}$ &
    \xmark &
    \xmark &
    Flow-based statistics
    \\

    \citet{abrishami2022label} &
    $2022$ &
    \xmark &
    -- &
    \xmark &
    \cmark &
    $\mathtt{DT} + \mathtt{AL~w/~US}$ &
    \fix &
    \xmark &
    \IOTNID &
    \xmark &
    \xmark &
    Flow-based statistics
    \\

    \rowcolor{Gray}
    \citet{yuan2023boau} &
    $2023$ &
    \xmark &
    -- &
    \xmark &
    \cmark &
    $\mathtt{AE} + \mathtt{KM} + \mathtt{BAL}$ &
    \rob &
    \xmark &
    \begin{tabular}[c]{@{}l@{}}$\mathtt{Malicious\_TLS}$, \\[\extraarrayvspace] $\mathtt{CIC}\text{-}\mathtt{IDS2017}$, \\[\extraarrayvspace] $\mathtt{CICMalDroid\text{-}2020}$, \\[\extraarrayvspace] $\mathtt{IoT\text{-}23}$ \end{tabular} &
    \cmark &
    \xmark &
    \begin{tabular}[c]{@{}l@{}}
    Per-packet informative\\[\extraarrayvspace]fields
    \end{tabular}
    \\

    \citet{yuan2023mcre} &
    $2023$ &
    \xmark &
    -- &
    \xmark &
    \cmark &
    $\mathtt{MCRe}$ &
    \fix &
    \xmark &
    \begin{tabular}[c]{@{}l@{}}$\mathtt{Malicious\_TLS}$, \\[\extraarrayvspace] $\mathtt{CIC}\text{-}\mathtt{IDS2017}$, \\[\extraarrayvspace] $\mathtt{CICMalDroid\text{-}2020}$, \\[\extraarrayvspace] $\mathtt{NSL\text{-}KDD}$, $\mathtt{LITNET\text{-}2020}$, \\[\extraarrayvspace]$\mathtt{UNSW\text{-}NB15}$, $\mathtt{IoT\text{-}23}$ \end{tabular} &
    \xmark &
    \xmark &
    Flow-based statistics
    \\

    \rowcolor{Gray}
    \citet{hallaji2023label} &
    $2023$ &
    \xmark &
    -- &
    \xmark &
    \cmark &
    $\mathtt{GAN}$ &
    \rob &
    \xmark &
    $\mathtt{UNSW\text{-}NB15}$, $\mathtt{NIMS}$ &
    \xmark &
    \xmark &
    Flow-based statistics
    \\

    \citet{FSCIL-NIDS1} &
    $2024$ &
    \cmark &
    $\mathtt{PN} + \mathtt{Mem} + \mathtt{PL}$  &
    \xmark &
    \xmark &
    -- &
    -- &
    \xmark &
    \begin{tabular}[c]{@{}l@{}} $\mathtt{CIC}\text{-}\mathtt{IDS2017}$, \\[\extraarrayvspace] \CIC \end{tabular} &
    \cmark &
    \cmark &
    Per-packet payload 
    \\

    \rowcolor{Gray}
    \citet{wang2024few} &
    $2024$ &
    \cmark &
    $\mathtt{MAML} + \mathtt{MMD}$ &
    \xmark &
    \xmark &
    -- &
    -- &
    \xmark &
    \begin{tabular}[c]{@{}l@{}} $\mathtt{CIC}\text{-}\mathtt{IDS2017}$, \\[\extraarrayvspace] \CIC \end{tabular} &
    \xmark &
    \xmark &
    Flow-based statistics
    \\

    \citet{hosseini2024intrusion} &
    $2024$ &
    \cmark &
    $\mathtt{FZ} + \mathtt{GAT}$ &
    \xmark &
    \xmark &
    -- &
    -- &
    \xmark &
    $\mathtt{NSL\text{-}KDD}$ &
    \xmark &
    \xmark &
    Flow-based statistics
    \\

    \rowcolor{Gray}
    \citet{cao2025study} &
    $2025$ &
    \cmark &
    $\mathtt{MAML} + \mathtt{Mem}$ &
    \xmark &
    \xmark &
    -- &
    -- &
    \xmark &
    $\mathtt{CIC}\text{-}\mathtt{IDS2017}$ &
    \xmark &
    \xmark &
    Flow-based statistics
    \\

    \citet{cerasuolo2025attack} &
    $2025$ &
    \cmark &
    \begin{tabular}[c]{@{}l@{}} $\mathtt{FT}$, $\mathtt{FT} + \mathtt{Mem}$, $\mathtt{LwF}$,\\[\extraarrayvspace]$\mathtt{EWC}$, $\mathtt{EEIL}$, $\mathtt{LUCIR}$,\\[\extraarrayvspace]$\mathtt{iCaRL\text{+}}$, $\mathtt{BiC}$ \end{tabular} &
    \xmark &
    \xmark &
    -- &
    -- &
    \xmark &
    \begin{tabular}[c]{@{}l@{}} $\mathtt{IoT\text{-}23}$, \IOTNID,\\[\extraarrayvspace]$\mathtt{TON\text{-}IoT}$, \EDGEIIOT\end{tabular} &
    \cmark &
    \cmark &
    \begin{tabular}[c]{@{}l@{}}
    Per-packet informative\\[\extraarrayvspace]fields
    \end{tabular}
    \\

    \rowcolor{Gray}
    \citet{meng2025unknown} &
    $2025$ &
    \xmark &
    -- &
    \xmark &
    \cmark &
    $\mathtt{PM}$ &
    \fix &
    \xmark &
    \begin{tabular}[c]{@{}l@{}}$\mathtt{Malicious\_TLS}$, \\[\extraarrayvspace]$\mathtt{CIC}\text{-}\mathtt{IDS2017}$, \\[\extraarrayvspace] \CIC, \\[\extraarrayvspace] $\mathtt{CIC\text{-}Darknet2020}$ \end{tabular} &
    \xmark &
    \xmark &
    Flow-based statistics
    \\

    \citet{zhan2025flowrefiner} &
    $2025$ &
    \xmark &
    -- &
    \xmark &
    \cmark &
    $\mathtt{TEO} + \mathtt{KM} + \mathtt{CC}$ &
    \rob &
    \xmark &
    \begin{tabular}[c]{@{}l@{}}$\mathtt{ISCXVPN2016}$, \\[\extraarrayvspace]$\mathtt{CrossPlatform}$, \\[\extraarrayvspace] $\mathtt{USTC\text{-}TFC2016}$, \\[\extraarrayvspace] $\mathtt{Malware2023}$ \end{tabular} &
    \cmark &
    \cmark &
    Per-packet L2 payload
    \\
    
    \hline
    \hline

    \rowcolor{Gray}
    \citet{FSCIL-NIDS2} &
    $2024$ &
    \cmark &
    \begin{tabular}[c]{@{}l@{}}$(\mathtt{MN}$, $\mathtt{PN}$, $\mathtt{MON}$, $\mathtt{FT}$,\\[\extraarrayvspace] $\mathtt{FZ}$, $\mathtt{RFS}) + \mathtt{Mem}$\end{tabular} &
    \cmark &
    \xmark &
    -- &
    -- &
    \xmark &
    \begin{tabular}[c]{@{}l@{}} \CIC, \\[\extraarrayvspace] \IOTNID \end{tabular} &
    \cmark &
    \cmark &
    \begin{tabular}[c]{@{}l@{}}
    Per-packet informative\\[\extraarrayvspace]fields
    \end{tabular}
    \\

    \hline

    \emph{This work} &
    $2026$ &
    \cmark &
    \begin{tabular}[c]{@{}l@{}} $(\mathtt{RS}, \mathtt{CS}, \mathtt{HS}) +$\\[\extraarrayvspace]$+~\mathtt{RFS} + \mathtt{Mem}\text{+}$ \end{tabular} &
    \cmark &
    \cmark &
    \begin{tabular}[c]{@{}l@{}} $(\mathtt{LOF}$, $\mathtt{IF}$, $\mathtt{E\text{-}CL}$,\\[\extraarrayvspace] $\mathtt{PM}) + \mathtt{Aug}$ \end{tabular} &
    \fix &
    \cmark &
    \begin{tabular}[c]{@{}l@{}}\CIC, \\[\extraarrayvspace] \IOTNID, \EDGEIIOT \end{tabular} &
    \cmark &
    \cmark &
    \begin{tabular}[c]{@{}l@{}}
    Per-packet informative\\[\extraarrayvspace]fields
    \end{tabular}
    \\
    
    \specialrule{.1em}{.05em}{.05em}
    \end{tabular}
\begin{tablenotes}
\footnotesize

\item
\textbf{Approach}---Few-Shot Class-Incremental Learning approach used;
\textbf{Denoising Approach}---Denoising approach used;
\textbf{Strat.}---Strategy related to the noise-aware capability;
\textbf{CN}---Whether the paper performs Cross-Network analyses;
\textbf{Raw}---Whether raw input data is used;
\textbf{Early}---Whether early traffic classification is used.
\textbf{Acronyms}:
Active Learning (AL) with Uncertainty Sampling (AL w/ US),
Autoencoder (AE),
Augmentation (Aug),
Bias Correction (BiC),
Boundary Augmentation Loss (BAL),
Confidence-based Correction (CC),
Contrastive Strategy (CS)
Decision Tree (DT),
Elastic Weights Consolidation (EWC),
End-to-End Incremental Learning (EEIL),
Example-Level Curriculum Learning (E-CL),
Fine-Tuning (FT),
Freezing (FZ),
Generative Adversarial Network (GAN),
Graph Attention Network (GAT),
Hybrid Strategy (HS),
Incremental Classifier and Representation Learning Plus (iCaRL+),
Isolation Forest (IF),
K-Means (KM),
Learning a Unified Classifier Incrementally via Re-balancing (LUCIR),
Learning Without Forgetting (LwF),
Local Outlier Factor (LOF),
Maximum Mean Discrepancy (MMD),
Memory (Mem),
MetaOptNet (MON),
Model-Agnostic Meta-Learning (MAML),
Model Merging (MM),
Multidimensional Constraint Representation (MCRe),
Matching Networks (MN),
Prototype-Margin (PM),
Prototypical Networks (PN),
Projection Layer (PL),
Reconstruction-based Strategy (RS),
Rethinking Few-Shot (RFS),
Transformer Encoder-Only (TEO).
\textbf{Symbols}:
\cmark~present, \xmark~lacking;
\rob~denoising approach that makes the model robust to label noise, \fix~denoising approach that corrects mislabeled samples;
\quotes{--}~indicates that the field is not relevant for that paper.
\end{tablenotes}
\end{threeparttable}
}
\end{table*}
\para{Background.}
%
In recent years, there has been growing research interest in applying ML, and later DL, as the core of AI-based NIDS~\cite{hasan2022securing}.
Following this first wave, the research community has shifted its focus to addressing the main limitations of ML \& DL~\cite{song20232, FSCIL-NIDS4, lu2023few}.
Two challenges are particularly relevant,
\begin{enumerate*}[label=(\emph{\roman*})]
    \item poor performance against network attacks that were not seen during training~\cite{song20232, CIL-NIDS2};
    \item difficulty of gathering and labeling high-quality and up-to-date datasets, which leads to class imbalance, and, consequently, unsatisfactory results~\cite{zhao2025MUFFIN,Fewshot2}.
\end{enumerate*}
To mitigate the first issue (\emph{i}), 
\cil has emerged as a promising direction.
Unlike conventional training, where models are retrained from scratch when new classes appear, \cil seeks to incrementally integrate new knowledge without 
forgetting.
\cil approaches typically rely on strategies such as rehearsal,
regularization, 
or dynamic architectures~\cite{masana2022class}. 
%
\fsl has shown potential in dealing with the second issue (\emph{ii}).
\fsl focuses on enabling models to generalize from a limited number of labeled samples by leveraging prior knowledge acquired from related tasks.
\fsl approaches typically rely on metric-learning, prototype-based classification, or meta-learning~\cite{wang2020generalizing}.
Efforts to integrate \cil and \fsl---\ie \fscil---remain relatively underexplored in current literature~\cite{FSCIL-NIDS2, FSCIL-NIDS1, cerasuolo2025attack}, despite its practical advantages (e.g., rapid adaptation to new threats).
In the computer vision domain, \fscil has been more extensively investigated~\cite{tao2020few, zhou2022forward, wang2023few}.
The common goal is to prepare models to recognize new classes by exploiting knowledge from previously learned ones.
Existing approaches mainly differ in how they allocate and adapt the representation space.
For instance, \citet{tao2020few} propose a growing \emph{Neural Gas Network} that dynamically expands its structure to accommodate new classes.
Similarly, \citet{zhou2022forward} reserve embedding space in advance by assigning multiple virtual prototypes.
More recently, \citet{wang2023few} observe that new-class samples are often misclassified into semantically similar old classes.
They leverage this property by calibrating new class prototypes through weighted combinations of old prototypes.
%
%
%
In the field of network intrusion detection, \fscil still faces critical challenges.
Particularly, their \emph{reliance on only a few labeled samples} for adapting to new traffic classes makes them highly vulnerable to \emph{label noise}.
Such noise may arise unintentionally (e.g., labeling in non-controlled environments) or through adversarial manipulation (e.g., label-flipping attacks).
Although the literature on label noise in non-\fscil contexts is growing~\cite{abrishami2022label, yuan2023boau}, 
\emph{the availability of larger datasets in these contexts partially mitigates its impact, making it less critical than in \fscil scenarios}.

\para{Literature Review.}
Table~\ref{tab:sota} categorizes the most relevant works that employ \fscil and address the label noise problem.
Research on \fscil is still in its early stages; the first contribution was proposed in $2021$~\cite{wang2021fscil}, while most subsequent works appeared between $2024$ and $2025$~\cite{wang2024few, hosseini2024intrusion, cao2025study, cerasuolo2025attack, FSCIL-NIDS2}.
%
Studies that address label noise in the context of NIDS remain relatively limited, reflecting the novelty of this research effort~\cite{abrishami2022label, yuan2023boau, yuan2023mcre, hallaji2023label, meng2025unknown, zhan2025flowrefiner}.
Notably, to the best of our knowledge, \emph{no existing work has yet tackled label noise within the \fscil setting}.

Among the \textbf{\texttt{FSCIL} approaches} explored in the literature, the most common is \emph{Model-Agnostic Meta-Learning} (MAML)~\cite{finn2017model}, originally designed for \fsl applications. 
Works adopting MAML~\cite{wang2021fscil, wang2024few, cao2025study} extend it with strategies to handle forgetting and add new classes. 
In particular, \citet{wang2021fscil} enable MAML for \fscil through a model growth technique.
\citet{wang2024few} use \emph{Maximum Mean Discrepancy} (MMD) to alleviate forgetting by selecting models with similar feature distributions.
\citet{cao2025study} adopt memory replay 
of previous classes during the incremental task.
%
Unlike the aforementioned works,
\citet{FSCIL-NIDS1} adapt \emph{Prototypical Networks} (PN)~\cite{snell2017proto} to \fscil using memory replay~\cite{cao2025study} and introduce session-specific projection layers to improve traffic-feature discriminability.
\citet{cerasuolo2025attack} evaluate several \cil approaches from computer vision and traffic classification under few-shot constraints.
%
In our previous work~\cite{FSCIL-NIDS2}, we adapt a set of \fsl approaches to \fscil by integrating memory replay to mitigate forgetting.

Regarding \textbf{denoising approaches}, state-of-the-art works can be grouped into two \textbf{strategies}.
%
The first aims to build models \emph{robust} to spurious labels.
\citet{yuan2023boau} design both the model architecture and the loss function to enhance robustness.
They employ an \emph{Autoencoder}
coupled with a \emph{K-Means}
to cluster samples in unsupervised fashion, thus reducing the impact of noisy annotations.
A \emph{Boundary Augmentation Loss} is then introduced to regularize decision boundaries.
Similarly, \citet{zhan2025flowrefiner} use K-Means to detect noisy labels and correct them based on prediction confidence. 
They use a \emph{Transformer encoder-only} for noise robustness.
\citet{hallaji2023label} adopt a generative approach to inject artificial noise, training classifiers on perturbed samples, \viz \emph{vaccination}.

The second strategy of works focuses on \emph{correcting mislabeled samples}.
Specifically, \citet{abrishami2022label} use \emph{Decision Trees} to detect potential label errors, assuming minority labels in a leaf are likely incorrect, and leverage \emph{Active Learning} to have humans verify uncertain cases.
\citet{yuan2023mcre} and \citet{meng2025unknown} both leverage \emph{local density} and \emph{feature similarity} to identify reliable prototypes for relabeling.
In both papers, samples are reassigned based on their similarity to class prototypes.
\citet{yuan2023mcre} introduce a custom loss to preserve class separability, while \citet{meng2025unknown} refine prototype selection by choosing high-density yet diverse samples.

Table~\ref{tab:sota} highlights that the current literature lacks systematic \textbf{cross-network} analyses~\cite{apruzzese2022cross, FSCIL-NIDS4} in \fscil and label noise scenarios (e.g., models are trained on one dataset, representing a network environment, and tested on another, or incremental classes belong to a different dataset than the one used for pretraining).
The only exception~\cite{FSCIL-NIDS2} performs a cross-network incremental analysis for \fscil approaches.
To the best of our knowledge, \emph{no existing work addresses cross-network studies under noisy-label conditions}.


Regarding the \textbf{datasets}, we observe that $\mathtt{CIC}\text{-}\mathtt{IDS2017}$ and \CIC are among the most popular choices~\cite{FSCIL-NIDS1, yuan2023boau, yuan2023mcre, wang2024few, cao2025study, meng2025unknown, FSCIL-NIDS2}.
Several IoT-oriented datasets are also frequently adopted (\ie $\mathtt{IoT\text{-}23}$, \IOTNID, and \EDGEIIOT~\cite{abrishami2022label, yuan2023boau, cerasuolo2025attack, FSCIL-NIDS2}).
While $\mathtt{NSL\text{-}KDD}$ remains popular~\cite{wang2021fscil, yuan2023mcre, hosseini2024intrusion}, despite being over two decades old, it no longer reflects modern network traffic and offers limited relevance for evaluating contemporary NIDS.

Table~\ref{tab:sota} also flags the studies that employ \textbf{raw input}~\cite{FSCIL-NIDS1, yuan2023boau, cerasuolo2025attack, FSCIL-NIDS2, zhan2025flowrefiner}, \ie unprocessed network data, in contrast to handcrafted or statistical features.
Raw input is naturally aligned with DL approaches, which are capable of automatically extracting meaningful representations.
On the other hand, works that rely on statistical features introduce additional pre-processing steps, increasing computational overhead.
Furthermore, they weaken the \quotes{promise} of end-to-end DL pipelines requiring little or no human intervention.

Finally, the table marks the works adopting an \textbf{early input strategy}~\cite{FSCIL-NIDS1, cerasuolo2025attack, FSCIL-NIDS2, zhan2025flowrefiner}.
This approach leverages only the initial portion of a traffic object (e.g., the first packets or bytes of the payload) as input, enabling \emph{early detection}.
Remaining works, instead, rely on the entire traffic object, making them suitable only for \emph{post-mortem detection}, which is effective for offline analysis but unsuitable for live network scenarios.

\para{Positioning.}
The last row of Table~\ref{tab:sota} positions our work against the reviewed literature.
Building on our previous work~\cite{FSCIL-NIDS2},
which compared multiple \fscil approaches,
we explore \emph{different training and adaptation optimizations for \fscil-based NIDS}, including improved pre-training stages and memory selection criteria (referred to as \memp).
Unlike the literature leveraging MAML for \fscil, which is known for its high computational cost and slow adaptation~\cite{raghu2019rapid}, our method builds upon the more efficient \rfs~\citep{RFS}.
In our prior study~\cite{FSCIL-NIDS2}, \rfs delivered the best performance and is better suited for real-world NIDS deployments.
Furthermore, it differs from~\citet{wang2024few}, which adopts a model selection strategy requiring the storage of multiple candidate models, thereby limiting scalability.
\rfs also differs from \fscil based on PN~\cite{FSCIL-NIDS1}, whose performance strongly depends on the quality of the learned prototypes.
%
Existing works reveals two main gaps concerning label noise:
\begin{enumerate*}[label=(\emph{\roman*})]
    \item the lack of studies addressing label noise within the \fscil setting, and
    \item the absence of comparative analyses across denoising strategies.
\end{enumerate*}
\emph{We address both gaps by systematically evaluating and comparing multiple denoising approaches under \fscil}.
Moreover, we ground our denoising strategy, based on filtering and synthesizing clean biflows, in the findings of~\cite{jiang2024more}.
In real-world NIDS deployments, the true noise ratio is unknown, making over-cleaning a realistic risk.
Under these conditions, \citet{jiang2024more} show that \emph{filtering significantly outperforms correction.}
The main limitation of filtering is the reduced sample size, which can increase the generalization error bound; our feature sampling module mitigates this effect. 
This distinguishes our approach from prior works based on label correction~\cite{meng2025unknown, yuan2023mcre, abrishami2022label}.
%
Equally important,
\emph{we conduct a cross-network incremental study} on both \fscil and label noise scenarios across three \emph{recent} and \emph{publicly-available} datasets.
%
All experiments are performed on \emph{raw} and \emph{early input} data to ensure realistic and low-latency NIDS.


\section{Problem Formalization \& Research Objectives}
\label{sec:problem}

In this section, we formalize \fscil and label noise in the context of NIDS. 
Next,
we present the research objectives of our proposed framework, \nafn.

\begin{figure*}[t]
    \centering
    
    \includegraphics[width=0.93\textwidth]{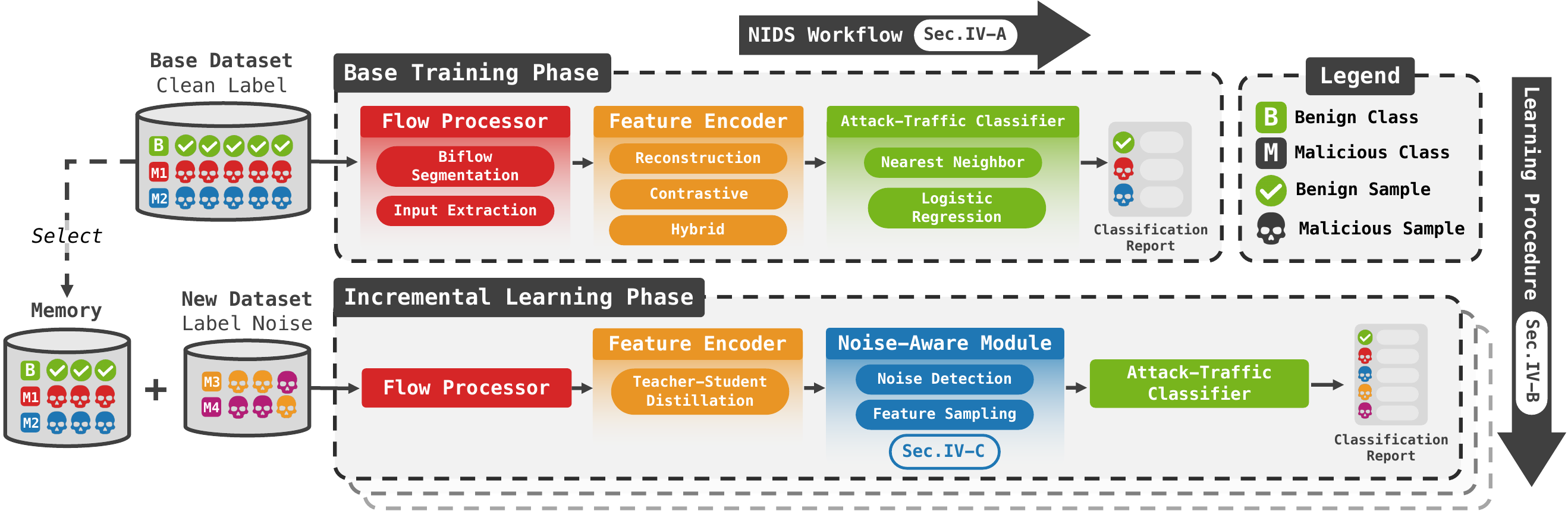}
    
    \caption{Overview of the \nafn framework. 
    It consists of three main perspectives:
    (\emph{i}) \textbf{NIDS Workflow (horizontal process)}: from flow processing and feature encoding to the attack-traffic classifier;
    (\emph{ii}) \textbf{Learning Procedure (vertical process)}: including base training and incremental learning phases with \memp-based memory selection;
    (\emph{iii}) \textbf{Noise-Aware Module}: operating during the incremental learning phase. It detects mislabeled traffic samples and synthesizes feature embeddings.
    }
    \label{fig:workflow}
\end{figure*}

\para{FSCIL Formalization.}
%
The goal of a \fscil-based NIDS is to continuously adapt to new attack classes
using only a few labeled samples, while preserving high classification performance on previously learned classes.
Formally, the \emph{base training phase} refers to training the model on the class set $\Cbase$, which includes benign traffic and a set of known attack classes.
Each base class contains an average of $\bar{N}_{\mathrm{base}}$ labeled samples for training, where $\bar{N}_{\mathrm{base}}$ denotes the average number of samples per base class.
This phase produces the initial (base) model used for subsequent incremental learning phases.
%
During each incremental phase $t \in \{ 1, 2, \ldots \}$, a new set of attack classes $\Cnewt$ appears, corresponding to previously unseen attacks:
\begin{equation}
\Cnewt \cap \Coldt = \emptyset, \quad 
\text{where } \Coldt = \Cbase \cup \bigcup_{i=1}^{t-1} \mathcal{C}^{(i)}_{\mathrm{new}}.
\end{equation}
This condition ensures that each incremental phase involves previously unseen attack classes.
Each new class $c \in \Cnewt$ contains only a small number of $K$ labeled samples, where $K \ll \bar{N}_{\mathrm{base}}$. 
\emph{The primary challenge of \fscil in the context of NIDS is to achieve rapid adaptation to newly emerging attacks in $\Cnewt$ while preserving high classification performance on previously learned classes in $\Coldt$.}

\para{Label Noise Formalization.}
%
In a real-world \fscil-based NIDS setting,
the limited samples available for new attacks $\Cnewt$ are often affected by label noise.
Let the attack dataset for incremental phase $t$ be
\begin{equation}
\mathcal{D}_{\text{new}}^{(t)}=\{(\x_i,\tilde{y}_i)\}_{i=1}^{N^{(t)}},
\end{equation}
with $\tilde{y}_i \in \Cnewt$ the observed (possibly noisy) label of attack sample $\x_i$.
For clarity, let $y_i$ denote the ground-truth label.
Since each new class $c \in \Cnewt$ contains only $K$ labeled samples, the total number of samples for the phase is $N^{(t)} = K \times |\Cnewt|$.
Label noise occurs when $\tilde{y}_i \neq y_i$, e.g., when a sample from attack class $c_1$ is mislabeled as attack $c_2$.
The noise rate $\eta$ is defined as
\begin{equation}
\eta = \frac{1}{N^{(t)}} \sum_{i=1}^{N^{(t)}} \mathbb{I}[\tilde{y}_i \neq y_i],
\end{equation}
where $\mathbb{I[\cdot]}$ is the indicator function.
%
Note that label noise is symmetrically distributed across classes in $\Cnewt$. 
%
Given the small number of samples for each new attack class, 
even a small noise rate $\eta$ can significantly degrade the discriminative power of the model.
Thus, \emph{the central challenge addressed in this paper lies in achieving stable incremental learning under the dual constraints of few-shot data and label noise.}

\para{Research Questions \& Objectives.}
Building upon the identified challenges, this work formulates the following research questions and their corresponding objectives, centered around the proposed Noise-Aware \fscil-based NIDS (\nafn).

\vspace{0.25em}
\noindent
\qone~\textit{Can \nafn address the primary challenge of \fscil by effectively adapting to new attack classes while retaining previously acquired knowledge?}  
\obj{~To design and evaluate \nafn for effective adaptation to novel attack classes from few labeled biflows, while retaining prior knowledge, as examined in Secs.~\ref{subsec:pretrain_Memory}, \ref{subsec:sen_shots}, and \ref{subsec:time}.
}

\noindent
\qtwo~\textit{How robust is \nafn to label noise, and how effective are its integrated noise detection and feature calibration mechanisms in mitigating the impact of mislabeled samples?}  
\obj{~To enhance robustness to mislabeled samples via noise detection and feature calibration, preserving stable, discriminative features under noisy few-shot conditions (Secs.~\ref{subsec:Noisy}, \ref{subsec:Visual}, and \ref{subsec:Time-to-Insight}).
}

\noindent
\qthree~\textit{Can \nafn maintain stable and generalizable performance when adapting across different networks, and remain robust when facing the combined challenges of cross-network adaptation and label noise?}  
\obj{~To assess \nafn’s generalization and stability across datasets, including cross-network adaptation and robustness under combined label noise and domain shift (Sec.~\ref{subsec:Crossnetwork}).
}


\section{Methodology}
\label{sec:methodology}













In this section, we present the overall methodology of \nafn, as illustrated in Fig.~\ref{fig:workflow}.
Particularly, we describe \nafn from \emph{three complementary perspectives}:
\begin{enumerate}[label=\textit{\roman*})]
    \item \textbf{NIDS Workflow (Horizontal Process).}  
    The end-to-end data pipeline---composed of \textit{Flow Processor}, \textit{Feature Encoder}, and \textit{Attack-Traffic Classifier}---transforms raw network traffic into feature embeddings and labels traffic in benign and attack classes (Sec.~\ref{subsec:NIDS_workflow}).

    \item \textbf{Learning Procedure (Vertical Process).}  
    Describes how \nafn learns across tasks via \textit{Base Training} and \textit{Incremental Learning} phases, following the teacher–student paradigm. The \textit{Memory Update mechanism} (\memp) preserves representative samples to reduce catastrophic forgetting (Sec.~\ref{subsec:learning_procedure}).

    \item \textbf{Noise-Aware Module.}  
    In addition to these two main processes, \nafn includes the Noise-Aware Module that operates exclusively during the incremental learning phase.
    It detects mislabeled samples of new attacks and synthesizes calibrated embeddings to maintain clean training data (Sec.~\ref{subsec:noise_aware}).
\end{enumerate}

By integrating these perspectives---the horizontal workflow, the vertical learning procedure, and the Noise-Aware Module---\nafn achieves robust and continual adaptation to emerging attacks
under a noisy few-shot regime.

\subsection{NIDS Workflow}
\label{subsec:NIDS_workflow}
The NIDS Workflow in Fig.~\ref{fig:workflow} outlines the end-to-end process through which raw network traffic is transformed into discriminative feature representations for attack detection.
It consists of three sequential modules---\emph{Flow Processor}, \emph{Feature Encoder}, and \emph{Attack-Traffic Classifier}---described below.

\para{Flow Processor.}
This module transforms raw network traffic into structured traffic data, which then serves as input to the Feature Encoder. 
The transformation occurs through two main steps:
\begin{enumerate*}[label=(\textit{\roman*})]
\item \emph{Biflow Splitting} and
\item \emph{Input Extraction}.
\end{enumerate*}
In the first step, raw network traffic is organized into bidirectional flows (\ie \emph{biflows}). A biflow is formed by merging two unidirectional flows that share the same five-tuple (\ie \emph{source IP}, \emph{destination IP}, \emph{source port}, \emph{destination port}, \emph{protocol}) but travel in opposite directions. 
This aggregation provides a unified representation of bidirectional communication
and avoids redundant flow entries~\cite{zhan2025flowrefiner, cerasuolo2025attack, FSCIL-NIDS2, FSCIL-NIDS1}.
In the second step, only the first $M_p$ packets of each biflow are retained.
This approach enables \emph{early detection} (Sec.~\ref{sec:related}).
By doing so, the NIDS can operate without waiting for the entire biflow to end, making it well-suited for real-time detection~\cite{cerasuolo2025attack, FSCIL-NIDS2, FSCIL-NIDS1, zhan2025flowrefiner}.
From the $M_p$ packets, $M_f$ per-packet features (e.g., packet length and direction) are extracted to form the structured representation.
Each biflow is thus represented as an $M_p \times M_f$ feature matrix $\x$.
If fewer than $M_p$ packets are available, zero-padding is applied to ensure $\x \in \mathbb{R}^{M_p \times M_f}$, 
and the resulting matrix is normalized to $[0,1]$ using Min--Max scaling.
%

\para{Feature Encoder.}
The Feature Encoder maps each biflow matrix $\x \in \mathbb{R}^{M_p \times M_f}$ into a $F$-dimensional feature embedding $\e = f_{\theta}(\x) \in \mathbb{R}^{F}$, where $f_{\theta}$ denotes the encoder parameterized by $\theta$.
It serves as the backbone representation extractor of \nafn.
Structurally, the encoder follows a \emph{symmetric U-Net–based architecture enhanced with residual connections}~\cite{UNet}.
The feature encoder consists of $L$ downsampling stages and the corresponding $L$ upsampling stages. Each downsampling stage $L_i$ is constructed using a \emph{Residual Convolution Block} (RCB) comprising two 2D convolutional layers, batch normalization, and \emph{LeakyReLU} activations, with an identity skip path to stabilize gradient propagation.
In the downsampling path, the input $\x$ is first reshaped to $[1, M_p, M_f]$ to match the 2D convolutional format, and processed by an initial RCB to yield a feature map of size $[Ch_0, M_p, M_f]$.
The feature map is passed through $L$ downsampling stages. Each stage progressively reduces the spatial resolution by pooling while doubling the number of feature channels. The output of the $i$-th downsampling stage ($i \in [1, L]$) has the shape $[Ch_{i}, M_P//2^{i}, M_f//2^{i}]$, 
where $//$ denotes floor division, \ie, $M_P // 2^i = \left\lfloor M_P / 2^i \right\rfloor$.
The resulting tensor is flattened and linearly projected to a fixed-length latent vector $\e \in \mathbb{R}^{F}$, which compactly encodes both local packet-level and global flow-level patterns.
In the upsampling path, each stage performs transposed convolution to restore the spatial resolution.
Upsampled feature maps are concatenated with the corresponding outputs from the downsampling stages through skip connections, restoring finer details. 
Symmetrically, the output of the $i$-th upsampling stage has the shape $[Ch_{L-i}, M_P//2^{L-i}, M_f//2^{L-i}]$.
Finally, a $1 \times 1$ convolution layer produces the reconstructed feature map of size $[1, M_p, M_f]$,
which is used as auxiliary outputs
for the reconstruction-based strategy in the base training phase.



\para{Attack-Traffic Classification.}
The Attack-Traffic Classifier is a decision-making component that operates on the feature embeddings generated by the encoder ($\mathbf{e} \in \mathbb{R}^{F}$).  
Its objective is to classify benign traffic and multiple attack classes (\viz multi-class misuse detection~\cite{liao2013intrusion}).
%
Formally, in a generic incremental phase $t$, we train the classifier using the embedded biflows from both the previously learned classes $\Coldt$ 
and new ones $\Cnewt$, where each class provides only $K$ labeled biflows. 
The classifier takes the embedding vector $\e$ as input and outputs a probability vector $\hat{\mathbf{y}}$ over all observed classes $\Callt = \Coldt \cup \Cnewt$.
This mapping can be formally expressed as $f_{\mathrm{cls}}: \mathbb{R}^{F} \rightarrow \mathbb{R}^{|\Callt|}$, where $|\Callt|$ denotes the total number of classes.
%
For \nafn, we consider two possible lightweight classifiers,
each offering complementary advantages in terms of stability and adaptability.
\begin{enumerate*}[label=(\textit{\roman*})]
    \item The \emph{Nearest-Neighbor} (\NN) classifier assigns each embedding to the class of its closest sample in the learned feature space.
    Here, we use a soft \NN rule, computing class posterior probabilities via a distance-based softmax over the nearest biflow from each class.
    This non-parametric approach relies on similarity measures and requires no trainable parameters, ensuring stability and low computational cost.
    \item The \emph{Logistic Regression} (\LR) classifier is a parametric linear model that learns decision boundaries by optimizing a cross-entropy objective.
    It provides greater discriminative power and flexibility, especially when sufficient labeled biflows are available.
\end{enumerate*}
These two classifiers can be selected depending on the task configuration and the desired trade-off between stability and plasticity.

\subsection{Learning Procedure}
\label{subsec:learning_procedure}

\begin{figure}[!tbp]
    \centering
    \includegraphics[width=1.0\hsize, trim=1.85cm 0 4.25cm 0, clip]{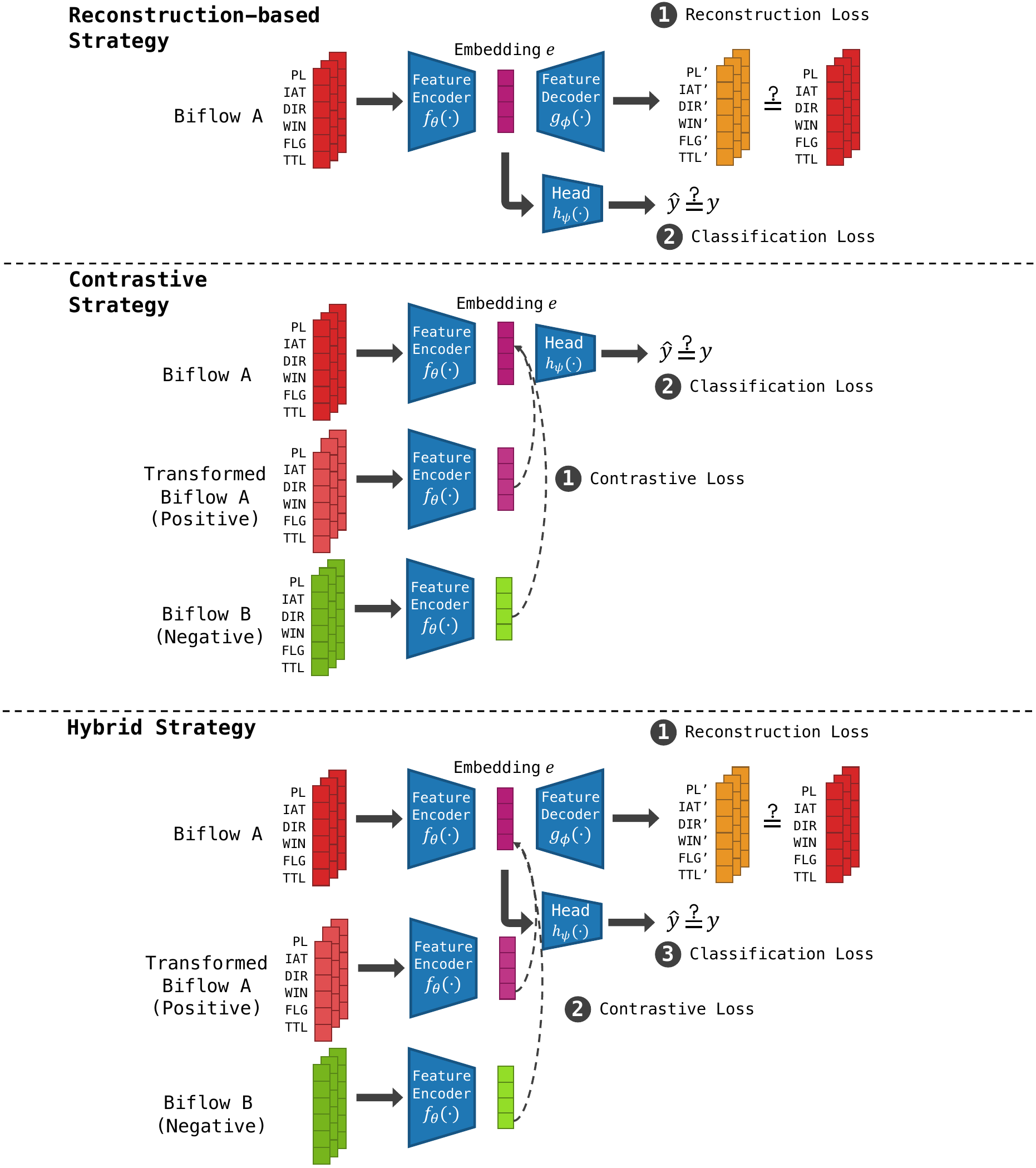}
    \caption{Overview of the three pretraining strategies.
    }
    \label{fig:pretrain}
\end{figure}

Fig.~\ref{fig:workflow} illustrates the learning procedure of \nafn, 
describing how the NIDS learns and adapts across sequential phases.
In particular, we consider two phases, \ie \emph{Base Training Phase} and \emph{Incremental Learning Phase}, outlined hereinafter.

\para{Base Training Phase.}
The goal of the base training phase is to construct an effective feature encoder on the base dataset, which consists of known attack classes and benign traffic classes (denoted as $\Cbase$) with ample samples, to learn robust and transferable feature representations from network traffic.
Once trained, 
this feature encoder serves as a teacher model that provides guidance for subsequent incremental phases.
%
To ensure that the encoder learns
discriminative and generalizable feature embeddings
on $\Cbase$, we jointly utilize supervised and self-supervised learning objectives and design three pretraining strategies 
(see Fig.~\ref{fig:pretrain}).
%
\begin{itemize}[leftmargin=*]
    \item \textit{Reconstruction-based Strategy.} 
    According to this strategy, the feature extractor is mirrored to form a decoder $g_{\phi}(\cdot)$ that reconstructs the input tensor $\x$ from the embedding vector $\e = f_{\theta}(\x)$.
    It corresponds to the upsampling path of the U-Net architecture (see Sec.~\ref{subsec:NIDS_workflow}), where transposed convolutions and skip connections are employed to reconstruct the spatial-temporal structure of a biflow.
    In addition, a temporary classification head $h_{\psi}(\cdot)$ outputs attack-traffic prediction using the embedding $\e$.
    The model is optimized by jointly minimizing the reconstruction loss \lone{} and the classification loss (\ie cross-entropy loss) \ltwo{} (see Fig.~\ref{fig:pretrain}):
    \begin{equation}
    \begin{aligned}
        \mathcal{L}_{\text{recon}} &= \| \x - \hat{\x} \|_2^2, \quad
        \mathcal{L}_{\text{CE}} = - \sum y_c \log(h_{\psi}(\mathbf{e})) \\
        \Ltotal
        &= \alpha \cdot \mathcal{L}_{\text{recon}}
        + (1 - \alpha) \cdot \mathcal{L}_{\text{CE}}
    \end{aligned}
    \end{equation}

    where $\hat{\x} = g_{\phi}(f_{\theta}(\x))$ is the reconstructed input,
    $y_c$ represents the true label for class $c$,
    and $\alpha$ is a balancing coefficient.
    \item \textit{Contrastive Strategy.} 
    The following strategy introduces a contrastive learning objective to encourage invariant and generalizable representations.  
    Specifically, we adopt the \textit{Momentum Contrast (MoCo)}~\cite{he2020momentum}, where 
    $\mathbf{e}_q$ denotes the query embedding produced by the encoder, 
    $\mathbf{e}_p$ is the positive embedding obtained from the momentum encoder (an augmented view of the same biflow), 
    and $\{\mathbf{e}_n^{(i)}\}_{i=1}^{N_{\mathrm{neg}}}$ are negative embeddings stored in a dynamic memory queue.  
    The contrastive loss is formulated as:
    \begin{equation}
    \Lcontrast = -\log \frac{\exp(\mathbf{e}_q \cdot \mathbf{e}_p / \tau)}{\exp(\mathbf{e}_q \cdot \mathbf{e}_p / \tau) + \sum_{i=1}^{N_{\mathrm{neg}}} \exp(\mathbf{e}_q \cdot \mathbf{e}_n^{(i)} / \tau)},
    \end{equation}
    where $\tau>0$ is a temperature parameter controlling the 
    softness of the probability distribution and $\mathbf{e}_q \cdot \mathbf{e}_p$ denotes the inner product between two embedding vectors.
    We generate positive samples via a biflow-specific augmentation, called \emph{permutation}, in which the original biflow is divided into $n$ segments and randomly reordered, simulating variations in packet sequencing~\cite{wang2024data}.
    %
    The overall loss combines the contrastive \lone{} and classification \ltwo{} losses:
    \begin{equation}
    \Ltotal = \beta \cdot \Lcontrast + (1 - \beta) \cdot \LCE,
    \end{equation}
    where $\beta$ is a balancing coefficient.
    This strategy enhances the invariance and generalization capability of the learned embeddings by contrasting positive and negative pairs across dynamic batches.
    \item \textit{Hybrid Strategy.}
    This strategy combines the reconstruction \lone{} and contrastive \ltwo{} losses with the classification loss \lthree{} to fully exploit complementary supervision from multiple objectives, thereby improving the discriminability and robustness of the feature encoder. 
    The overall loss function is formulated as:
    \begin{equation}
    \Ltotal = \alpha \cdot \Lrecon + \beta \cdot \Lcontrast + (1-\alpha-\beta) \cdot \LCE.
    \end{equation}
\end{itemize}

\para{Incremental Learning Phase.}
At each incremental phase $t$, we adopt the \rfs approach~\cite{RFS}, using the feature extractor from the previous $(t\!-\!1)^{th}$ phase as a teacher model to distill knowledge into the new student feature extractor $f_{\theta^{(t)}}$. 
In this phase, training relies exclusively on the biflows of previously learned classes stored in the memory, with the goal of preserving the stability of their feature representations.
Formally, given the teacher model $f_{\theta^{(t-1)}}$ from the previous phase $(t\!-\!1)$, we update the student $f_{\theta^{(t)}}$ using only the classes $\Coldt$ from phase $(t\!-\!1)$.
Let $\mathcal{D}_{\text{old}}^{(t)}$ denote the training set containing $\Coldt$ biflows selected by \memp from the memory bank, the student is optimized by a weighted sum of the classification and KL distillation objectives:
\begin{align}
\theta^{(t)}
= \arg\min_{\theta^{(t)}}\;\mathbb{E}_{(\x,y)\sim \mathcal{D}_{\text{old}}^{(t)}}\Big[
&\,\lambda_1 \cdot \LCE\!\big(h_{\psi^{(t)}}(f_{\theta^{(t)}}(\x)),\, y\big) \notag \\
+ \lambda_2 \cdot \LKL\!\big(h_{\psi^{(t)}}&(f_{\theta^{(t)}}(\x)),\, h_{\psi^{(t-1)}}(f_{\theta^{(t-1)}}(\x))\big)
\Big].
\end{align}

After convergence, during the $t^{th}$ phase, we freeze the feature extractor $f_{\theta^{(t)}}$ of the student to avoid catastrophic forgetting. 
The frozen model is then used to extract biflow embeddings $\e$ for the attack-traffic classifier (see Sec.~\ref{subsec:NIDS_workflow}).
Finally, the updated student parameters $\theta^{(t)}$
replace
the previous teacher for the next incremental phase $(t\!+\!1)$.

\para{Memory Update (Mem+).}
After each incremental phase, the exemplar memory is updated to retain the most representative biflows for all learned classes.
We denote the proposed enhanced exemplar selection as \memp. 
For each class, we compute feature embeddings using the current model and select the $K$ most representative samples. 
\memp supports multiple strategies: \emph{Herding}~\cite{iCaRL}, \emph{Rainbow}~\cite{Rainbow}, and \emph{random sampling}. 
The memory size is bounded by a fixed budget, ensuring balanced representation across old and new classes. 
The updated memory bank is then used for the incoming incremental phase to support student training.

\subsection{Noise-Aware Module}
\label{subsec:noise_aware}
The Noise-Aware Module is the core component of our \nafn framework.
It operates only during the incremental learning phase to handle potential 
spurious labels 
in newly introduced attack classes, improving robustness in the \fscil-NIDS setting.
It consists of two key components:
\begin{enumerate*}[label=(\textit{\roman*})]
    \item a \emph{Noise Detection Module}, which detects and removes mislabeled samples, and 
    \item a \emph{Feature Sampling Module}, which synthesizes additional clean feature embeddings to maintain the original data scale and class balance.
\end{enumerate*}

\para{Noise Detection Module.}
During the $t^{th}$ incremental learning phase, for each new attack class $c \in \Cnewt$ with feature embeddings $\Ec = \{\e_1, \dots, \e_{K}\}$, we obtain a denoised subset $\Echat \subset \Ec$ by filtering out suspected mislabeled items
using a denoising strategy.
We remove the top $\kappa = \lfloor \rho K \rfloor$ anomaly samples, where $\rho \in [0,1]$ is a hyperparameter that controls the denoising ratio (\ie the proportion of samples regarded as noisy).
To achieve this goal, we explore four noise detection approaches, each designed to identify noisy samples from different statistical or representation-based perspectives.

\begin{enumerate}[label=\textit{\roman*})]
    \item \emph{Local Outlier Factor} (\LOF) evaluates the local density deviation of a biflow relative to its neighbors:
    \begin{equation}
    \mathrm{LOF}_r(\e_i)
    = \frac{1}{|N_r(\e_i)|}
    \sum_{\e_j\in N_r(\e_i)}
    \frac{\mathrm{lrd}_r(\e_j)}{\mathrm{lrd}_r(\e_i)}.
    \end{equation}
    Here, $N_r(\e_i)$ denotes the $r$-nearest neighbors of $\e_i$, where $r$ controls the neighborhood size, 
    and $\mathrm{lrd}_r(\cdot)$ is the local reachability density. 
    
    \item \emph{Isolation Forest} (\IF) measures the expected isolation path length of each biflow:
    \begin{equation}
    s(\e_i) = 2^{-\frac{E[L(\e_i)]}{\zeta(K)}},
    \end{equation}
    where $E[L(\e_i)]$ is the expected path length of $\e_i$ in the isolation trees, and $\zeta(K)$ is a normalization factor depending on the sample size $K$.

    \item \emph{Prototype-Margin} (\PM) iteratively refines the class prototype and measures each biflow’s similarity margin between intra- and inter-class prototypes. 
    The prototype is updated as:
    \begin{align}
    p_c^{(0)} &= \frac{1}{K}\sum_{i=1}^{K}\e_i,\\ 
    p_c^{(t+1)} &= \mean\big(\Topr\{\e_i \mid \simop(\e_i, p_c^{(t)})\}\big),
    \end{align}
    where $r = (1-\rho)K$, and $\simop(\cdot,\cdot)$ denotes the similarity function (e.g., cosine similarity). 
    After $I$ iterations, the score of each biflow is computed as:
    \begin{equation}
    s(\e_i) = \simop(\e_i,p_c) 
    - \max_{c'\neq c}\simop(\e_i,p_{c'}).
    \end{equation}
    The top $(1-\rho)K$ biflows
    with the highest scores are retained as clean samples.

    \item \emph{Example-Level Curriculum Learning} (\ECL) trains a temporary classifier $f_{\phi}$ for a few epochs across all observed classes $\Callt$~\cite{E-CL}.
    The cross-entropy loss for a biflow $\e_i$ with label $\tilde{y}_i$ is:
    \begin{equation}
    \ell_i = -\log
    \frac{\exp(f_{\phi}(\e_i)[\tilde{y}_i])}{
    \sum_{j=1}^{|\Callt|}\exp(f_{\phi}(\e_i)[j])}.
    \end{equation}
    Here, $\exp(f_{\phi}(\e_i)[\tilde{y}_i])$ represents the probability score of sample $\e_i$ on its observed label $\tilde{y}_i$ in the classifier output.
    The biflows with the highest $\ell_i$ values are regarded as noisy and removed from the training set.

\end{enumerate}

\para{Feature Sampling Module.}
Since denoising reduces the number of training biflows, the Feature Sampling Module compensates for this loss by synthesizing
new and clean feature embeddings.
%
For each new attack class $c$, it estimates and calibrates a Gaussian distribution from the denoised subset $\Echat$:
\begin{equation}
\mu_c = \frac{1}{|\Echat|}\sum_{\e\in \Echat}\e, \quad
\Sigma_c = \Cov(\Echat).
\end{equation}
To enhance distributional stability, we adapt the estimated Gaussian parameters by fusing the distributional statistics of base classes $\Cbase$ into the attack class $c$. 
Let $(\mu_b, \Sigma_b)$ denote the mean and covariance of a base class $b \in \Cbase$, and
let $N_b$ be the number of samples.
The similarity between class $b$ and $c$ is measured by the Euclidean distance $d_b = \|\mu_c - \mu_b\|_2$. 
We assign normalized weights to these base classes as:
\begin{equation}
\omega_b = \frac{1/(N_b \cdot d_b)}{\sum_{b\in \Cbase} 1/ (N_b \cdot d_b)}.
\end{equation}
The calibrated Gaussian parameters are then computed as:
\begin{align}
\mu_c' &= \varepsilon \sum_{b\in \Cbase}\omega_b\,\mu_b + (1-\gamma)\mu_c, \label{eq:fusion_mu}\\
\Sigma_c' &= \varepsilon \sum_{b\in \Cbase}\omega_b\,\Sigma_b + (1-\gamma)\Sigma_c + \mathbf{I},
\label{eq:fusion_Sigma}
\end{align}
where $\varepsilon \in [0,1]$ and $\gamma \in [0,1]$ control the fusion ratio.
Finally, $\kappa = \lfloor \rho K \rfloor$ synthetic embeddings are sampled from $\Normal(\mu_c', \Sigma_c')$ and added to the denoised subset $\Echat$ to restore a total of $K$ samples per class. 
This calibrated Gaussian sampling strategy effectively preserves discriminative and stable feature representations under both label noise and few-shot conditions.

\section{Experimental Setup}
\label{sec:exp_setup}

In this section, we present the experimental setup adopted in our evaluation. Sec.~\ref{sec:dataset} describes the threat model and the related dataset. Next, Sec.~\ref{sec:fscil_setup} reports the main characteristics defining the \fscil setup. 
Finally, Sec.~\ref{sec:metrics} introduces the performance metrics.

\begin{figure}
    \centering

    \includegraphics[width=1.0\hsize]{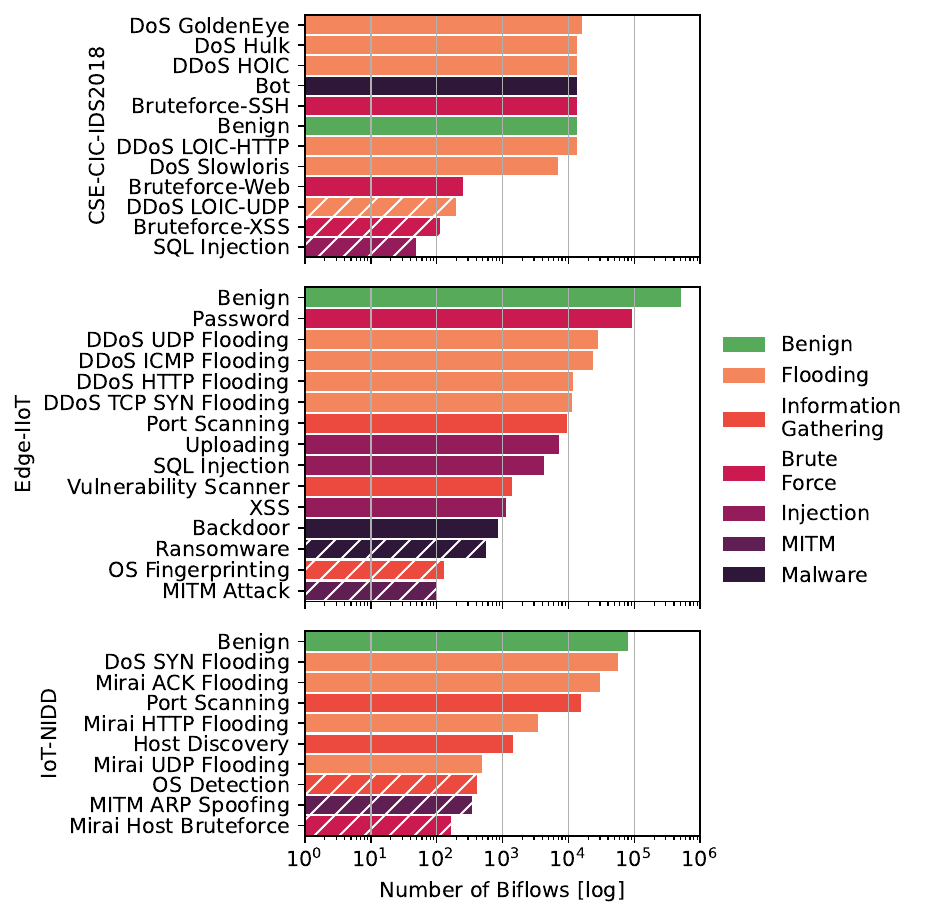}
    
    \caption{Statistics of the Number of biflows per class for \CIC (top), \EDGEIIOT(middle) and \IOTNID (bottom) datasets.
    Bars with diagonal hatching represent the classes used in the incremental phase ($\Cnewt$), while the plain bars correspond to the base classes ($\Cbase$).
    }
    \label{fig:dataset}
\end{figure}

\subsection{Threat Model \& Datasets}
\label{sec:dataset}

In this paper, we tackle a \textbf{threat model} based on \emph{flooding}, \emph{information gathering}, \emph{brute-force}, \emph{injection}, \emph{man-in-the-middle}, and \emph{malware} attacks. 
\begin{itemize}[leftmargin=*]
    \item \emph{Flooding} attacks aim to make targeted resources unavailable by overwhelming them with a large volume of traffic.
    Flooding is particularly dangerous in IoT environments, where compromised devices can be organized into \emph{botnets} that generate massive coordinated traffic.

    \item \emph{Information Gathering} includes activities aimed at discovering vulnerabilities, weak points, and sensitive information about a target resource that can be later exploited. 

    \item \emph{Brute-Force} attacks refer to an effort to gain unauthorized access to a system by systematically attempting multiple username and password combinations.

    \item \emph{Injection} involves uploading malicious input to a system in order to manipulate its execution logic, access sensitive information, or execute unauthorized commands.

    \item \emph{Man-In-The-Middle} (MITM) attacks occur when an adversary secretly intercepts the communication between two parties, allowing the attacker to eavesdrop on sensitive data or manipulate transmitted information.

    \item \emph{Malware} refers to malicious code that, once installed on a victim system, can perform harmful actions such as stealing sensitive information, disrupting system operations, or enabling remote control by an attacker.
\end{itemize}

To represent this threat model, we selected three network intrusion detection datasets: \CIC~\cite{CIC2018}, \EDGEIIOT~\cite{Edge-IIoT}, and \IOTNID~\cite{IoT-NID}. 
This selection is motivated by the fact that they include attack classes aligned with the threat categories described above.
Moreover,
\begin{enumerate*}[label=(\textit{\roman*})]
    \item they cover a wide range of attack scenarios across enterprise, IoT, and Industrial IoT (IIoT) environments;
    \item they are well-established and widely used in current literature~\cite{abrishami2022label, FSCIL-NIDS2, wang2024few, FSCIL-NIDS4};
    \item they are publicly available, ensuring reproducibility; and
    \item they are relatively recent, with the oldest dataset (\CIC) captured in $2018$ and the newest (\EDGEIIOT) in $2022$.
\end{enumerate*}

The \textbf{\texttt{CSE-CIC-IDS2018}} dataset
is a widely used benchmark for intrusion detection~\cite{FSCIL-NIDS1, wang2024few, meng2025unknown, FSCIL-NIDS2}, released by the Canadian Institute for Cybersecurity. 
It contains realistic benign and attack traffic captured from a simulated enterprise network. 
The dataset includes $13$ classes ($1$ benign and $12$ attacks), covering attacks such as DoS, DDoS, Brute Force, Web attacks, and infiltration.
Rather than relying on the released pre-computed features, we extract biflows directly from the PCAP traces and subsequently revise and clean the resulting dataset following the procedure described in~\cite{FSCIL-NIDS4}.
The resulting dataset contains $\approx109k$ labeled biflows.

The \textbf{\texttt{Edge-IIoT}} dataset 
is designed for intrusion detection in industrial IoT environments. 
It captures traffic from edge computing scenarios with various industrial protocols (e.g., Modbus, MQTT). 
The dataset includes $16$ classes ($1$ benign and $15$ attack classes), covering botnets, ransomware, scanning, and various flooding attacks. 
%
Starting from the PCAPs,
we perform a cleaning phase on \EDGEIIOT as in~\cite{cerasuolo2025adaptable}.
The final dataset consists of $\approx731k$ labeled biflows.

The \textbf{\texttt{IoT-NID}} dataset 
is collected from a realistic smart home environment, containing benign traffic and attacks on IoT devices such as SKT NUGU and EZVIZ Wi-Fi Camera. It includes 10 classes ($1$ benign and $9$ attack classes), with attacks such as Mirai botnet, DoS, MITM, and port scanning. 
%
After the PCAP cleaning phase~\cite{cerasuolo2025adaptable},
it contains $\approx197k$ labeled biflows.

Fig.~\ref{fig:dataset} shows the number of biflows per class for each dataset, colored according to the attack categories defined in the threat model and ranked in decreasing order.


\subsection{FSCIL Setup} 
\label{sec:fscil_setup}

This section describes the experimental setup used to evaluate our proposed approach. We detail the class-splitting criteria, evaluation protocol, hyperparameter settings, and implementation specifics to ensure reproducibility.

\para{Class Splitting.}
All classes in each dataset are divided into base ($\Cbase$) and new classes ($\Cnewt$) according to the number of per-class biflows.
Base classes, including benign traffic and the most frequent attacks, are used for base training, while the three classes with the fewest biflows (\ie the rarest attacks) are treated as $\Cnewt$ during incremental learning.
For simplicity, we consider a single incremental phase in the main experiments, while Appendix~\ref{appx:multi_incr} extends the analysis to multiple incremental phases.
Fig.~\ref{fig:dataset} illustrates the class split via the hatched bars.

\para{Evaluation Scenario.}
Following common practice in the literature~\cite{FSCIL-NIDS1, FSCIL-NIDS4, FSCIL-NIDS2, cao2025study}, we evaluate \nafn in the \fscil setting over $100$ classification tests.
%
Each test simulates a few-shot classification task involving base and new classes.
For each task, we construct a training set containing few labeled per-class biflows (\ie from $5$ to $10$) and a test set with $40$ biflows per class.
%
The training biflows of the base classes are retrieved from the memory buffer maintained by \memp, while the training biflows of the novel classes are drawn from the corresponding dataset.
%
This setup evaluates \nafn's ability to adapt to new attack classes while preserving previously learned knowledge in the \fscil setting.

\para{Hyperparameter Setup.}
For the Flow Processor (see Sec.~\ref{subsec:NIDS_workflow}),
in line with previous literature~\cite{FSCIL-NIDS2, monda2023bigdata},
we use the first $20$ packets of each biflow ($M_P = 20$) and extract $6$ features from each packet ($M_F=6$).
These features include: 
\begin{enumerate*}[label=(\textit{\roman*})]
    \item packet length, 
    \item packet direction, 
    \item TCP window size, 
    \item inter-arrival time, 
    \item time-to-live (TTL), and 
    \item TCP flags.
\end{enumerate*}
%
For the feature encoder, we set the feature embedding size $F$ to $320$.
The number of downsampling operation is configured as $L = 2$, with the corresponding channel sizes specified as $Ch_{0} = 16$, $Ch_{1} = 32$, and $Ch_{2} = 64$.
For training, we use a batch size of $64$ and the \emph{Adam} optimizer with a learning rate of $1\times10^{-4}$. Each model is trained for up to $100$~epochs with an early-stopping strategy (patience of $20$~epochs) to prevent overfitting. 
Hyperparameters are selected based on preliminary tuning analyses (omitted for brevity).
All random seeds are fixed to $0$ for reproducibility, and unless otherwise noted, the same hyperparameter configuration is applied across all datasets and models.%
\footnote{\para{Implementation Details.}
All experiments are conducted on a workstation running Debian Linux, equipped with an Intel® 13th Gen Core i7-13700KF CPU ($24$ threads, $16$ cores, base frequency $800$~MHz, turbo frequency up to $5.4$~GHz) and an NVIDIA® GeForce RTX~$4090$~D GPU, with sufficient main memory under a single NUMA node.
The software environment is based on Python~3.10 with the following key libraries: PyTorch~1.12.1, PyTorch Lightning~1.0.2, Learn2Learn~0.1.7, scikit-learn~1.3.2, and Matplotlib/Seaborn for visualization. All packages are installed from PyPI unless otherwise specified, and all experiments are executed within a controlled \texttt{conda} environment to ensure reproducibility. 
The source code is publicly available at \url{https://github.com/Luffy-2000/NAFN}.}



\definecolor{NNLow}{RGB}{255, 255, 255}
\definecolor{NNHigh}{RGB}{200, 80, 80}
\definecolor{LRLow}{RGB}{255, 255, 255}
\definecolor{LRHigh}{RGB}{70, 130, 180}

\newcommand{\setheatrange}[2]{%
  \pgfmathsetmacro{\HEATMIN}{#1}%
  \pgfmathsetmacro{\HEATMAX}{#2}%
  \xdef\HEATMIN{\HEATMIN}%
  \xdef\HEATMAX{\HEATMAX}%
}

\newcommand{\heatNN}[2]{%
  \begingroup
  \pgfmathsetmacro{\praw}{((#1-\HEATMIN)/max(0.0001,(\HEATMAX-\HEATMIN)))*100}%
  \pgfmathtruncatemacro{\pmix}{min(100, max(0, round(\praw)))}%
  \xdef\mixp{\pmix}
  \endgroup
  \cellcolor{NNHigh!\mixp!NNLow}#2%
}

\newcommand{\heatLR}[2]{%
  \begingroup
  \pgfmathsetmacro{\praw}{((#1-\HEATMIN)/max(0.0001,(\HEATMAX-\HEATMIN)))*100}%
  \pgfmathtruncatemacro{\pmix}{min(100, max(0, round(\praw)))}%
  \xdef\mixp{\pmix}%
  \endgroup
  \cellcolor{LRHigh!\mixp!LRLow}#2%
}

\begin{table*}[!ht]
\setlength{\tabcolsep}{4pt}
\renewcommand{\arraystretch}{1.22}
\centering
\caption{Evaluation results on (a) \CIC, (b) \EDGEIIOT, and (c) \IOTNID under different pretraining strategies and memory selection methods. \NN (left, \textcolor{NNHigh}{\textbf{red heatmap}}) and \LR (right, \textcolor{LRHigh}{\textbf{blue heatmap}}) are shown side-by-side. 
For each dataset and each classifier, the entry with the highest mean \textit{F1-New}, \textit{F1-Old}, and \textit{F1-All} is highlighted in bold; the corresponding standard deviation is bolded as well.}

\label{tab:Evaluation Results}
\resizebox{0.70\textwidth}{!}{%
\begin{tabular}{ll|cccc|cccc}
\toprule
\multicolumn{10}{c}{\textbf{(a) \texttt{CSE-CIC-IDS2018}}}\\
\midrule
\multirow{2}{*}{\textbf{Metric}} &
\multirow{2}{*}{\textbf{\texttt{Mem+}}} &
\multicolumn{4}{c|}{\textbf{\texttt{NN} Classifier}} &
\multicolumn{4}{c}{\textbf{\texttt{LR} Classifier}}\\
\cmidrule(lr){3-6}\cmidrule(lr){7-10}
 &  & W/o Pretrain. & Reconstr. & Contras. & Hybrid & W/o Pretrain. & Reconstr. & Contras. & Hybrid\\
\midrule
\setheatrange{84}{99}
\multirow{3}{*}{F1-New}
& Random  & \heatNN{93.11}{93.11\std{2.14}} & \heatNN{92.81}{92.81\std{2.15}} & \heatNN{92.02}{92.02\std{2.51}} & \heatNN{93.44}{\textbf{93.44\std{2.03}}} &
             \heatLR{90.02}{90.02\std{1.91}} & \heatLR{90.53}{90.53\std{1.95}} & \heatLR{84.15}{84.15\std{2.90}} & \heatLR{90.17}{90.17\std{2.38}}\\
& Rainbow & \heatNN{90.74}{90.74\std{2.36}} & \heatNN{92.96}{92.96\std{2.46}} & \heatNN{93.29}{93.29\std{2.26}} & \heatNN{93.35}{93.35\std{2.22}} &
             \heatLR{89.70}{89.70\std{2.04}} & \heatLR{90.25}{90.25\std{1.81}} & \heatLR{90.24}{90.24\std{1.68}} & \heatLR{90.64}{\textbf{90.64\std{1.60}}}\\
& Herding & \heatNN{90.51}{90.51\std{2.14}} & \heatNN{90.31}{90.31\std{2.25}} & \heatNN{91.22}{91.22\std{2.43}} & \heatNN{90.31}{90.31\std{2.23}} &
             \heatLR{85.31}{85.31\std{3.05}} & \heatLR{90.49}{90.49\std{1.13}} & \heatLR{89.01}{89.01\std{1.61}} & \heatLR{90.26}{90.26\std{1.08}}\\
\cline{1-2}\addlinespace[2pt]
\multirow{3}{*}{F1-Old}
& Random  & \heatNN{98.25}{98.25\std{0.53}} & \heatNN{98.26}{98.26\std{0.51}} & \heatNN{98.03}{98.03\std{0.57}} & \heatNN{98.40}{\textbf{98.40\std{0.46}}} &
             \heatLR{95.38}{95.38\std{1.75}} & \heatLR{96.00}{96.00\std{0.63}} & \heatLR{90.07}{90.07\std{2.89}} & \heatLR{95.95}{95.95\std{0.66}}\\
& Rainbow & \heatNN{97.23}{97.23\std{0.62}} & \heatNN{98.31}{98.31\std{0.53}} & \heatNN{98.23}{98.23\std{0.55}} & \heatNN{98.38}{98.38\std{0.52}} &
             \heatLR{95.90}{95.90\std{0.60}} & \heatLR{95.94}{95.94\std{0.58}} & \heatLR{96.14}{\textbf{96.14\std{0.63}}} & \heatLR{95.95}{95.95\std{0.62}}\\
& Herding & \heatNN{96.70}{96.70\std{0.61}} & \heatNN{96.99}{96.99\std{0.60}} & \heatNN{97.84}{97.84\std{0.60}} & \heatNN{96.49}{96.49\std{0.70}} &
             \heatLR{91.19}{91.19\std{3.53}} & \heatLR{95.94}{95.94\std{0.53}} & \heatLR{95.51}{95.51\std{0.67}} & \heatLR{95.86}{95.86\std{0.67}}\\
\cline{1-2}\addlinespace[2pt]
\multirow{3}{*}{F1-All}
& Random  & \heatNN{96.97}{96.97\std{0.81}} & \heatNN{96.89}{96.89\std{0.77}} & \heatNN{96.53}{96.53\std{0.95}} & \heatNN{97.16}{\textbf{97.16\std{0.74}}} &
             \heatLR{94.04}{94.04\std{1.71}} & \heatLR{94.63}{94.63\std{0.84}} & \heatLR{88.59}{88.59\std{2.83}} & \heatLR{94.50}{94.50\std{0.95}}\\
& Rainbow & \heatNN{95.61}{95.61\std{0.96}} & \heatNN{96.97}{96.97\std{0.92}} & \heatNN{97.00}{97.00\std{0.87}} & \heatNN{97.12}{97.12\std{0.87}} &
             \heatLR{94.35}{94.35\std{0.69}} & \heatLR{94.52}{94.52\std{0.67}} & \heatLR{94.66}{\textbf{94.66\std{0.64}}} & \heatLR{94.62}{94.62\std{0.66}}\\
& Herding & \heatNN{95.15}{95.15\std{0.87}} & \heatNN{95.32}{95.32\std{0.89}} & \heatNN{96.19}{96.19\std{0.97}} & \heatNN{94.95}{94.95\std{0.94}} &
             \heatLR{89.72}{89.72\std{3.34}} & \heatLR{94.58}{94.58\std{0.58}} & \heatLR{93.88}{93.88\std{0.79}} & \heatLR{94.46}{94.46\std{0.62}}\\
\midrule
\multicolumn{10}{c}{\textbf{(b) \texttt{Edge-IIoT}}}\\ \midrule
\setheatrange{70}{83}
\multirow{3}{*}{F1-New}
& Random  & \heatNN{79.07}{79.07\std{3.48}} & \heatNN{77.16}{77.16\std{3.55}} & \heatNN{74.35}{74.35\std{3.79}} & \heatNN{75.62}{75.62\std{3.57}} &
             \heatLR{82.33}{82.33\std{5.45}} & \heatLR{81.76}{81.76\std{5.71}} & \heatLR{48.84}{48.84\std{4.32}} & \heatLR{82.82}{\textbf{82.82\std{4.27}}}\\
& Rainbow & \heatNN{76.76}{76.76\std{4.41}} & \heatNN{77.46}{77.46\std{3.70}} & \heatNN{73.70}{73.70\std{3.91}} & \heatNN{75.44}{75.44\std{3.49}} &
             \heatLR{53.01}{53.01\std{11.99}} & \heatLR{80.03}{80.03\std{7.29}} & \heatLR{79.72}{79.72\std{7.19}} & \heatLR{79.97}{79.97\std{6.77}}\\
& Herding & \heatNN{82.19}{\textbf{82.19\std{1.93}}} & \heatNN{77.40}{77.40\std{1.82}} & \heatNN{77.50}{77.50\std{3.52}} & \heatNN{74.97}{74.97\std{2.49}} &
             \heatLR{82.20}{82.20\std{1.24}} & \heatLR{82.37}{82.37\std{1.05}} & \heatLR{82.37}{82.37\std{1.08}} & \heatLR{81.43}{81.43\std{1.22}}\\
\cline{1-2}\addlinespace[2pt]
\multirow{3}{*}{F1-Old}
& Random  & \heatNN{79.48}{79.48\std{0.87}} & \heatNN{79.51}{79.51\std{0.83}} & \heatNN{79.58}{79.58\std{0.81}} & \heatNN{79.41}{79.41\std{0.86}} &
             \heatLR{76.04}{76.04\std{1.32}} & \heatLR{75.74}{75.74\std{1.58}} & \heatLR{73.51}{73.51\std{1.26}} & \heatLR{76.53}{76.53\std{1.26}}\\
& Rainbow & \heatNN{79.40}{79.40\std{0.83}} & \heatNN{79.55}{79.55\std{0.80}} & \heatNN{79.45}{79.45\std{0.87}} & \heatNN{79.80}{79.80\std{0.84}} &
             \heatLR{75.87}{75.87\std{2.19}} & \heatLR{75.82}{75.82\std{1.64}} & \heatLR{78.44}{78.44\std{1.28}} & \heatLR{78.04}{78.04\std{1.71}}\\
& Herding & \heatNN{80.82}{\textbf{80.82\std{1.10}}} & \heatNN{78.87}{78.87\std{1.09}} & \heatNN{79.95}{79.95\std{1.05}} & \heatNN{79.06}{79.06\std{1.07}} &
             \heatLR{79.77}{\textbf{79.77\std{1.11}}} & \heatLR{79.69}{79.69\std{1.12}} & \heatLR{79.66}{79.66\std{1.11}} & \heatLR{79.35}{79.35\std{1.12}}\\
\cline{1-2}\addlinespace[2pt]
\multirow{3}{*}{F1-All}
& Random  & \heatNN{79.40}{79.40\std{1.11}} & \heatNN{79.04}{79.04\std{1.08}} & \heatNN{78.53}{78.53\std{1.12}} & \heatNN{78.65}{78.65\std{1.15}} &
             \heatLR{77.30}{77.30\std{1.16}} & \heatLR{76.94}{76.94\std{0.71}} & \heatLR{68.58}{68.58\std{0.76}} & \heatLR{77.79}{77.79\std{0.74}}\\
& Rainbow & \heatNN{78.88}{78.88\std{1.18}} & \heatNN{79.13}{79.13\std{1.11}} & \heatNN{78.30}{78.30\std{1.19}} & \heatNN{78.93}{78.93\std{1.03}} &
             \heatLR{71.30}{71.30\std{3.19}} & \heatLR{76.66}{76.66\std{0.85}} & \heatLR{78.70}{78.70\std{1.29}} & \heatLR{78.42}{78.42\std{0.91}}\\
& Herding & \heatNN{81.09}{\textbf{81.09\std{1.11}}} & \heatNN{78.58}{78.58\std{1.01}} & \heatNN{79.46}{79.46\std{1.27}} & \heatNN{78.24}{78.24\std{1.11}} &
             \heatLR{80.25}{\textbf{80.25\std{0.98}}} & \heatLR{80.22}{80.22\std{1.00}} & \heatLR{80.20}{80.20\std{0.97}} & \heatLR{79.77}{79.77\std{1.00}}\\
\midrule
\multicolumn{10}{c}{\textbf{(c) \texttt{IoT-NID}}}\\ \midrule
\setheatrange{77}{97}
\multirow{3}{*}{F1-New}
& Random  & \heatNN{84.83}{84.83\std{3.15}} & \heatNN{84.89}{84.89\std{3.21}} & \heatNN{85.03}{85.03\std{3.34}} & \heatNN{84.08}{84.08\std{3.51}} &
             \heatLR{83.79}{83.79\std{3.39}} & \heatLR{82.98}{82.98\std{5.01}} & \heatLR{86.19}{\textbf{86.19\std{2.58}}} & \heatLR{85.43}{85.43\std{3.04}}\\
& Rainbow & \heatNN{84.20}{84.20\std{3.57}} & \heatNN{83.71}{83.71\std{3.72}} & \heatNN{86.05}{\textbf{86.05\std{2.84}}} & \heatNN{84.40}{84.40\std{3.48}} &
             \heatLR{83.26}{83.26\std{3.49}} & \heatLR{83.25}{83.25\std{4.60}} & \heatLR{84.51}{84.51\std{3.98}} & \heatLR{84.48}{84.48\std{3.44}}\\
& Herding & \heatNN{79.86}{79.86\std{3.85}} & \heatNN{77.78}{77.78\std{5.05}} & \heatNN{81.01}{81.01\std{3.83}} & \heatNN{83.46}{83.46\std{3.04}} &
             \heatLR{80.39}{80.39\std{4.72}} & \heatLR{81.24}{81.24\std{4.83}} & \heatLR{81.45}{81.45\std{4.59}} & \heatLR{83.31}{83.31\std{4.11}}\\
\cline{1-2}\addlinespace[2pt]
\multirow{3}{*}{F1-Old}
& Random  & \heatNN{96.19}{96.19\std{0.92}} & \heatNN{96.46}{96.46\std{0.94}} & \heatNN{96.03}{96.03\std{1.08}} & \heatNN{96.87}{96.87\std{1.00}} &
             \heatLR{96.03}{96.03\std{0.77}} & \heatLR{96.34}{96.34\std{0.77}} & \heatLR{96.33}{96.33\std{0.73}} & \heatLR{96.23}{96.23\std{0.80}}\\
& Rainbow & \heatNN{96.19}{96.19\std{0.89}} & \heatNN{96.59}{96.59\std{0.92}} & \heatNN{97.11}{\textbf{97.11\std{0.78}}} & \heatNN{96.95}{96.95\std{0.72}} &
             \heatLR{95.94}{95.94\std{0.91}} & \heatLR{96.59}{\textbf{96.59\std{0.73}}} & \heatLR{96.54}{96.54\std{0.67}} & \heatLR{96.56}{96.56\std{0.78}}\\
& Herding & \heatNN{92.96}{92.96\std{2.20}} & \heatNN{91.64}{91.64\std{2.92}} & \heatNN{88.25}{88.25\std{2.44}} & \heatNN{94.50}{94.50\std{1.33}} &
             \heatLR{94.57}{94.57\std{2.81}} & \heatLR{95.62}{95.62\std{2.03}} & \heatLR{91.32}{91.32\std{4.53}} & \heatLR{95.88}{95.88\std{1.67}}\\
\cline{1-2}\addlinespace[2pt]
\multirow{3}{*}{F1-All}
& Random  & \heatNN{92.78}{92.78\std{1.39}} & \heatNN{92.99}{92.99\std{1.36}} & \heatNN{92.73}{92.73\std{1.53}} & \heatNN{93.03}{93.03\std{1.55}} &
             \heatLR{92.36}{92.36\std{1.41}} & \heatLR{92.33}{92.33\std{1.68}} & \heatLR{93.29}{\textbf{93.29\std{1.12}}} & \heatLR{92.99}{92.99\std{1.32}}\\
& Rainbow & \heatNN{92.59}{92.59\std{1.48}} & \heatNN{92.73}{92.73\std{1.46}} & \heatNN{93.79}{\textbf{93.79\std{1.22}}} & \heatNN{93.18}{93.18\std{1.29}} &
             \heatLR{92.13}{92.13\std{1.54}} & \heatLR{92.59}{92.59\std{1.66}} & \heatLR{92.93}{92.93\std{1.50}} & \heatLR{92.93}{92.93\std{1.33}}\\
& Herding & \heatNN{89.03}{89.03\std{2.40}} & \heatNN{87.49}{87.49\std{3.36}} & \heatNN{86.08}{86.08\std{2.37}} & \heatNN{91.19}{91.19\std{1.69}} &
             \heatLR{90.31}{90.31\std{3.05}} & \heatLR{91.30}{91.30\std{2.41}} & \heatLR{88.36}{88.36\std{3.86}} & \heatLR{92.11}{92.11\std{2.04}}\\
\bottomrule
\end{tabular}
}
\end{table*}

\subsection{Performance Metrics}
\label{sec:metrics}

In this paper, we use two performance metrics: 
the
\emph{macro F1-score} (F1 in short) 
and \emph{pAUC} (Partial Area Under the Curve).
The definitions are as follows.

\para{Macro F1-score.}
It evaluates the attack classification capabilities
(\ie multi-class classification of benign and attack-specific traffic) of the NIDS.
It is the harmonic mean of per-class $\mathtt{Precision}$ and $\mathtt{Recall}$ (or True Positive Rate, $\mathtt{TPR}$), averaged equally across both benign and malicious classes.
This ensures a fair evaluation that is not skewed by class imbalance. 
It is defined as:
\begin{equation}
\mathtt{F1}\text{-}\mathtt{score} = \frac{1}{|\mathcal{C}|} \sum_{i=1}^{|\mathcal{C}|} \frac{2 \times \mathtt{Precision}_i \times \mathtt{Recall}_i}{\mathtt{Precision}_i + \mathtt{Recall}_i},
\end{equation}
where $\mathcal{C}$ denotes the set of all classes.
%

\para{pAUC.}
The Area Under the Curve (AUC) evaluates the detection capabilities of the NIDS, \ie its ability to discriminate between benign and malicious traffic in a binary classification task, by summarizing performance across all possible decision thresholds.
The Receiver Operating Characteristic (ROC) curve plots the $\mathtt{TPR}$ 
against the False Positive Rate ($\mathtt{FPR}$, which represents false alarms), and the AUC is the area under this curve.
While the AUC provides a comprehensive view of classifier performance, it includes high-$\mathtt{FPR}$ regions, which are undesirable in practical NIDS deployments where excessive false positives may lead to blocking benign traffic.
To better reflect realistic operational constraints, we therefore adopt the standardized \emph{pAUC}, which restricts the AUC to a low-$\mathtt{FPR}$ range.
Specifically, we consider a maximum $\mathtt{FPR}$ of $r = 0.01~(1\%)$, viz.:
\begin{equation}    
\mathtt{pAUC} = \frac{1}{2}\left(1 + \frac{\int_0^{r} \mathtt{TPR}(\mathtt{FPR})\,d\mathtt{FPR} - r^2/2}{r - r^2/2}\right),
\end{equation}
which standardizes the partial area such that $\mathtt{pAUC}=0.5$ indicates random performance and $\mathtt{pAUC}=1.0$ indicates perfect classification.
In the following, F1 and pAUC are averaged over $100$ classification tasks and reported with standard deviations to assess stability.

\section{Experimental Results}
\label{sec:exp_results}

In this section, we present the empirical evaluation of \nafn. 
Specifically, Sec.~\ref{subsec:pretrain_Memory} investigates different pretraining and memory selection strategies.
Sec.~\ref{subsec:sen_shots} analyzes the sensitivity of the NIDS to the number of shots (\ie labeled biflows from novel attacks).
Sec.~\ref{subsec:time} examines training and inference efficiency.
Secs.~\ref{subsec:Noisy}, \ref{subsec:Visual} and \ref{subsec:Time-to-Insight} study robustness and feature separability under label noise.
Finally, Sec.~\ref{subsec:Crossnetwork} evaluates the cross-network incremental detection performance of \nafn under label noise.


\subsection{Evaluation Results under Different Pretraining Strategies and Memory Selection Methods}
\label{subsec:pretrain_Memory}


This experiment addresses \qone~(Sec.~\ref{sec:problem}), namely \textit{whether \nafn can adapt to new attack classes while retaining previously learned knowledge.}
To this end, we evaluate different combinations of pretraining and memory selection strategies (\memp) across three datasets to determine the optimal configuration 
under few-shot conditions.
%
In the incremental learning phase, new attack classes from the same dataset as the base classes are introduced, each containing $K = 10$ labeled biflows.
Table~\ref{tab:Evaluation Results} reports F1-scores for new (\textit{F1-New}), old (\textit{F1-Old}), and all classes (\textit{F1-All}).

Overall, the \NN classifier outperforms the \LR by an average of \textbf{1.37\%} \textit{F1-All}.  
Therefore, the following analyses focus on the results obtained with
the \NN classifier, 
while observations regarding \LR will only be briefly mentioned.

\para{Performance on \texttt{CSE-CIC-IDS2018}.}
%
%
\emph{Hybrid + Random} achieves
the best overall performance, with \textit{F1-New}=$93.44\%$, \textit{F1-Old}=$98.40\%$, and \textit{F1-All}=$97.16\%$ (see Table~\ref{tab:Evaluation Results}(a)).  
\CIC contains complex and fine-grained traffic patterns, where Hybrid pretraining provides richer representations by combining reconstruction and contrastive objectives, while Random memory selection avoids the bias of ranking-based heuristics and keeps the retained biflows sufficiently diversified for adaptation.


\para{Performance on \texttt{Edge-IIoT}.}
The best performance is obtained through \emph{W/o pretraining + Herding} configuration, achieving \textit{F1-New}=$82.19\%$, \textit{F1-Old}=$80.82\%$, and \textit{F1-All}=$81.09\%$ (see Table~\ref{tab:Evaluation Results}(b)). 
The dataset contains diverse and noisy traffic behaviors, so selecting representative biflows via Herding preserves overall stability.
Other pretraining strategies show limited gains or even degradation in \textit{F1-New}.

\para{Performance on \texttt{IoT-NID}.}
%
\emph{Contrastive + Rainbow} achieves the best overall performance, 
with \textit{F1-New}=$86.05\%$, \textit{F1-Old}=$97.11\%$, and \textit{F1-All}=$93.79\%$ (see Table~\ref{tab:Evaluation Results}(c)).  
Contrastive pretraining yields discriminative features, while Rainbow memory selection enhances exemplar diversity, jointly supporting both new-class adaptation and old-class retention.

\takeaway{The optimal configuration varies across datasets due to differences in traffic characteristics.
On \CIC, \textbf{Hybrid + Random} (\NN) and \textbf{Contrastive + Rainbow} (\LR) perform best, benefiting from richer representations and informative memory selection.
On \EDGEIIOT, \textbf{W/o pretraining + Herding} dominates for both classifiers, confirming that exemplar diversity outweighs feature initialization on this dataset.
On \IOTNID, contrastive pretraining is the key driver, with \textbf{Contrastive + Rainbow} (\NN) and \textbf{Contrastive + Random} (\LR) yielding the best trade-off between new- and old-class performance.
In the following experiments, we will adopt these optimal configurations.
}


\subsection{Sensitivity Analysis to the Number of Shots}
\label{subsec:sen_shots}

\begin{figure}[!t]
    \centering
    \caption{Comparison of \textit{F1-All} [\%] for \NN and \LR classifiers across different numbers of shots ($K$) on the three datasets.
    The y-axis is zoomed in between $60\%$ and $100\%$ for clarity.
    }
    \begin{minipage}[t]{1.0\hsize}
        \centering
        \includegraphics[width=\linewidth]{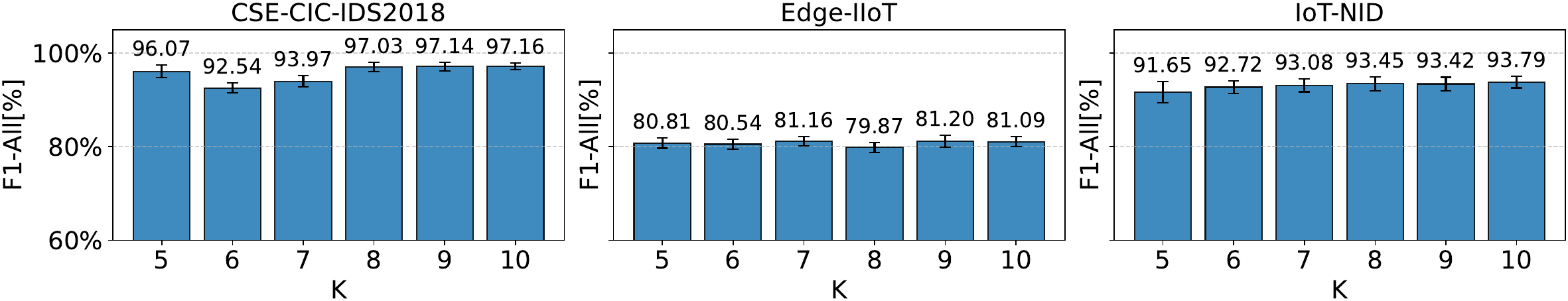}
        \caption*{(a) \NN classifier}
        \label{fig:shot_analysis_nn}
    \end{minipage}
    \begin{minipage}[t]{1.0\hsize}
        \centering
        \includegraphics[width=\linewidth]{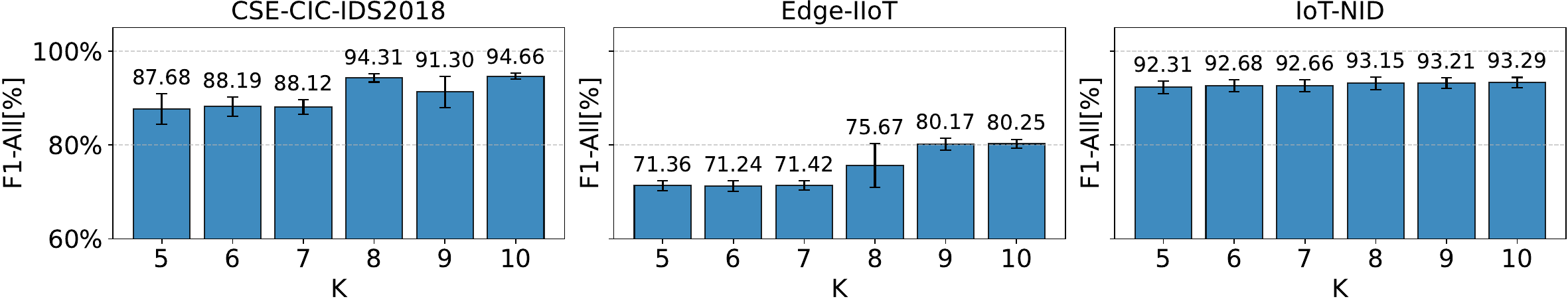}
        \caption*{(b) \LR classifier}
        \label{fig:shot_analysis_lr}
    \end{minipage}
    \label{fig:shot_analysis}
\end{figure}

The current section further explores \qone{} by investigating \emph{how the number of shots} (\ie the number of labeled biflows per new attack class) \emph{affects the adaptation ability of \nafn.}
We assess the stability and scalability of the model as $K$ increases from $5$ to $10$ shots,
providing insight into the trade-off
between sample availability and attack classification performance.
As in the previous section, new and base classes belong to the same dataset, and we adopt the configuration selected in Sec.~\ref{subsec:pretrain_Memory} for each dataset.
%
Fig.~\ref{fig:shot_analysis}(a) and (b) show the \textit{F1-All} achieved by \nafn using the \NN and \LR classifiers, respectively, on the three datasets for varying numbers of shots.
%

For the \NN classifier (Fig.~\ref{fig:shot_analysis}(a)), performance remains stable across different $K$, with a slight improvement as $K$ increases.
On \CIC, \textit{F1-All} stays above $92\%$, rising marginally from $5$ to $10$ shots.
A similar trend appears on \IOTNID ($92\%$–$94\%$), while \EDGEIIOT shows the lowest results ($\approx80\%$), peaking at $81.2\%$ at $K=9$.

For the \LR classifier (Fig.~\ref{fig:shot_analysis}(b)), the benefit of larger $K$ is more evident.
On \CIC, performance improves from $87\%$–$88\%$ ($5$–$7$ shots) to nearly $94\%$ ($9$–$10$ shots), a gain of $\approx7\%$.
On \IOTNID, the \textit{F1-All} remains steady ($92\%$–$93\%$), while on \EDGEIIOT, the effect is stronger: \textit{F1-All} increases from $\approx71\%$ ($5$–$7$ shots) to $80\%$ ($9$–$10$ shots), a gain exceeding $9\%$ \textit{F1-All}.

\takeaway{As expected, the sensitivity analysis shows that increasing the number of shots improves performance for both classifiers, though the degree of improvement varies depending on the dataset.
%
For \NN, the gains are relatively small yet consistent, whereas for \LR, the effect is more substantial, especially on \CIC and \EDGEIIOT. 
These findings suggest that while more shots are beneficial overall, the sensitivity to sample size is highly dataset- and classifier-dependent.
}

\subsection{Training and Inference Efficiency}
\label{subsec:time}

\begin{table}[!ht]
\centering
\caption{Computational overhead across datasets. 
\textcolor{lr}{\textbf{Base}} and \textcolor{lb}{\textbf{incremental}} training times are in seconds; \textcolor{dg}{\textbf{inference}} time is in microseconds per biflow.}
\label{tab:training_times}
\setlength{\tabcolsep}{4pt}
\renewcommand{\arraystretch}{1.05}
\resizebox{\linewidth}{!}{%
\begin{tabular}{lccc}
\toprule
& \textbf{\texttt{CSE-CIC-IDS2018}} & \textbf{\texttt{Edge-IIoT}} & \textbf{\texttt{IoT-NID}} \\
\midrule
\rowcolor{lr}
\multicolumn{4}{l}{\textbf{\textit{Base Training Time ($\bm{s}$)}}} \\
\midrule
\quad W/o Pretrain   & 139.41\std{0.46}       & 740.52\std{1.97}       & 456.39\std{1.92} \\
\quad Reconstruction & 214.95\std{0.54}       & 832.19\std{2.41}       & 490.81\std{1.27} \\
\quad Contrastive    & 1{,}352.22\std{155.49} & 4{,}669.52\std{969.00} & 1{,}066.13\std{167.92} \\
\quad Hybrid         & 1{,}400.56\std{113.37} & 4{,}125.98\std{752.35} & 1{,}055.94\std{152.43} \\
\midrule
\rowcolor{lb}
\multicolumn{4}{l}{\textbf{\textit{Incremental Training Time ($\bm{s}$~/~increment)}}} \\
\midrule
\quad Per increment  & 117.21\std{8.50}       & 666.75\std{193.44}     & 346.19\std{92.81} \\
\midrule
\rowcolor{lg}
\multicolumn{4}{l}{\textbf{\textit{Inference Time ($\bm{\mu s}$~/~biflow)}}} \\
\midrule
\quad \NN            & 8.45\std{10.20}        & 7.45\std{5.73}         & 8.62\std{9.42} \\
\quad \LR            & 32.47\std{15.21}       & 25.54\std{17.42}       & 33.25\std{16.72} \\
\bottomrule
\end{tabular}%
}
\end{table}

\begin{figure*}[!t]
    \centering
    \caption{Performance comparison of different noise detection methods and noisy-label learning baselines under a fixed denoising ratio $\rho=50\%$ and varying noise rates $\eta = 0\%–50\%$ on \CIC, \EDGEIIOT, and \IOTNID with \NN and \LR classifiers. Plots report the F1-score for all classes (\textit{F1-All}), new classes (\textit{F1-New}), and old classes (\textit{F1-Old}).
    }
    \begin{minipage}[t]{0.93\linewidth}
        \centering
        \includegraphics[width=\linewidth]{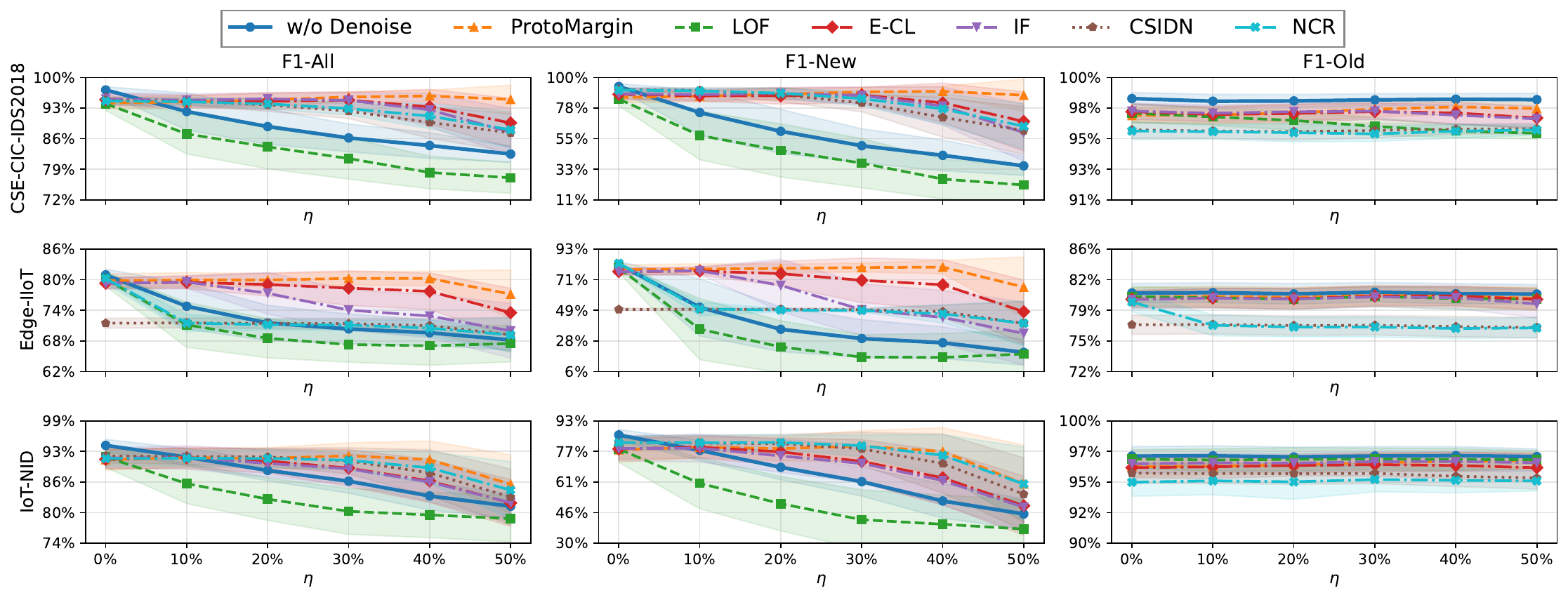}
        \caption*{(a) \NN classifier}
    \end{minipage}
    \hfill
    
    \begin{minipage}[t]{0.93\linewidth}
        \centering
        \includegraphics[width=\linewidth, trim={0 0 0 1.25cm},clip]{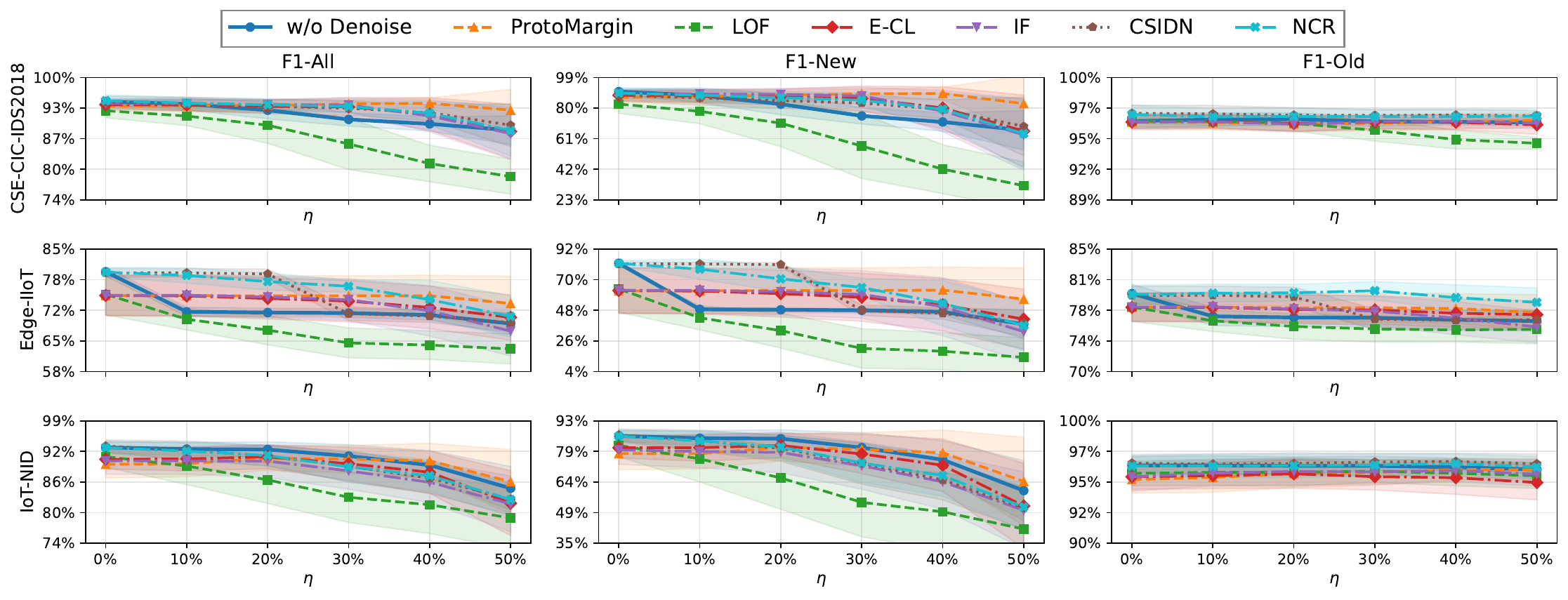}
        \caption*{(b) \LR classifier}
    \end{minipage}
    \label{fig:noiselabel_comparison}
\end{figure*}

To complement the analyses on \qone, we measure the \emph{computational efficiency of \nafn during both the base training and incremental phases.}
All timings are obtained with the per-dataset configuration selected in Sec.~\ref{subsec:pretrain_Memory}.
Furthermore, we evaluate per-biflow inference latency to assess \nafn's suitability for real-time NIDS deployment.
%
%
We report the \emph{mean \(\pm\) standard deviation} in seconds (for training) and microseconds (for inference) across the three considered datasets.

Table~\ref{tab:training_times} shows that
the incremental training phase consistently requires far less time than any of the pretraining procedures, regardless of the dataset or strategy. 
For instance, on \CIC, the heaviest pretraining strategy (Hybrid) requires over $1,400.00s$, while student training completes in just $117.21s$.
%
On \EDGEIIOT, which is the largest dataset, contrastive pretraining takes approximately $4,669.00s$, whereas incremental training finishes in only $666.75s$,
which is less than $15\%$ of the cost. 
%
A similar pattern is observed on \IOTNID, where even the lightest pretraining (W/o Pretraining, $456.39s$) still exceeds the adaptation time ($346.19s$).
%
Beyond the training efficiency, the bottom section of Table~\ref{tab:training_times} highlights the real-time operational capacity of \nafn. 
Specifically, classifying a single incoming biflow requires merely $7.45$--$8.62\mu s$ using the \NN classifier, and $25.54$--$33.25\mu s$ with the \LR classifier. 
This microsecond-level latency yields a theoretical throughput capable of processing tens to hundreds of thousands of biflows per second,
ensuring that \nafn will not become a bottleneck in high-speed networks.

\takeaway{These findings underscore the practicality of the teacher–student incremental framework, \rfs.
Once a teacher model has been pretrained, adapting a student model to new attack classes in the incremental training phase
is both fast and computationally lightweight. 
%
Coupled with the low inference latency per biflow, this dual efficiency makes the \nafn suitable for real-world NIDS deployments.
}

\begin{figure*}[!tbp]
\centering
\caption{Confusion matrix of \NN and \LR classifiers on \CIC, \EDGEIIOT, and \IOTNID with $50\%$ label noise. Each row corresponds to one dataset. The first two columns show
results for the \NN classifier
without and with \nafn (\PM as noise detection method), respectively; the last two columns present the corresponding results for the \LR classifier. 
We report $\mathtt{TPR}$ at a 1\% $\mathtt{FPR}$
constraint above each subfigure. 
The theoretical upper bound (\ie the maximum achievable $\mathtt{TPR}_{\mathtt{FPR}=1\%}$ when trained on 0\% noise) is provided in the sub-captions.}

\begin{minipage}[t]{0.22\textwidth}
\centering
{\small $\mathtt{TPR}_{\mathtt{FPR}=1\%} = 78.02\%$}
\includegraphics[width=\linewidth]{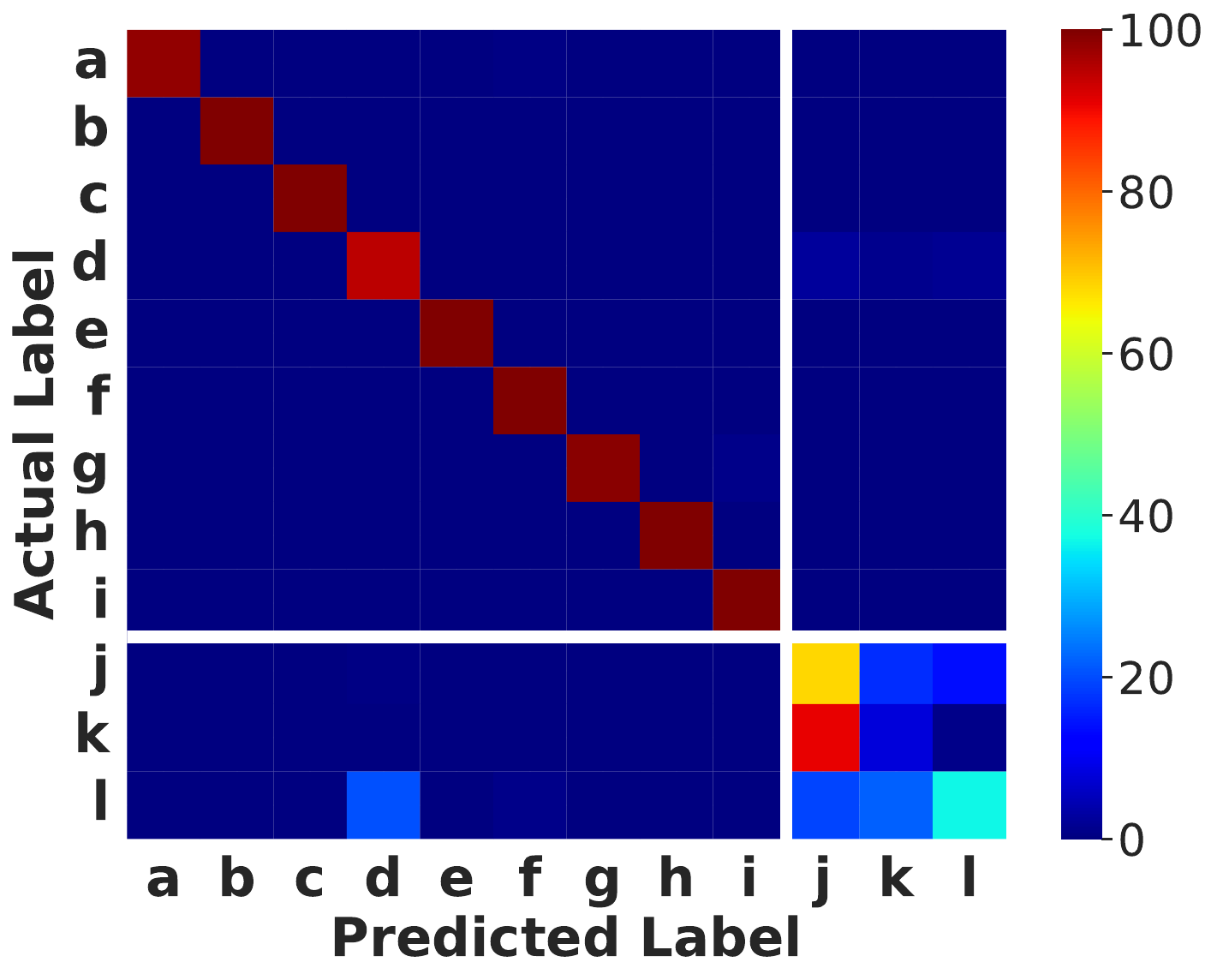}
\caption*{\NN - W/o \nafn}
\end{minipage}
\begin{minipage}[t]{0.22\textwidth}
\centering
{\small $\mathtt{TPR}_{\mathtt{FPR}=1\%} = 79.68\%$}
\includegraphics[width=\linewidth]{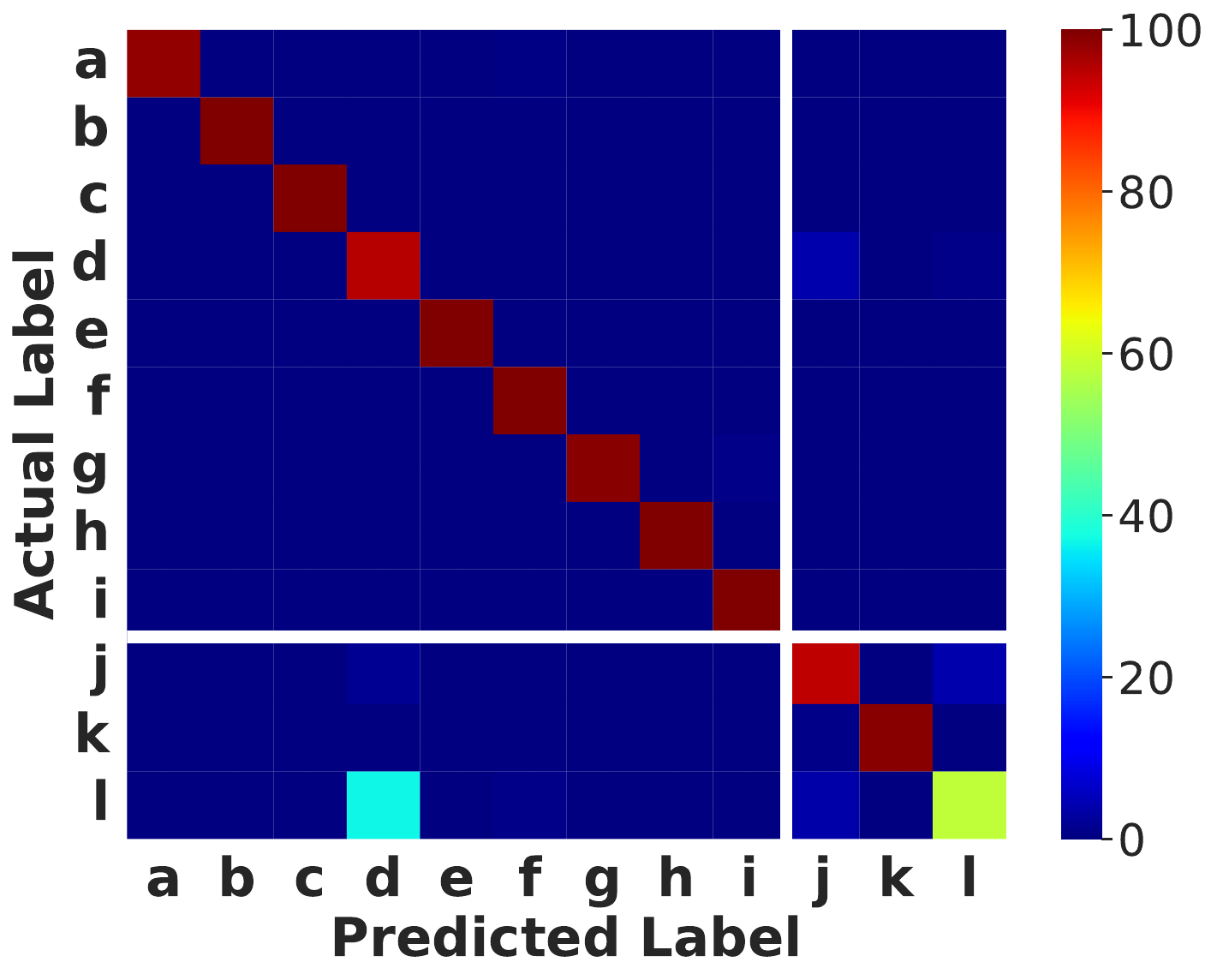}
\caption*{\NN - \nafn (\PM)}
\end{minipage}
\begin{minipage}[t]{0.22\textwidth}
\centering
{\small $\mathtt{TPR}_{\mathtt{FPR}=1\%} = 93.61\%$}
\includegraphics[width=\linewidth]{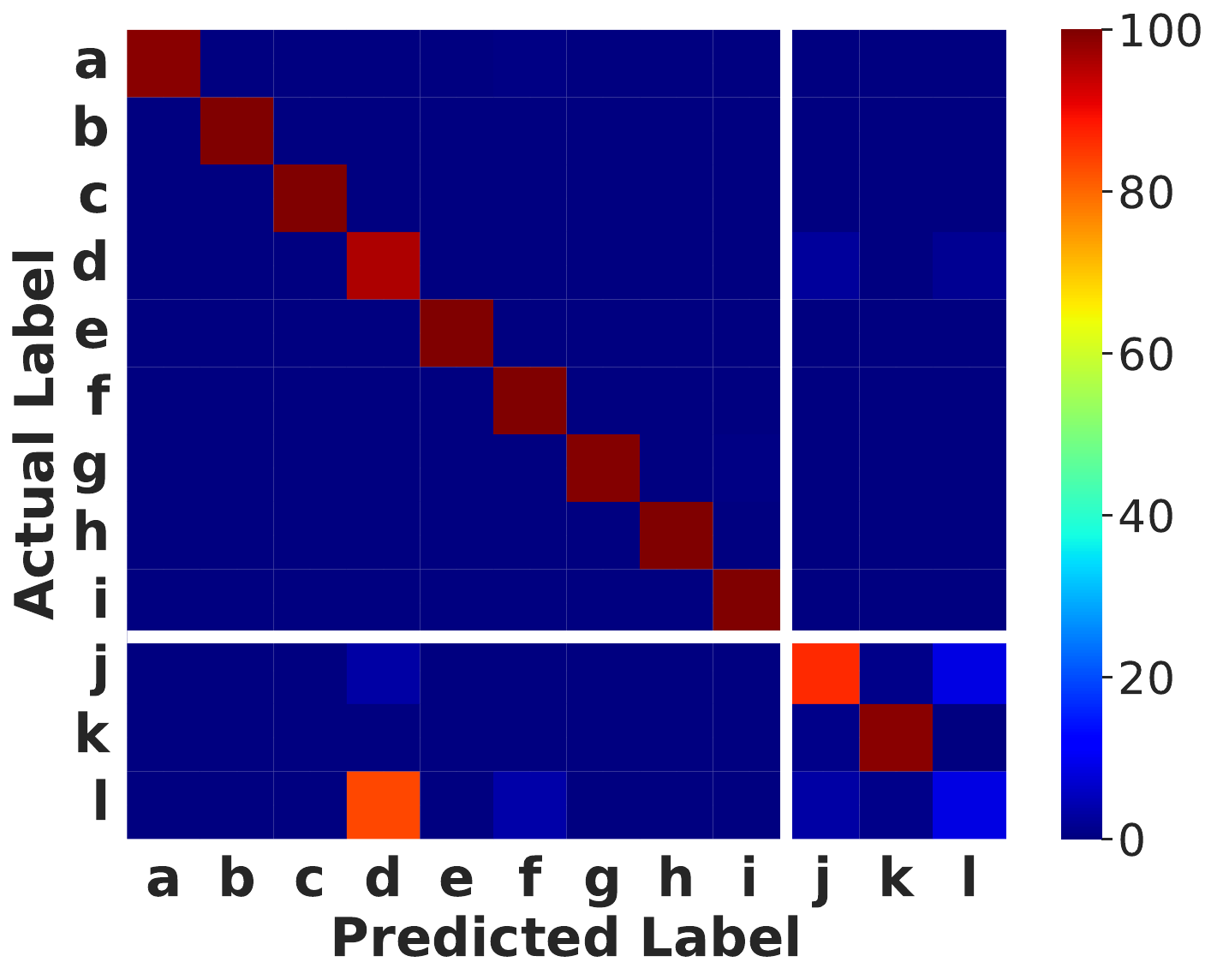}
\caption*{\LR - W/o \nafn}
\end{minipage}
\begin{minipage}[t]{0.22\textwidth}
\centering
{\small $\mathtt{TPR}_{\mathtt{FPR}=1\%} = 93.76\%$}
\includegraphics[width=\linewidth]{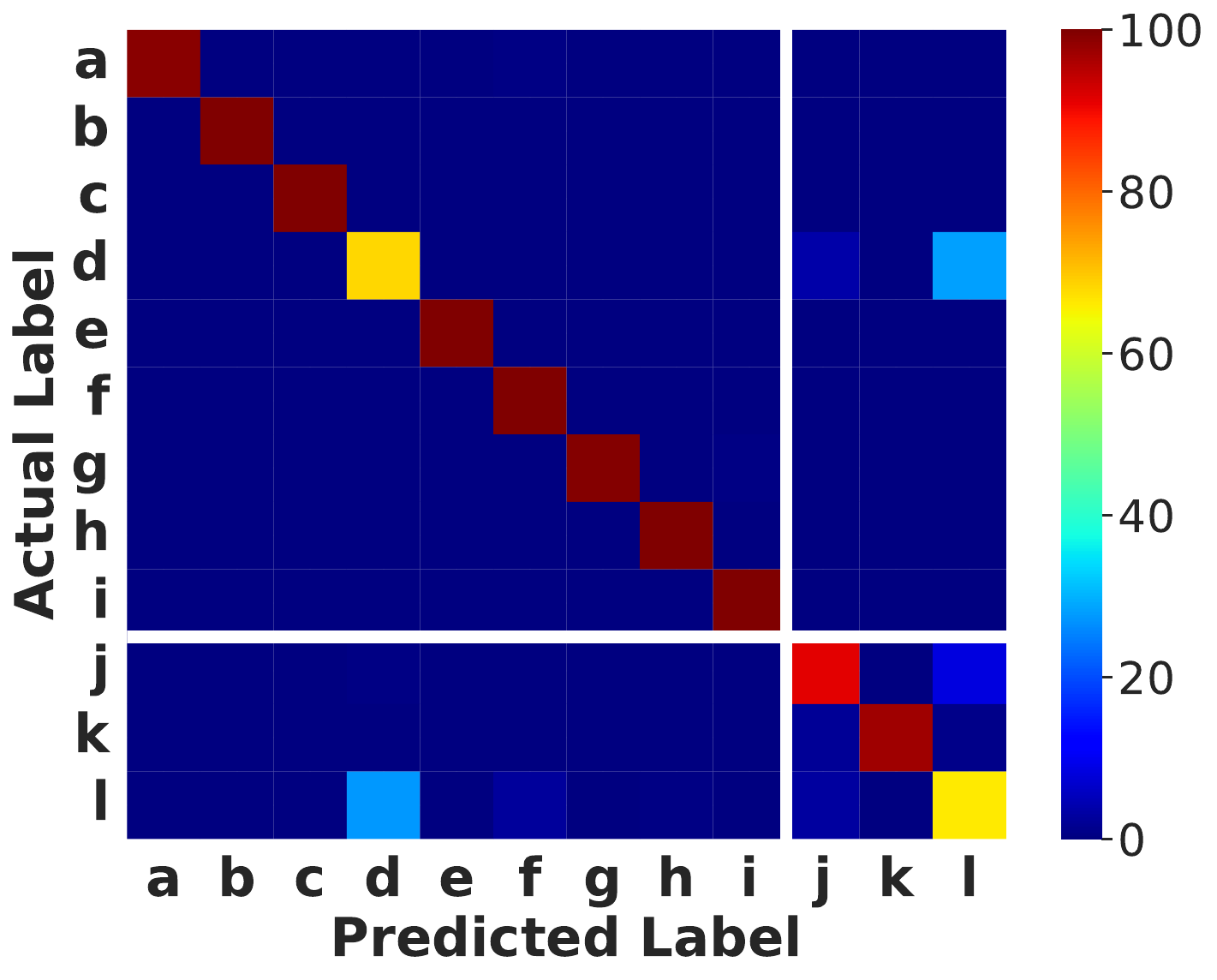}
\caption*{\LR - \nafn (\PM)}
\end{minipage}
\caption*{(a) \CIC. 0\% noise bounds: \NN: 82.13\%, \LR: 96.64\% \scriptsize{\textbf{Label encoding:} \(\mathbf{a}\) = DoS GoldenEye, \(\mathbf{b}\) = DoS Hulk, \(\mathbf{c}\) = DDoS HOIC, \(\mathbf{d}\) = Bot, \(\mathbf{e}\) = Bruteforce-SSH, \(\mathbf{f}\) = Benign, \(\mathbf{g}\) = DDoS LOIC-HTTP, \(\mathbf{h}\) = DOS Slowloris,  \(\mathbf{i}\) = Bruteforce-Web, \(\mathbf{j}\) = \textit{DDoS LOIC-UDP}, \(\mathbf{k}\) = \textit{Bruteforce-XSS},  \(\mathbf{l}\) = \textit{SQL Injection}.}}
\vspace{0.5em}

\begin{minipage}[t]{0.22\textwidth}
\centering
{\small $\mathtt{TPR}_{\mathtt{FPR}=1\%} = 76.09\%$}
\includegraphics[width=\linewidth]{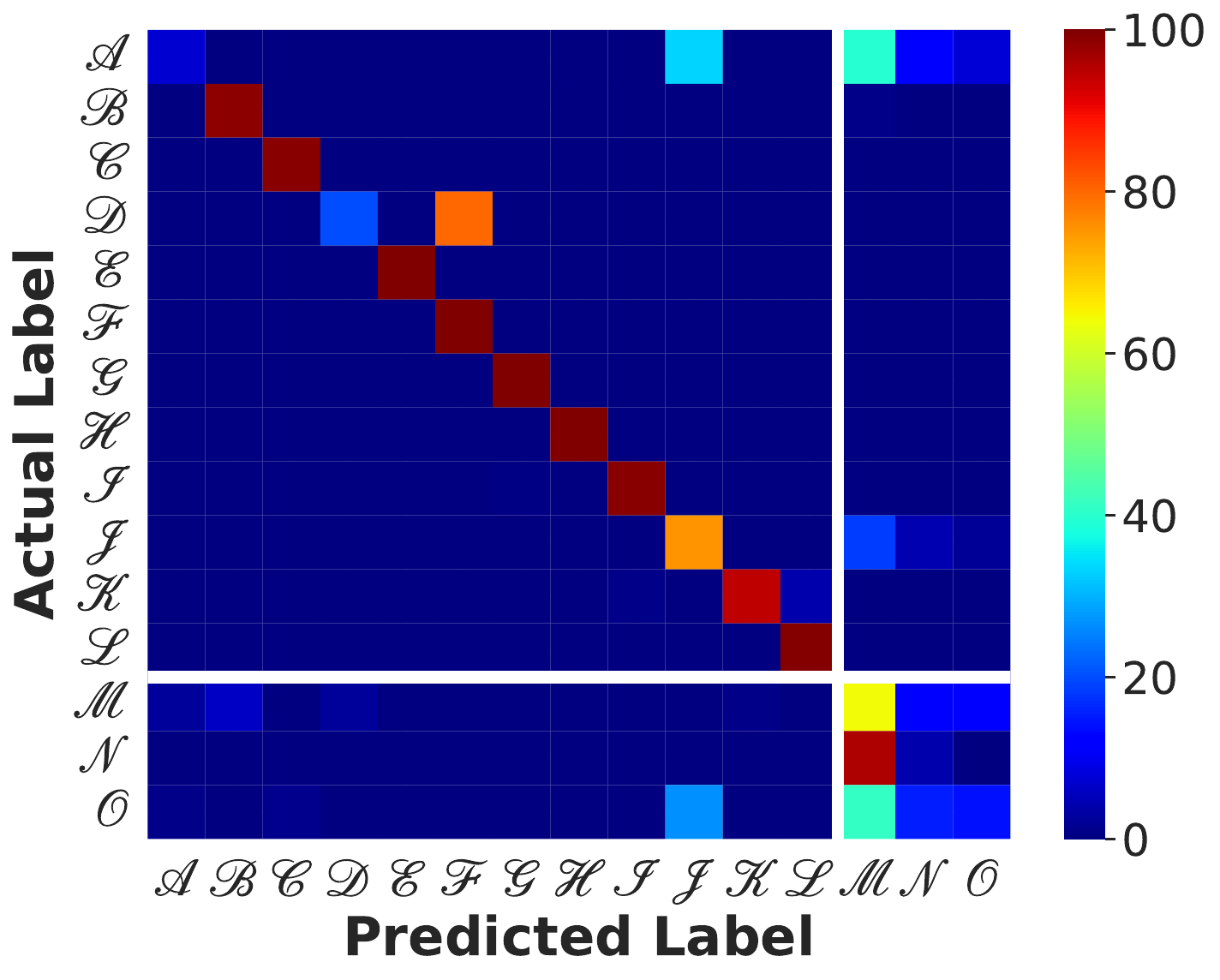}
\caption*{\centering \NN - W/o \nafn} 
\end{minipage}
\begin{minipage}[t]{0.22\textwidth}
\centering
{\small $\mathtt{TPR}_{\mathtt{FPR}=1\%} = 84.63\%$}
\includegraphics[width=\linewidth]{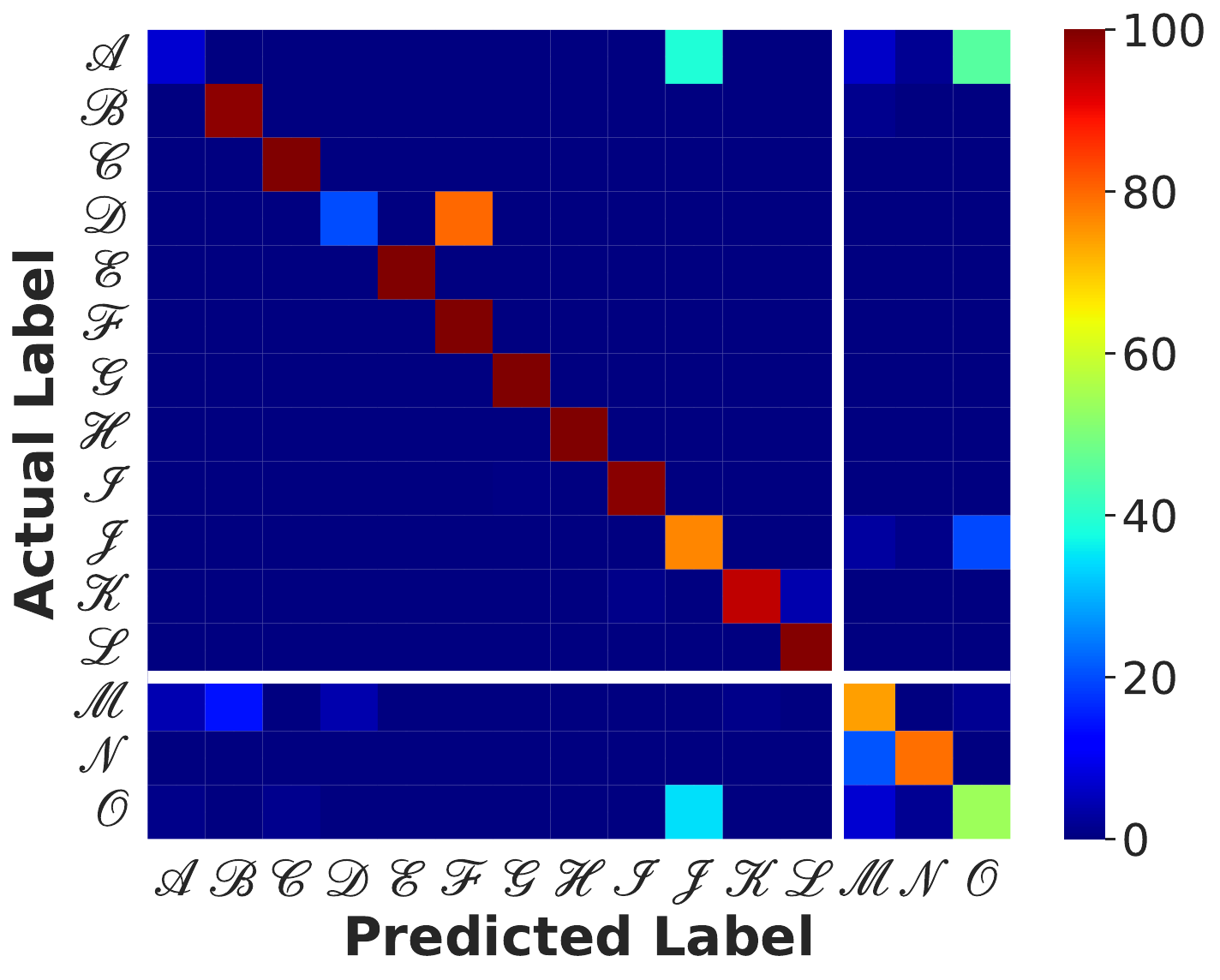}
\caption*{\NN - \nafn (\PM)}
\end{minipage}
\begin{minipage}[t]{0.22\textwidth}
\centering
{\small $\mathtt{TPR}_{\mathtt{FPR}=1\%} = 98.00\%$}
\includegraphics[width=\linewidth]{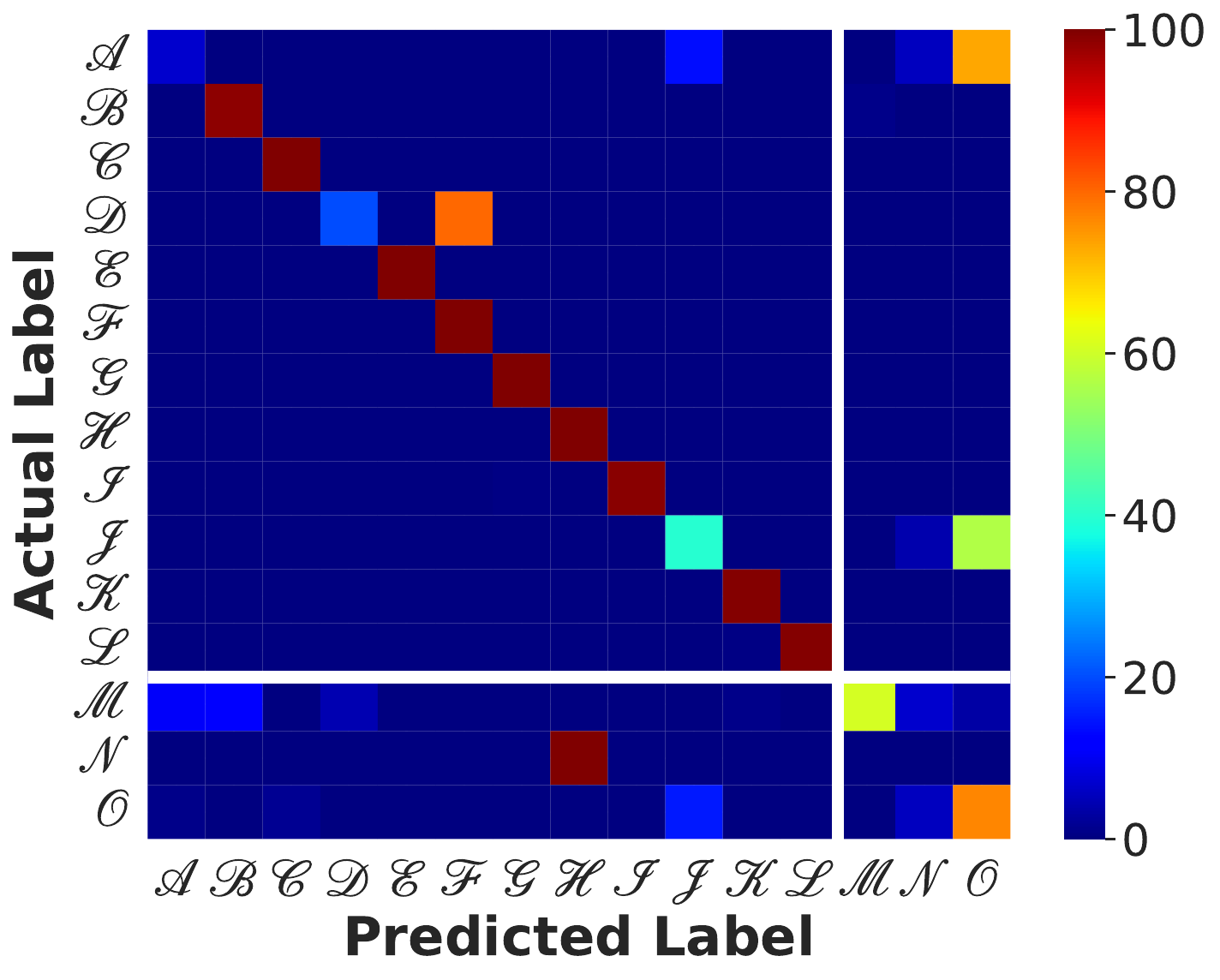}
\caption*{\LR - W/o \nafn}
\end{minipage}
\begin{minipage}[t]{0.22\textwidth}
\centering
{\small $\mathtt{TPR}_{\mathtt{FPR}=1\%} = 98.56\%$}
\includegraphics[width=\linewidth]{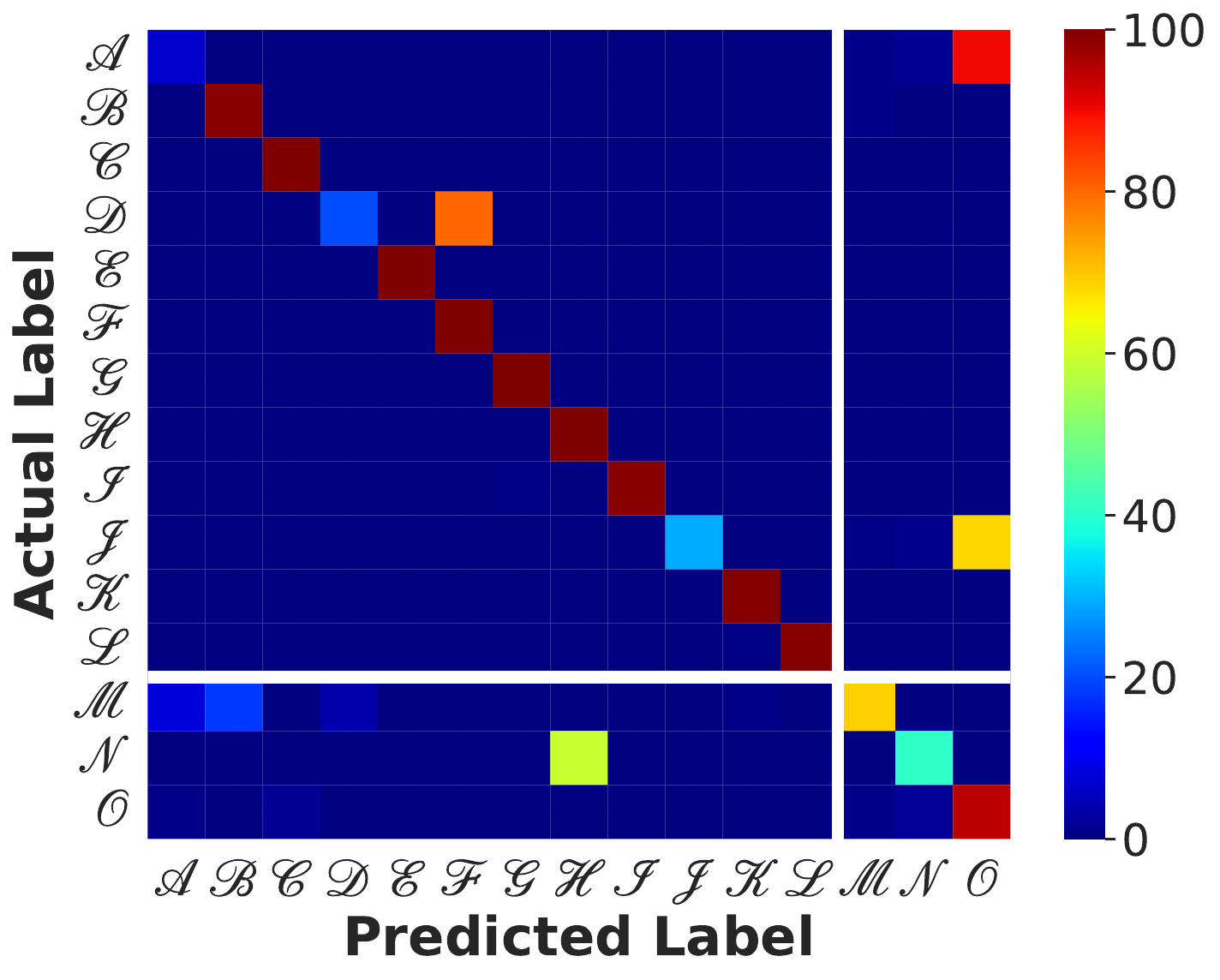}
\caption*{\LR - \nafn (\PM)}
\end{minipage}
\caption*{(b) \EDGEIIOT. 0\% noise bounds: \NN: 84.68\%, \LR: 98.81\%. \scriptsize{\textbf{Label encoding:} \(\mathscr{A}\) = Benign, \(\mathscr{B}\) = Password Attack, \(\mathscr{C}\) = DDoS UDP Flooding, \(\mathscr{D}\) = DDoS ICMP Flooding, \(\mathscr{E}\) = DDoS HTTP Flooding, \(\mathscr{F}\) = DDoS TCP SYN Flooding, \(\mathscr{G}\) = Port Scanning, \(\mathscr{H}\) = Uploading Attack, \(\mathscr{I}\) = SQL Injection, \(\mathscr{J}\) = Vulnerability Scanner, \(\mathscr{K}\) = XSS, \(\mathscr{L}\) = Backdoor, \(\mathscr{M}\) = \textit{Ransomware Attack}, \(\mathscr{N}\) = \textit{OS Fingerprinting}, \(\mathscr{O}\) = \textit{MiTM Attack}.}}
\vspace{0.5em}

\begin{minipage}[t]{0.22\textwidth}
\centering
    {\small $\mathtt{TPR}_{\mathtt{FPR}=1\%} = 65.78\%$}
\includegraphics[width=\linewidth]{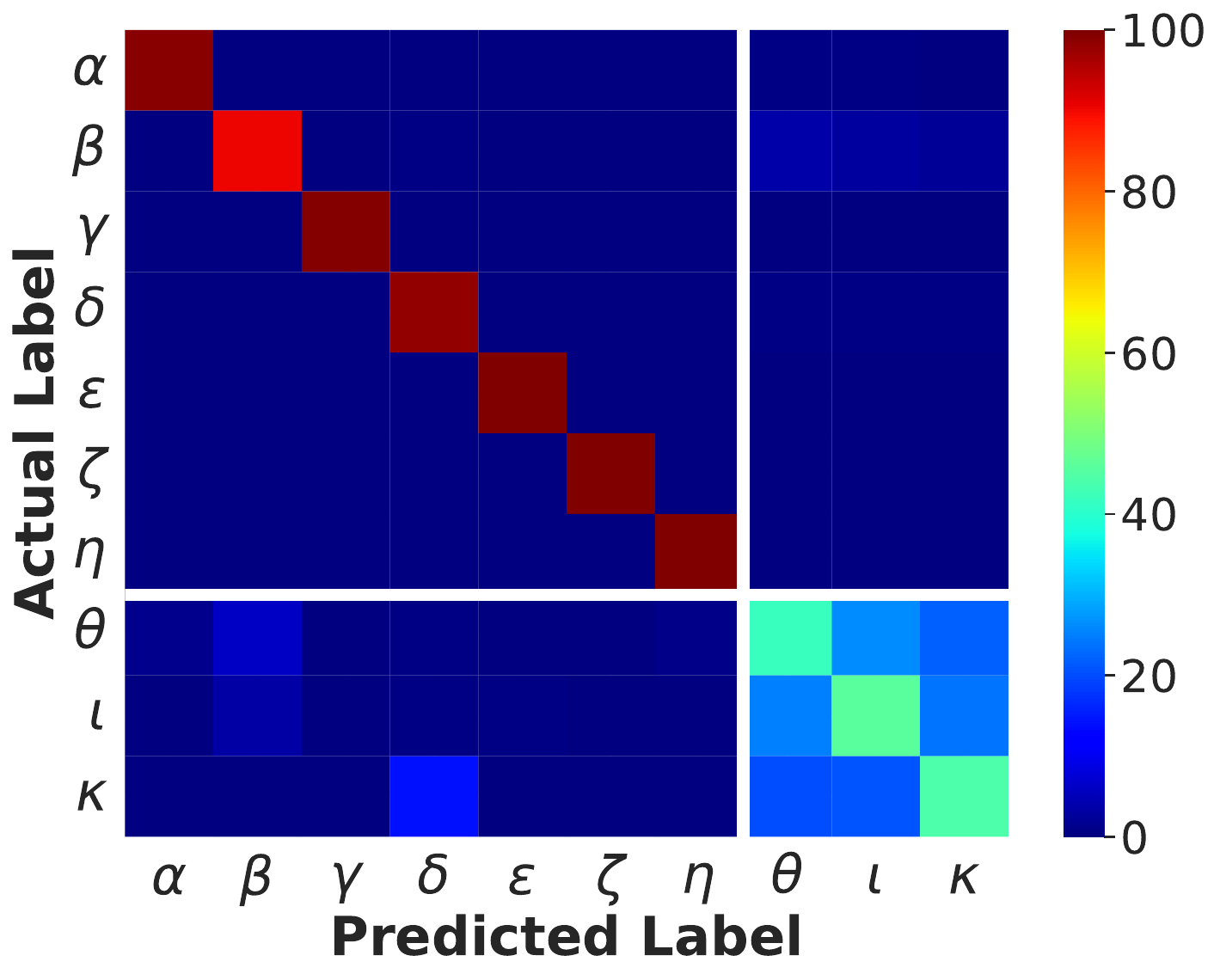}
\caption*{\NN - W/o \nafn}
\end{minipage}
\begin{minipage}[t]{0.22\textwidth}
\centering
{\small $\mathtt{TPR}_{\mathtt{FPR}=1\%} = 68.22\%$}
\includegraphics[width=\linewidth]{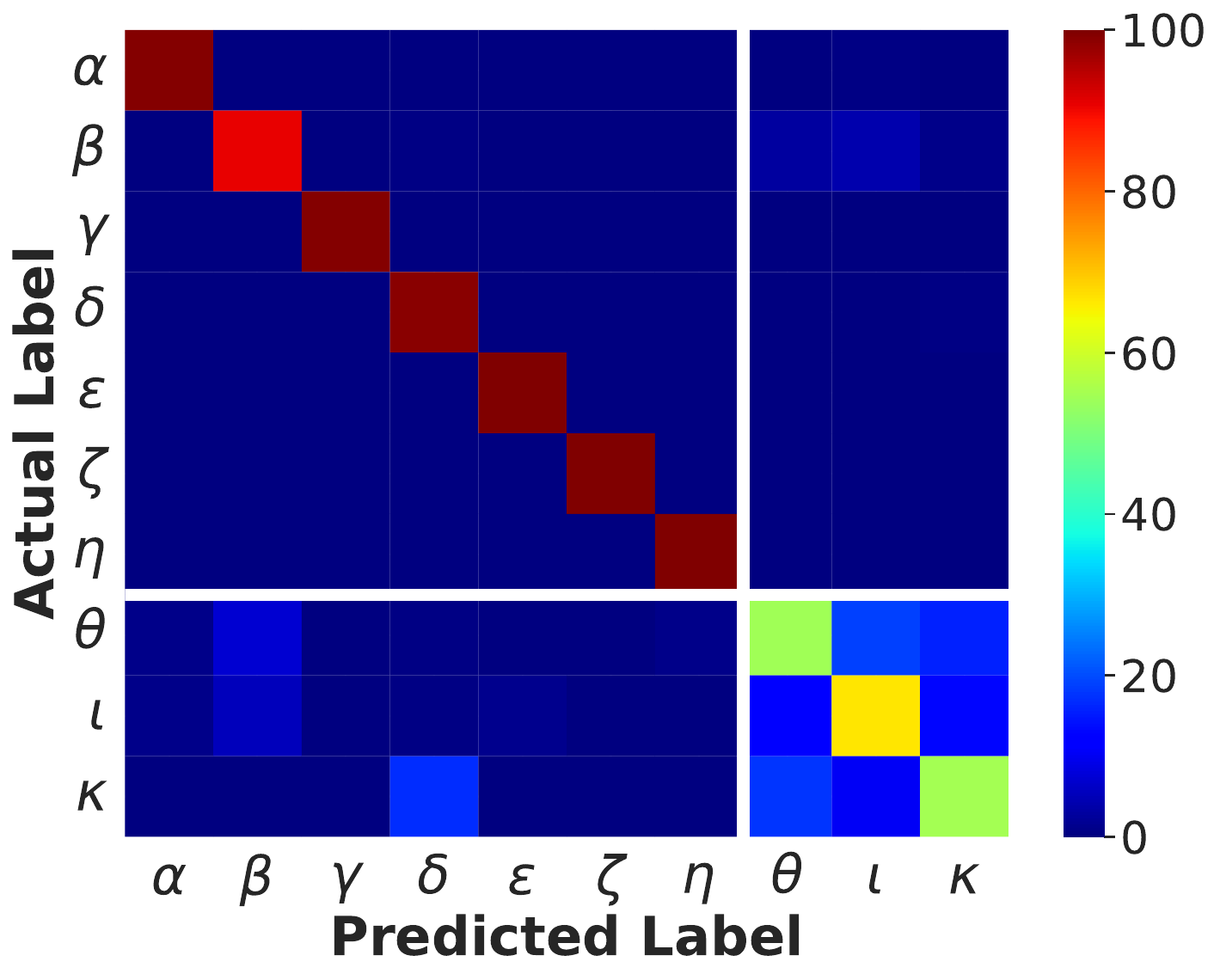}
\caption*{\NN - \nafn (\PM)}
\end{minipage}
\begin{minipage}[t]{0.22\textwidth}
\centering
{\small $\mathtt{TPR}_{\mathtt{FPR}=1\%} = 61.05\%$}
\includegraphics[width=\linewidth]{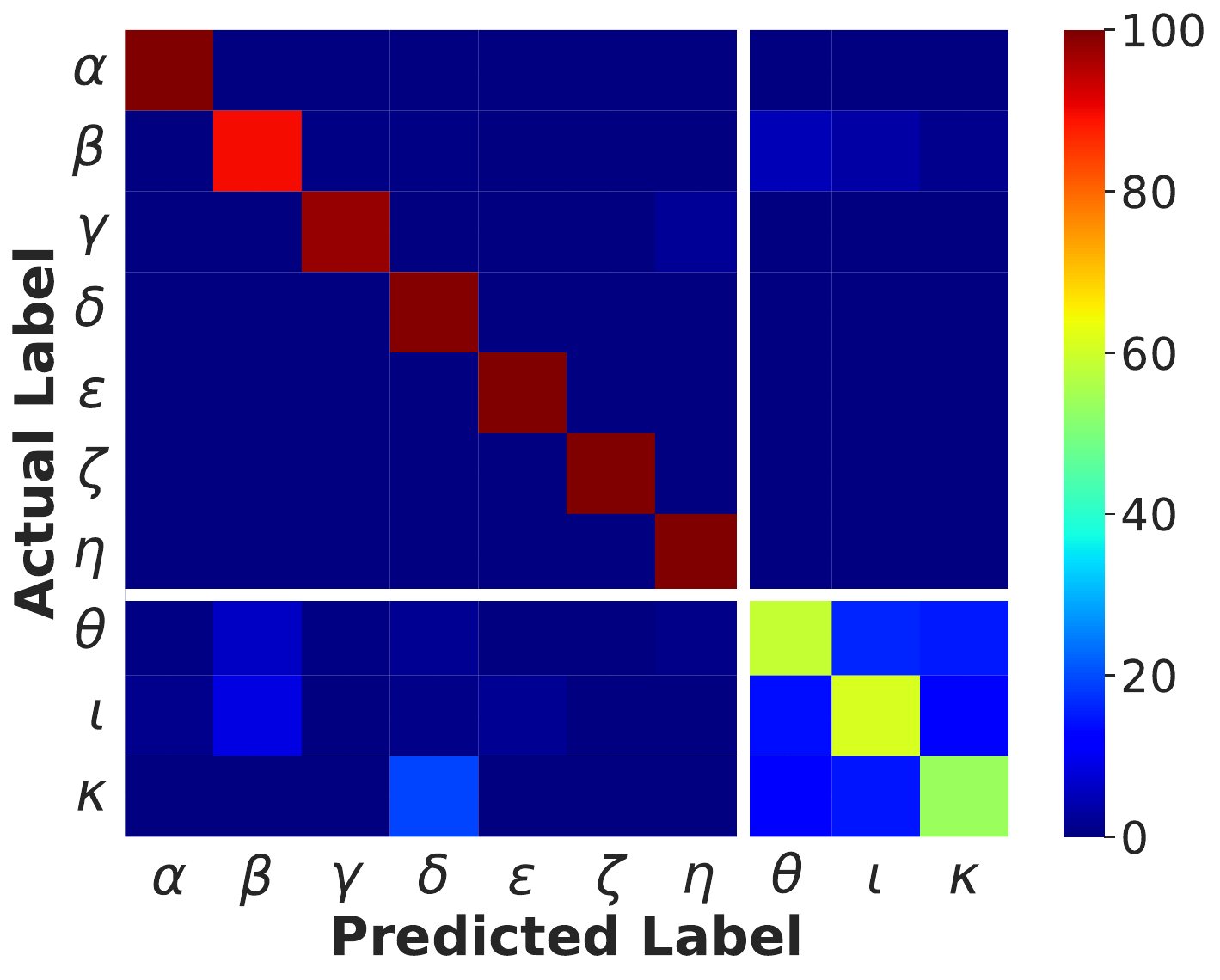}
\caption*{\LR - W/o \nafn}
\end{minipage}
\begin{minipage}[t]{0.22\textwidth}
\centering
    {\small $\mathtt{TPR}_{\mathtt{FPR}=1\%} = 61.23\%$}
\includegraphics[width=\linewidth]{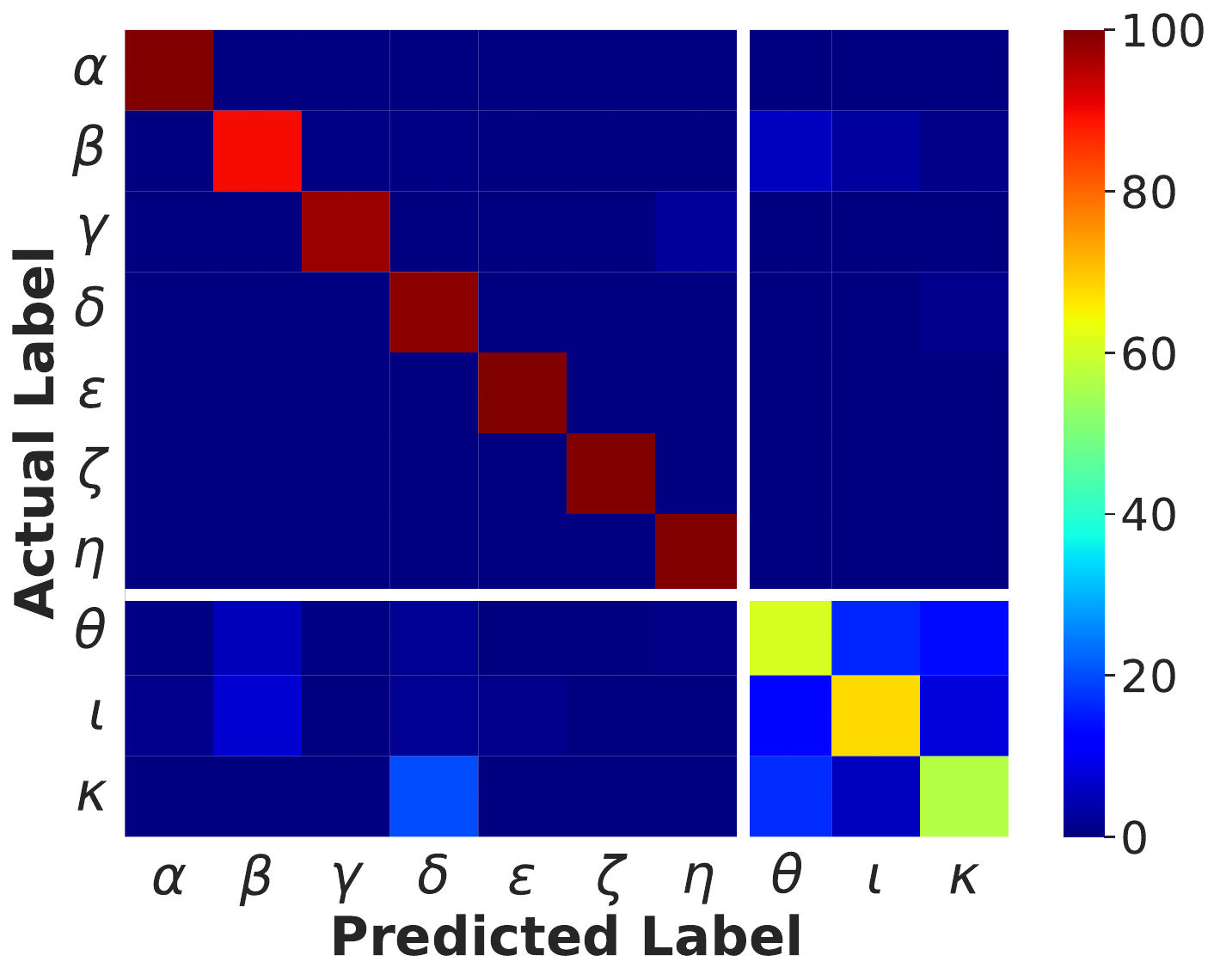}
\caption*{\LR - \nafn (\PM)}
\end{minipage}
\caption*{(c) \IOTNID. 0\% noise bounds: \NN: 69.60\%, \LR: 62.68\%. \scriptsize{\textbf{Label encoding:} \(\alpha\) = Benign, \(\beta\) = DoS SYN Flooding, \(\gamma\) = Mirai ACK Flooding, \(\delta\) = Port Scanning, \(\varepsilon\) = Mirai HTTP Flooding, \(\zeta\) = Host Discovery, \(\eta\) = Mirai UDP Flooding, \(\theta\) = \textit{OS Detection}, \(\iota\) = \textit{MiTM ARP Spoofing}, \(\kappa\) = \textit{Mirai Host Bruteforce}.}}

\label{fig:cm_grid}
\end{figure*}


\subsection{NAFN under Noisy Labels}
\label{subsec:Noisy}

This section addresses \qtwo~by assessing the \emph{robustness of \nafn to mislabeled biflows in the incremental data.}
Unless otherwise stated, each dataset uses the configuration selected in Sec.~\ref{subsec:pretrain_Memory}.
We introduce controlled label noise in the new attack classes, with $\eta=0\%$–$50\%$, where $\eta=0\%$ represents clean labels and higher values simulate increasingly severe label corruption.
Unless otherwise specified, the denoising ratio is fixed at $\rho=50\%$.
%
We compare four noise detection methods (Sec.~\ref{subsec:noise_aware})---\PM, \LOF, \IF, and \ECL---against a baseline without denoising approaches (briefly, \wodenoise).
In addition, we include two noisy-label learning baselines: \CSIDN~\cite{Berthon2021confidence} and \NCR~\cite{Ahmet2022Neighbor}.
\CSIDN assigns a reliability weight to each biflow based on predictive confidence, down-weighting samples with consistently low probabilities as potential noise.
\NCR addresses label noise through $k$-NN topological consistency in the latent space, weighting samples according to agreement with their neighborhood and reducing the influence of isolated, inconsistent biflows.
Herein, both new and base classes are drawn from the same dataset.
Fig.~\ref{fig:noiselabel_comparison} displays the performance in terms of \textit{F1-All}, \textit{F1-New}, and \textit{F1-Old}.

Overall, performance declines as noise increases from $\eta=10\%$ to $50\%$, with the largest degradation on new attacks due to spurious labels.
%
While \wodenoise shows high performance
under low noise, its \textit{F1-New} drops rapidly when noise exceeds $30\%$, often dropping below $60\%$ and in some cases below $40\%$ at $\eta=50\%$. 
By contrast, \PM substantially improves robustness to noisy labels across all datasets and classifiers.
For example, on \CIC with \NN, \PM maintains an \textit{F1-New} of $87.03\%$ at $50\%$ noise, compared with only $35.34\%$ of \wodenoise.
Similarly, on \EDGEIIOT, \PM doubles the \textit{F1-New} of the baseline under high noise conditions. 
Even with \LR, where improvements are generally smaller, 
\PM delivers consistent gains,
highlighting that its effect is not limited
to the \NN classifier.
%
Importantly, \PM filters labels of noisy biflows from $\Cnewt$, enhancing new-attack detection without significantly affecting knowledge of existing classes.


Compared with other noise detection methods, \PM exhibits superior stability and effectiveness.
\ECL shows some advantages with \LR, particularly under high noise, but it does not achieve the same gains on new attacks as \PM.
\LOF and \IF, on the other hand, perform poorly, with \textit{F1-New} scores often dropping below those of \wodenoise, indicating that generic outlier detection approaches are ill-suited for incremental scenarios where noise is structured and class-specific. 
It is worth noting that all methods yield lower \textit{F1-Old} scores \wrt \wodenoise. 
Such a decline occurs because the denoising process inevitably introduces small perturbations to the decision boundaries of old classes, causing minor performance degradation.
%
However, this drop is relatively modest and acceptable when weighed against the gains in \textit{F1-All} scores.
%
Furthermore, when evaluated against noisy-label learning baselines (\CSIDN and \NCR),
\PM continues to show clear advantages in the incremental setting.
Although \CSIDN and \NCR exhibit some resilience against 
spurious labels
\wrt \wodenoise, they struggle to adapt to new attacks under severe noise (e.g., $\eta \ge 40\%$), resulting in sub-optimal \textit{F1-New}. 
%
Moreover, lacking a denoising mechanism to preserve past knowledge, these baselines show a larger drop in \textit{F1-Old}.
%
Conversely, \PM better isolates noisy samples without distorting the decision boundaries, yielding superior \textit{F1-All}.
%
Overall, \PM achieves the best trade-off,
\emph{it mitigates the effects of label noise on new attacks, sustains competitive overall F1-scores, and preserves stability on base classes.}

At $\eta=0\%$ and $\rho=50\%$, denoising causes a limited overall performance loss.
Across all methods, the reduction in \textit{F1-All} remains within $\approx 10\%$, while \textit{F1-New} and \textit{F1-Old} decrease by up to approximately $32\%$ and $4\%$, respectively, with occasional improvements in some settings.
For \PM, \textit{F1-All} typically decreases by only $1\%$-$5\%$ and \textit{F1-Old} by less than $2\%$, while \textit{F1-New} shows a larger, dataset- and classifier-dependent drop of up to $\approx 20\%$.
Thus, even under the $\rho=50\%$ setting, \PM incurs a limited cost on clean incremental data, mainly affecting new-class adaptation.

\takeaway{Addressing label noise in incremental learning is critical, as noise in new attacks can degrade overall performance.
\PM is more stable than \wodenoise, other noise detection methods, and noisy-label baselines, especially at medium to high noise levels.
Moreover, it preserves accuracy on new attacks while keeping base-class performance stable.
Therefore, we adopt \PM as the core of
the Noise-Aware Module.
}

\begin{figure}[!t]
    \centering
    \caption{KDE plot of new attacks affected by label noise, obtained via 2D t-SNE projection. Label encoding is the same as Fig.~\ref{fig:cm_grid}.
    }
    \begin{minipage}[t]{1.00\hsize}
        \centering
        \includegraphics[width=\linewidth]{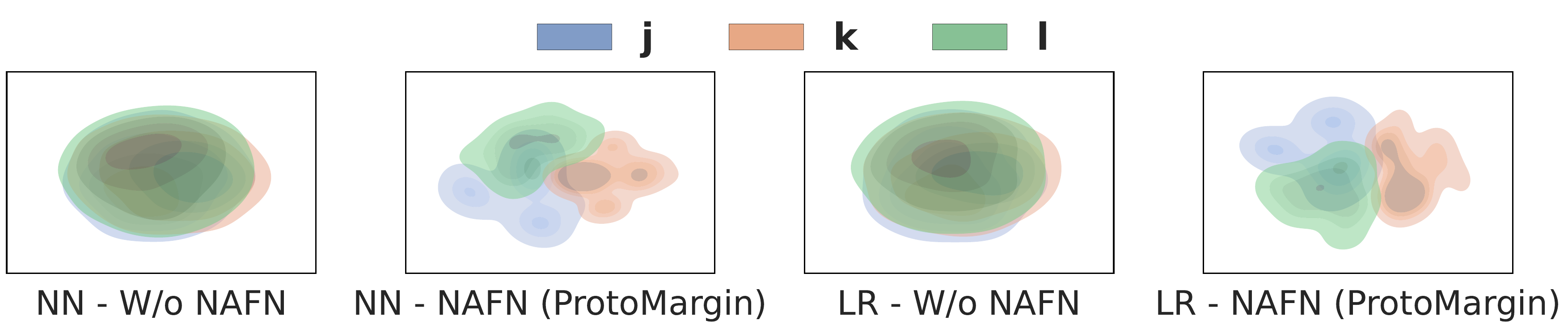}
        \caption*{(a) \CIC}
    \end{minipage}
    \hfill
    
    \begin{minipage}[t]{1.00\hsize}
        \centering
        \includegraphics[width=\linewidth]{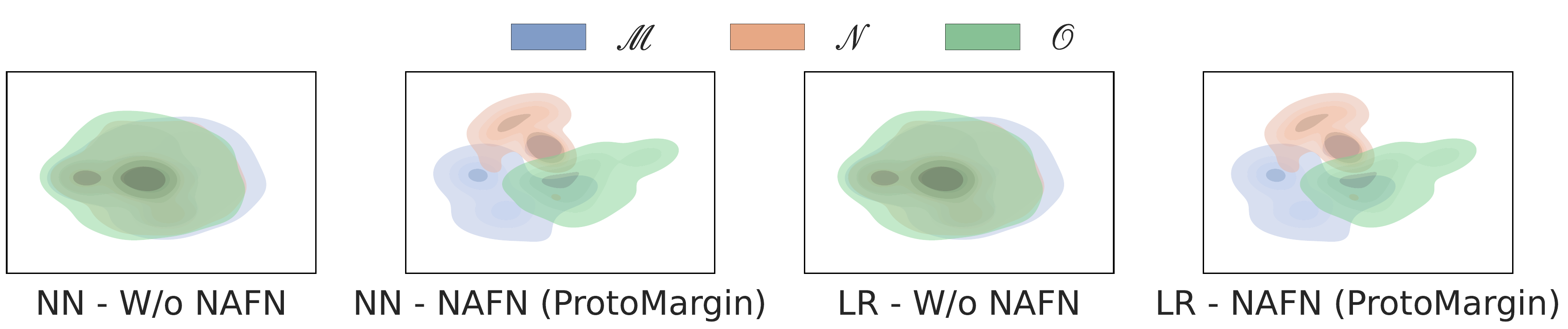}
        \caption*{(b) \EDGEIIOT}
    \end{minipage}
    \hfill
    
    \begin{minipage}[t]{1.00\hsize}
        \centering
        \includegraphics[width=\linewidth]{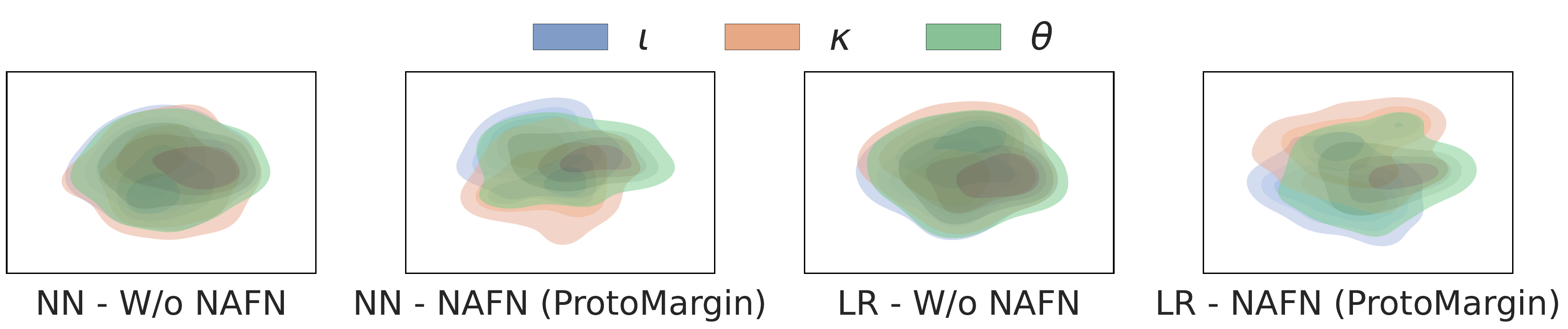}
        \caption*{(c) \IOTNID}
    \end{minipage}
    \label{fig: T-SNE}
\end{figure}

\subsection{Visual Analysis of NAFN}
\label{subsec:Visual}
The following visualization experiment further supports \qtwo~by illustrating \emph{how \nafn enhances feature separability under noisy conditions.}
The analysis uses \PM as the noise detector, together with the per-dataset configuration selected in Sec.~\ref{subsec:pretrain_Memory}.
Specifically, we visualize the confusion matrices and t-SNE projections of \nafn using the \NN and \LR classifiers under $\eta = 50\%$ label noise, 
as well as the feature distributions of new attack classes.

\para{Confusion Matrices.}
As shown in Fig.~\ref{fig:cm_grid}, each row corresponds to a dataset (\CIC, \EDGEIIOT, and \IOTNID).
The four columns report results for \NN (left) and \LR (right), both \textit{with} and \textit{without} \nafn (with \PM).
Across all datasets, label noise mainly affects new attacks located in the lower-right region of each confusion matrix.  
Without \nafn, classifiers exhibit greater confusion, as reflected by the fading of diagonal elements (which is particularly pronounced in \EDGEIIOT and \IOTNID). 
On the other hand, with \nafn, misclassifications are greatly reduced, as can be seen from the diagonal.
This outcome underscores that the Noise-Aware Module effectively filters mislabeled biflows and enhances intra-class consistency.  
%
Moreover, to address strict operational constraints, we report the binary detection performance ($\mathtt{TPR}_{\mathtt{FPR}=1\%}$) above each subfigure. 
These metrics confirm that \nafn not only recovers the fine-grained semantic boundaries of specific attacks, but also maintains a highly robust detection rate that approaches the theoretical upper bound achieved in the noiseless setting.
Finally, the upper-left regions ($\Cbase$) remain stable under both noisy and denoised conditions, confirming that label noise primarily affects $\Cnewt$.

\para{t-SNE.} 
Fig.~\ref{fig: T-SNE} reports the Kernel Density Estimation (KDE) plots generated from 2D t-SNE projection, illustrating the feature distributions of new attack biflows under noisy conditions. 
Consistent with the setup in Fig.~\ref{fig:cm_grid}, each row corresponds to a dataset (\CIC, \EDGEIIOT, and \IOTNID), while the four columns show the results of \NN and \LR classifiers, both with and without \nafn.
Projections show that \nafn substantially enhances feature separability across all three datasets. 
Specifically, when not using \nafn (columns $1$ and $3$), the feature distributions of the three new classes almost completely overlap, making them indistinguishable. 
By contrast, with \nafn (columns $2$ and $4$), the class-specific clusters are much more distinct, highlighting an improved discriminative structure of the feature space.

\takeaway{Confusion matrices and t-SNE provide a visual corroboration of the classification results: label noise mainly harms the detection of new attacks, while \nafn effectively restores their separability, thereby improving robustness 
under noisy-label incremental-learning scenarios.}

\begin{table*}[!t]
\centering
\renewcommand{\arraystretch}{1.2}
\caption{
Cross-network incremental detection performance measured by $\text{pAUC}_{\text{base-mal}}[\%]$, $\text{pAUC}_{\text{inc-mal}}[\%]$, and $\text{pAUC}_{\text{unk-inc}}[\%]$, comparing benign samples against, respectively, base malicious classes, newly introduced incremental malicious classes, and incremental-dataset base malicious classes treated as unknown. 
The setting uses $\Cbase$ from one dataset and $\Cnewt$ from a different network environment. 
\textcolor{lr}{\textbf{Red}} cells indicate results without \nafn, while \textcolor{lb}{\textbf{blue}} cells indicate results with \nafn.
}
\label{tab:crossnetwork}
\begin{minipage}{0.49\linewidth}
\centering
\small{(a) \NN Classifier}\\
\vspace{0.1in}
\resizebox{\linewidth}{!}{
    \begin{tabular}{llc|ccc}
    \toprule
    \textbf{Base Dataset} & \textbf{Incr.~Dataset} & \textbf{Label Noise} & \textbf{$\text{pAUC}_{\text{base-mal}}$} & \textbf{$\text{pAUC}_{\text{inc-mal}}$} & \textbf{$\text{pAUC}_{\text{unk-inc}}$} \\
    \midrule
    
    \multirow{6}{*}{\CIC} 
     & \multirow{3}{*}{\EDGEIIOT} 
     & 0\% & \rc88.12  & \rc97.46  & \rc60.85 \\
     & & 50\% & \rc85.07 & \rc82.26 & \rc61.83 \\
     & & 50\% & \bc88.12 & \bc95.68 & \bc64.10 \\
    \cline{2-3}\addlinespace[2pt]
     & \multirow{3}{*}{\IOTNID}
     & 0\% & \rc90.01 & \rc58.89 & \rc50.72 \\
     & & 50\% & \rc89.68 & \rc57.21 & \rc52.22 \\
     & & 50\% & \bc93.45 & \bc63.45 & \bc56.49 \\
    \cline{1-3}\addlinespace[2pt]
    
    \multirow{6}{*}{\EDGEIIOT} 
     & \multirow{3}{*}{\CIC}
     & 0\% & \rc99.92 & \rc99.98 & \rc93.57 \\
     & & 50\% & \rc99.58 & \rc98.41 & \rc92.62 \\
     & & 50\% & \bc99.90 & \bc99.95 & \bc94.68 \\
    \cline{2-3}\addlinespace[2pt]
     & \multirow{3}{*}{\IOTNID}
     & 0\% & \rc99.77 & \rc94.07 & \rc88.71 \\
     & & 50\% & \rc99.28 & \rc90.03 & \rc83.91 \\
     & & 50\% & \bc99.36 & \bc91.97 & \bc84.30 \\
    \cline{1-3}\addlinespace[2pt]
    
    \multirow{6}{*}{\IOTNID} 
     & \multirow{3}{*}{\CIC}
     & 0\% & \rc94.64 & \rc95.96 & \rc72.05 \\
     & & 50\% & \rc92.47 & \rc84.00 & \rc62.45 \\
     & & 50\% & \bc92.49 & \bc96.68 & \bc73.13 \\
    \cline{2-3}\addlinespace[2pt]
     & \multirow{3}{*}{\EDGEIIOT}
     & 0\% & \rc98.42 & \rc93.92 & \rc66.96 \\
     & & 50\% & \rc99.36 & \rc76.53 & \rc65.34 \\
     & & 50\% & \bc99.42 & \bc90.06 & \bc66.11 \\
    \bottomrule
    \end{tabular}
}
\end{minipage}
\hfill
\begin{minipage}{0.49\linewidth}
\centering
\small{(b) \LR Classifier}\\
\vspace{0.1in}
\resizebox{\linewidth}{!}{
    \begin{tabular}{llc|ccc}
    \toprule
    \textbf{Base Dataset} & \textbf{Incr.~Dataset} & \textbf{Label Noise} & \textbf{$\text{pAUC}_{\text{base-mal}}$} & \textbf{$\text{pAUC}_{\text{inc-mal}}$} & \textbf{$\text{pAUC}_{\text{unk-inc}}$} \\
    \midrule
    
    \multirow{6}{*}{\CIC} 
     & \multirow{3}{*}{\EDGEIIOT} 
     & 0\% & \rc98.32 & \rc91.55 & \rc64.31 \\
     & & 50\% & \rc96.29 & \rc91.31 & \rc63.08 \\
     & & 50\% & \bc96.34 & \bc92.54 & \bc63.28 \\
    \cline{2-3}\addlinespace[2pt]
     & \multirow{3}{*}{\IOTNID}
     & 0\% & \rc95.27 & \rc76.67  & \rc64.91 \\
     & & 50\% & \rc95.03 & \rc73.89 & \rc61.92 \\
     & & 50\% & \bc95.05 & \bc74.81 & \bc65.01 \\
    \cline{1-3}\addlinespace[2pt]
    
    \multirow{6}{*}{\EDGEIIOT} 
     & \multirow{3}{*}{\CIC}
     & 0\% & \rc99.79 & \rc100.00 & \rc93.95 \\
     & & 50\% & \rc99.79 & \rc98.08 & \rc93.44 \\
     & & 50\% & \bc99.79 & \bc100.00 & \bc94.93 \\
    \cline{2-3}\addlinespace[2pt]
     & \multirow{3}{*}{\IOTNID}
     & 0\% & \rc99.93 & \rc99.62 & \rc97.56 \\
     & & 50\% & \rc99.82 & \rc94.15 & \rc96.97 \\
     & & 50\% & \bc99.82 & \bc95.93 & \bc96.99 \\
    \cline{1-3}\addlinespace[2pt]
    
    \multirow{6}{*}{\IOTNID} 
     & \multirow{3}{*}{\CIC}
     & 0\% & \rc88.12 & \rc95.19 & \rc63.69 \\
     & & 50\% & \rc87.40 & \rc85.44 & \rc59.54 \\
     &  & 50\% & \bc88.10 & \bc89.16 & \bc59.70 \\
    \cline{2-3}\addlinespace[2pt]
     & \multirow{3}{*}{\EDGEIIOT}
     & 0\% & \rc92.38 & \rc80.50 & \rc63.93 \\
     & & 50\% & \rc92.17 & \rc71.19 & \rc62.18 \\
     & & 50\% & \bc92.20 & \bc78.06 & \bc62.51 \\
    \bottomrule
    \end{tabular}
}
\end{minipage}

\end{table*}

\begin{figure*}[!ht]
    \centering
    \caption{
    Analysis of \textit{Time-to-Insight} (TTI) versus detection performance under label noise ($\eta = 50\%$) on \CIC. The plots illustrate the \textit{F1-All} scores for \NN and \LR classifiers as a function of the time threshold $T_s$. \scriptsize{\textbf{Label encoding:} \(\mathbf{a}\) = DoS GoldenEye, \(\mathbf{b}\) = DoS Hulk, \(\mathbf{c}\) = DDoS HOIC, \(\mathbf{d}\) = Bot, \(\mathbf{e}\) = Bruteforce-SSH, \(\mathbf{f}\) = Benign, \(\mathbf{g}\) = DDoS LOIC-HTTP, \(\mathbf{h}\) = DOS Slowloris, \(\mathbf{i}\) = Bruteforce-Web, \(\mathbf{j}\) = \textit{DDoS LOIC-UDP}, \(\mathbf{k}\) = \textit{Bruteforce-XSS}, \(\mathbf{l}\) = \textit{SQL Injection}.}}

    \begin{minipage}[t]{0.90\linewidth}
        \centering
        \includegraphics[width=\linewidth]{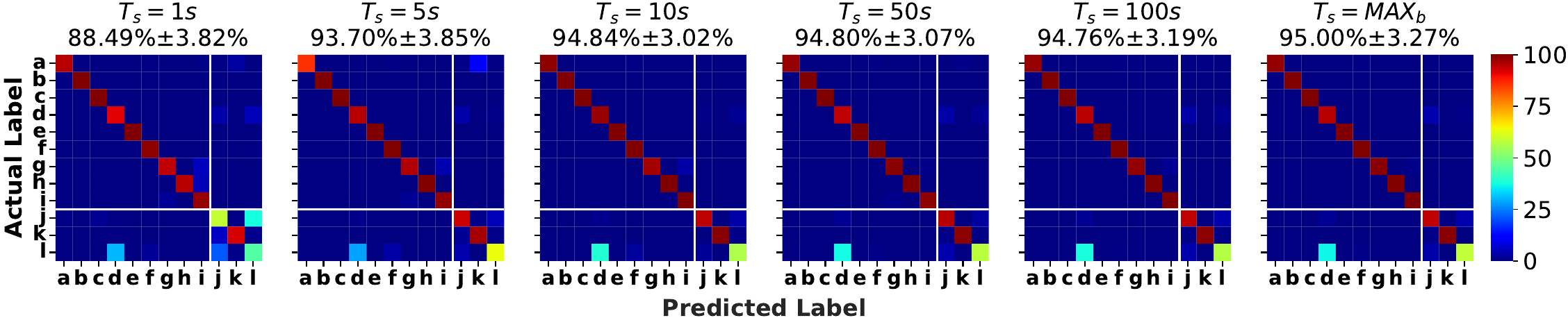}
        \caption*{(a) \NN classifier}
    \end{minipage}
    \hfill
    
    \begin{minipage}[t]{0.90\linewidth}
        \centering
        \includegraphics[width=\linewidth]{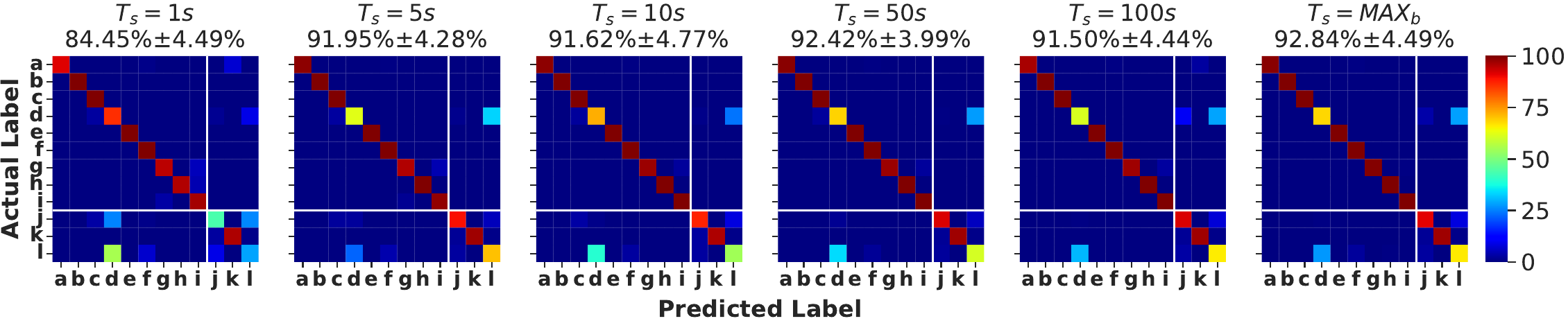}
        \caption*{(b) \LR classifier}
    \end{minipage}
    \label{fig:Time_to_Insight}
\end{figure*}


\subsection{Analysis of Early Detection and Time-to-Insight of NAFN}
\label{subsec:Time-to-Insight}

To further address \qtwo, we evaluate the real-time detection ability of \nafn under label noise ($\eta=50\%$), confirming its effectiveness for early detection.
This analysis focuses on \CIC and uses the corresponding configuration selected in Sec.~\ref{subsec:pretrain_Memory}.
In detail, we investigate the performance under varying time thresholds $T_s$, which directly dictate the \textit{Time-to-Insight} (TTI). 
%
TTI is defined as the duration from the arrival of the initial packet of a biflow to its final classification~\cite{FSCIL-NIDS2}, 
and it dominates inference latency, far exceeding the overhead of feature extraction and embedding.
While lower $T_s$ values enhance real-time responsiveness, they increase vulnerability to slow-rate flooding attacks; conversely, higher $T_s$ values improve detection accuracy but delay biflow classification.
%
In this analysis, we focus on \CIC due to the presence of slow-rate flooding attacks.

Fig.~\ref{fig:Time_to_Insight} illustrates the confusion matrices and corresponding \textit{F1-All} scores across $T_s \in \{1s, 5s, 10s, 50s, 100s, \mathtt{MAX}_b\}$, where $\mathtt{MAX}_b = 330s$ denotes the maximum flow duration. 
Despite noise interference, we observe a significant performance leap as $T_s$ increases from $1$s to $10$s (\ie $\approx+7\%$ \textit{F1-All} for \NN and \LR). 
Beyond the $10$s mark, the detection performance enters a stable plateau, with gains below $1\%$.
%
Although the highest \textit{F1-All} ($95.00\%$ for \NN and $92.84\%$ for \LR) values are achieved at $T_s = \mathtt{MAX}_b$, results show that $T_s = 50s$ offers the best balance.

\takeaway{\nafn reaches near-peak accuracy within $\mathit{10s}$ and achieves the best trade-off at $\mathit{T_s=50s}$, reducing detection latency by $\mathit{6\times}$ while remaining robust under $\mathit{50\%}$ label noise and effectively mitigating slow-rate attacks.
}


\subsection{Cross-Network Incremental Detection Performance}
\label{subsec:Crossnetwork}





This section tackles \qthree, focusing on \emph{whether \nafn can consistently sustain robust and generalizable detection performance across diverse network environments}.
To this end, we evaluate the \emph{cross-network incremental detection capability} of \nafn by training on traffic from one dataset and incrementally introducing new attack classes from a different dataset, while using the same base and new classes defined in Sec.~\ref{sec:fscil_setup}.
Base training uses, for each base dataset, the configuration selected in Sec.~\ref{subsec:pretrain_Memory}, and the incremental phase reuses the same encoder and classifier setup with the new classes drawn from the other dataset.
%
To analyze performance under this cross-network setup, we report three distinct pAUCs.
Each metric is defined based on the origin and role of the malicious classes involved.
Specifically,
\begin{enumerate*}[label=(\emph{\roman*})]
    \item $\text{pAUC}_{\text{base-mal}}$: evaluates the ability to retain discrimination between benign and \textbf{base} malicious traffic, where all attacks from the base dataset are grouped into a single malicious class (as explained in Sec.~\ref{sec:metrics});
    \item $\text{pAUC}_{\text{inc-mal}}$: measures the ability to detect \textbf{newly introduced} malicious traffic from a different dataset during the incremental learning phase;
    \item $\text{pAUC}_{\text{unk-inc}}$: assesses the detection of \textbf{unseen} malicious traffic from a different network environment.
    It considers the \emph{base classes of the incremental dataset} which are observed neither during base training nor during the incremental phase, and are therefore unknown at test time.
\end{enumerate*}
Table~\ref{tab:crossnetwork} reports the three pAUCs for each base–incremental dataset pair, using the \NN and \LR classifiers.
Results are shown without \nafn---both without noise ($\eta=0\%$) and with noise ($\eta=50\%$)---as well as for \nafn with noise ($\eta=50\%$).

Regarding the \NN classifier (Table~\ref{tab:crossnetwork}(a)), \nafn exhibits strong detection performance according to all three pAUCs. Detection of base malicious classes remains consistently high, with $\text{pAUC}_{\text{base-mal}}$ values often exceeding $95\%$, indicating that the NIDS effectively retains previous knowledge even in the presence of domain shifts. 
In contrast, the performance on $\text{pAUC}_{\text{inc-mal}}$ and $\text{pAUC}_{\text{unk-inc}}$ varies noticeably across dataset pairs.
Particularly, combinations involving \CIC and \IOTNID show greater degradation, 
reflecting a larger distributional gap.
Nonetheless, \nafn
mitigates the impact of label noise and domain shift, improving the detection of new and unknown attacks. 
For example, when adapting from \IOTNID to \CIC, \nafn improves $\text{pAUC}_{\text{inc-mal}}$ from $84.00\%$ to $96.68\%$, and $\text{pAUC}_{\text{unk-inc}}$ from $62.45\%$ to $73.13\%$. 
Such gains highlight the effectiveness of \nafn in noise filtering and stabilizing incremental adaptation.

The \LR classifier (Table~\ref{tab:crossnetwork}(b)) exhibits a more stable yet slightly more conservative performance profile. 
Although $\text{pAUC}_{\text{base-mal}}$ is consistently high (above $\approx95\%$), the metrics involving cross-network generalization, namely
$\text{pAUC}_{\text{inc-mal}}$ and $\text{pAUC}_{\text{unk-inc}}$, 
show moderate drops, especially when \IOTNID serves as either the base or incremental dataset. 
Despite this, \nafn again provides consistent improvements in all settings, albeit with smaller margins \wrt \NN. 
These outcomes indicate that \LR benefits from noise-aware filtering, 
although its feature representation is inherently less sensitive to label noise than the \NN classifier.

\takeaway{Cross-network experiments show that both \NN and \LR can adapt to unseen attack classes from heterogeneous environments, with generalization varying by dataset pair.
Across all three pAUCs, \nafn consistently improves performance by mitigating label noise and enhancing robustness to unseen malicious biflows.
These results indicate that \nafn supports reliable incremental learning and strengthens cross-network transferability under both clean and noisy label conditions.}

\section{Conclusions \& Future Perspectives}
\label{sec:conclusion}

In this paper, we propose \nafn, a Noise-Aware Few-Shot Class-Incremental Learning framework for Network Intrusion Detection Systems (NIDSs). 
\nafn introduces a teacher–student adaptation approach and a Noise-Aware Module to enable continual adaptation to new attack classes while mitigating the detrimental effects of noisy labels. 
Through extensive evaluations across three benchmark datasets, we validate the effectiveness of \nafn by addressing a comprehensive set of research questions (Sec.~\ref{sec:problem}).
With regard to \qone, our experiments show that 
\emph{\nafn can efficiently adapt to new attacks under a few-shot regime} while preserving performance on previously learned classes.
For \qtwo, \emph{the proposed Noise-Aware Module, via the Prototype-Margin method, significantly improves robustness against mislabeled biflows}. 
At label-noise ratios ranging from $10\%$ to $50\%$, 
the F1-score on new classes improves 
by up to $+52\%$ \wrt the baseline without denoising, 
consistently outperforming alternative noise detection methods while preserving the separability of new-class representations.
Finally, regarding \textbf{\qthree}, \emph{our cross-network experiments reveal that \nafn maintains stable and transferable detection capabilities across heterogeneous network environments.} 
Even under severe label noise and pronounced domain shifts, \nafn consistently improves cross-network incremental detection, with observed pAUC gains of up to $13\%$ and $11\%$ for new and previously unseen attacks, respectively.

%

\para{Limitations \& Future Avenues.}
The findings of this manuscript suggest several promising directions for future research to address the current limitations of \nafn.
\begin{enumerate*}[label=(\textit{\roman*})]
\item Retransmissions and packet loss in dynamic network environments may alter traffic features, potentially affecting the robustness of \nafn. 
Future work could address this limitation through adaptive feature extraction or more robust representation learning techniques~\cite{jiang2024sequence}.
\item Traffic drift and memory bias accumulation during long-term incremental updates may be reduced by developing scalable memory management strategies or continual prototype calibration methods~\cite{zhang2024AOCIDS}, improving model stability in online or extended deployments.
\item Explainable AI can increase the transparency and interpretability of the model~\cite{cerasuolo2025adaptable}, enabling security operators to better understand detection decisions and enhancing usability in real-world scenarios.
\item \nafn assumes that the presence of label noise is known; future work could detect label corruption through confidence monitoring or prototype consistency, automatically triggering the denoising pipeline.
\item The paper assumes a symmetric distribution of noise across the new classes. 
Future work could investigate asymmetric noise distributions based on attack-type similarity.
\end{enumerate*}

\appendix
\section{Appendix}

\subsection{Hyperparameter Sensitivity Analysis}

\begin{figure}[!h]
    \centering
            \includegraphics[width=1.0\hsize]{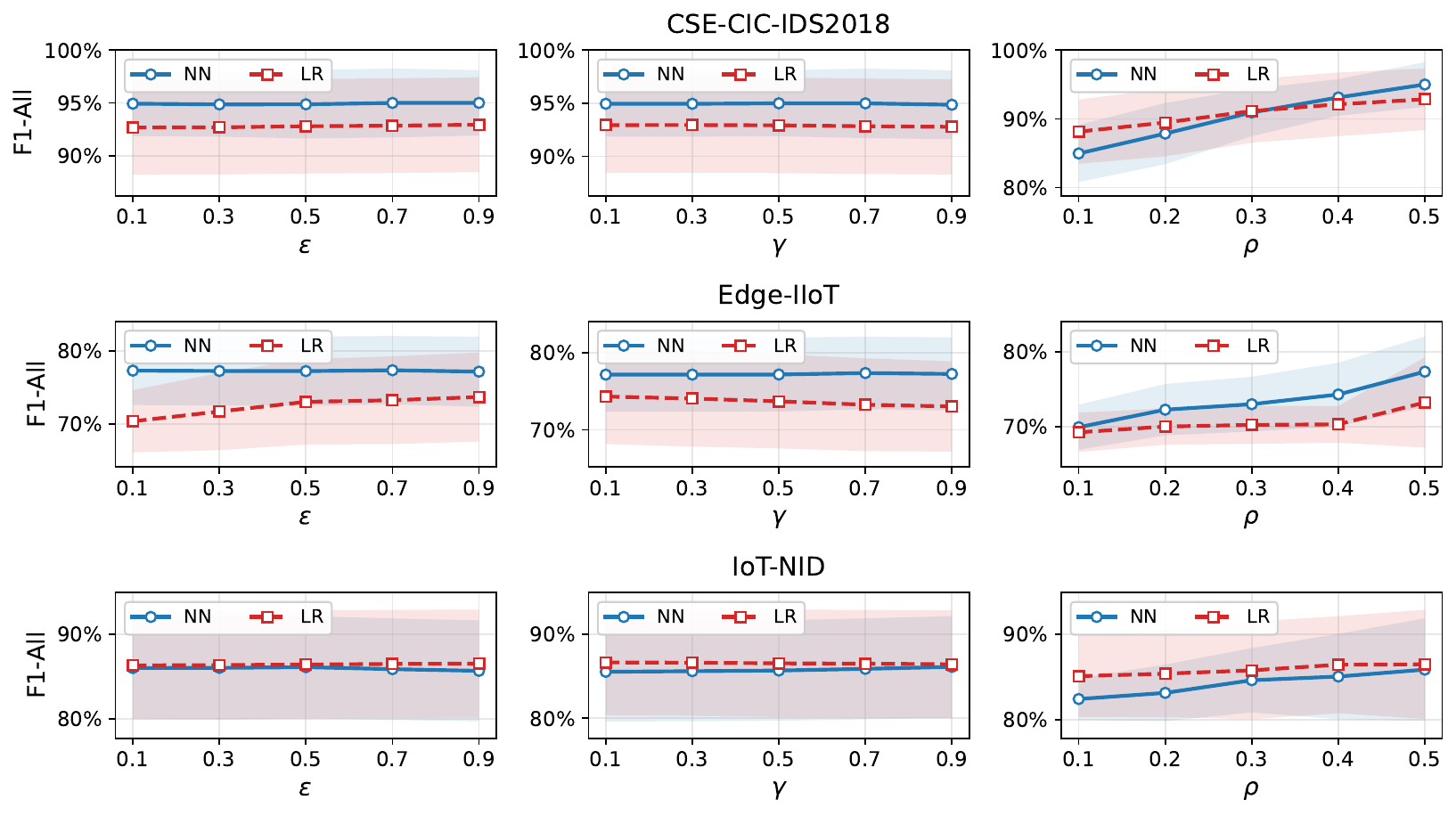}

    \caption{
    Sensitivity analysis of the Feature Sampling Module hyperparameters ($\epsilon$, $\gamma$, $\rho$) on \CIC, \EDGEIIOT, and \IOTNID with \NN and \LR under $50\%$ label noise. Columns correspond to $\epsilon$, $\gamma$, and $\rho$; rows to \CIC, \EDGEIIOT, and \IOTNID.
    }  
    \label{fig:Sensitive}
\end{figure}



In this section, we conduct a sensitivity analysis of the hyperparameters $(\epsilon, \gamma, \rho)$ related to the feature sampling module (Sec.~\ref{subsec:noise_aware}).
Specifically, $\epsilon$ and $\gamma$ control the fusion rates for the base classes $C_{\text{base}}$ and the novel classes $C_{\text{new}}$ (cf. Eqs.~\ref{eq:fusion_mu} and~\ref{eq:fusion_Sigma}), respectively, while $\rho$ controls the denoising ratio. 
%
We vary $\epsilon$ and $\gamma$ over the interval $[0.1, 0.9]$, while setting $\rho$ within $[0.1, 0.5]$.
Unless otherwise specified, the default setting is $\epsilon=\gamma=0.7$ and $\rho=0.5$. 
%
To isolate the effect of each factor, we adopt a one-at-a-time protocol: when tuning one hyperparameter, the others are fixed at their default values.

Fig.~\ref{fig:Sensitive} shows the \textit{F1-All} curves of \nafn over the three hyperparameters on \CIC, \EDGEIIOT, and \IOTNID, using two classifiers (\NN and \LR) under label noise with $\eta=50\%$.
%
Overall, \nafn remains stable across $\epsilon$ and $\gamma$. Changing the fusion rates leads to only marginal performance fluctuations, suggesting that the model is weakly sensitive to the mixture weights between $C_{\text{base}}$ and $C_{\text{new}}$.
%
In contrast, $\rho$ has a stronger effect since it controls the denoising strength.
As $\rho$ increases, \nafn achieves better detection performance under $\eta=50\%$ label noise, implying that stronger denoising is beneficial when supervision is heavily corrupted.
Notably, the best performance is obtained when $\rho$ is close to the true label-noise ratio $\eta$, which aligns with the intuition that the denoising ratio should match the corruption level and further corroborates the effectiveness of the Noise-Aware Module. 
Meanwhile, \nafn remains robust even with conservative denoising (e.g., $\rho=0.1$). Despite $50\%$ noise, performance does not drop significantly, showing that the framework does not rely on aggressive denoising to work well.

In summary, \nafn is largely insensitive to $(\epsilon,\gamma,\rho)$.
%
In particular, $\epsilon$ and $\gamma$ can be chosen from a broad range without noticeable impact, while $\rho$ provides a controllable knob that adaptively improves performance as the noise level increases.



\subsection{Ablation of Feature Sampling Module}

\begin{table}[h]
    \centering
    \caption{
    Ablation of post-denoising compensation strategies. \emph{Denoise-only} is compared with embedding-space compensation via MixUp, SMOTE, and \textcolor{lb}{\textbf{Ours}} (Gaussian embedding synthesis fitted on the remaining samples). We report \textit{F1-All}, \textit{F1-New}, \textit{F1-Old} [\%].
    }
    \label{tab:AblationCompensation}
    \resizebox{\hsize}{!}{
        \begin{tabular}{lcl|ccc}
            \toprule
            \textbf{Dataset} & \textbf{Classifier} & \textbf{Setting} & \textbf{F1-All} & \textbf{F1-New} & \textbf{F1-Old} \\ \midrule
            \multirow{8}{*}{\CIC}      & \multirow{4}{*}{\NN} & Denoise-only  & 94.96\% & 87.04\% & 97.59\% \\
                                       &                      & Denoise $+$ MixUp & 94.98\% & 87.00\% & 97.65\% \\
                                       &                      & Denoise $+$ SMOTE & 95.00\% & 86.91\% & 97.65\% \\ 
                                       &                      & \cellcolor{lb}Denoise $+$ Ours  & \cellcolor{lb}95.01\% \cellcolor{lb}& \cellcolor{lb}87.02\% \cellcolor{lb}& \cellcolor{lb}97.66\% \\ \cmidrule(lr){2-6}        
                                       & \multirow{4}{*}{\LR} & Denoise-only  & 78.66\% & 32.74\% & 93.97\% \\            
                                       &                      & Denoise $+$ MixUp & 92.81\% & 82.94\% & 96.18\% \\
                                       &                      & Denoise $+$ SMOTE & 92.75\% & 82.17\% & 96.19\% \\ 
                                       &                      & \cellcolor{lb}Denoise $+$ Ours  & \cellcolor{lb}92.84\% & \cellcolor{lb}82.74\% & \cellcolor{lb}96.21\% \\ \midrule
            
            \multirow{8}{*}{\EDGEIIOT} & \multirow{4}{*}{\NN} & Denoise-only  & 77.15\% & 64.93\% & 80.20\% \\
                                       &                      & Denoise $+$ MixUp & 77.56\% & 65.65\% & 80.35\% \\
                                       &                      & Denoise $+$ SMOTE & 77.46\% & 65.81\% & 80.38\% \\
                                       &                      & \cellcolor{lb}Denoise $+$ Ours  & \cellcolor{lb}77.35\% & \cellcolor{lb}65.91\% & \cellcolor{lb}80.21\% \\ \cmidrule(lr){2-6}
                                       & \multirow{4}{*}{\LR} & Denoise-only  & 64.26\% & 17.08\% & 76.06\% \\
                                       &                      & Denoise $+$ MixUp & 73.58\% & 56.48\% & 78.86\% \\
                                       &                      & Denoise $+$ SMOTE & 73.39\% & 56.75\% & 78.80\% \\ 
                                       &                      & \cellcolor{lb}Denoise $+$ Ours  & \cellcolor{lb}73.25\% & \cellcolor{lb}56.19\% & \cellcolor{lb}77.51\% \\ \midrule                                       
            \multirow{8}{*}{\IOTNID}   & \multirow{4}{*}{\NN} & Denoise-only  & 85.53\% & 59.90\% & 96.52\% \\
                                       &                      & Denoise $+$ MixUp & 85.90\% & 60.41\% & 96.76\% \\
                                       &                      & Denoise $+$ SMOTE & 85.93\% & 60.80\% & 96.70\% \\ 
                                       &                      & \cellcolor{lb}Denoise $+$ Ours  & \cellcolor{lb}85.87\% & \cellcolor{lb}60.50\% & \cellcolor{lb}96.74\% \\ \cmidrule(lr){2-6}
                                       & \multirow{4}{*}{\LR} & Denoise-only  & 78.27\% & 48.79\% & 90.90\% \\
                                       &                      & Denoise $+$ MixUp & 86.43\% & 64.21\% & 95.96\% \\
                                       &                      & Denoise $+$ SMOTE & 86.45\% & 64.20\% & 95.98\% \\ 
                                       &                      & \cellcolor{lb}Denoise $+$ Ours  & \cellcolor{lb}86.48\% & \cellcolor{lb}64.36\% & \cellcolor{lb}95.97\% \\ \bottomrule

        \end{tabular}
    }
\end{table}

In the following, we ablate the post-denoising compensation stage.
Specifically, after the noise detection module, we compare \emph{Denoise-only} (no compensation) with three embedding-space compensation strategies: MixUp, SMOTE, and \textbf{Ours} (Gaussian embedding synthesis fitted on the remaining samples).
%
Experiments are conducted on the three datasets,
using \NN and \LR. 
Table~\ref{tab:AblationCompensation} reports \textit{F1-All}, \textit{F1-New}, and \textit{F1-Old}.

Overall, \textit{Denoise-only} is insufficient and can significantly degrade performance after samples are filtered out.
This issue is particularly pronounced for \LR, which relies on adequate coverage of the feature space.
For instance, on \CIC with \LR, \textit{Denoise-only} yields only $78.66\%$ \textit{F1-All} and $32.74\%$ \textit{F1-New}, resulting in weaker generalization to new classes due to the lack of compensation.
On the other hand, biflow compensation restores and often improves performance, suggesting that denoising and compensation play complementary roles: denoising improves sample reliability, while compensation mitigates the distribution gaps caused by filtering.
Moreover, our Gaussian embedding synthesis (\textit{Denoise+Ours}) performs on par with widely-used alternatives such as MixUp and SMOTE in embedding space.
%
Notably, on \CIC with \NN, \textit{Denoise+Ours} achieves the best results ($95.01\%$ \textit{F1-All} and $87.02\%$ \textit{F1-New}), slightly outperforming the baselines.
These results confirm that post-denoising compensation is crucial for maintaining performance under few-shot filtering, and our approach provides competitive (and occasionally superior) gains relative to standard embedding-space augmentation methods.

\subsection{Multiple Incremental Results}
\label{appx:multi_incr}

To complement the single-session protocol used in the main evaluation, we further assess \nafn under a multi-step incremental setting.
For each dataset, the rare attack classes originally assigned to $\Cnewt$ are introduced sequentially over three incremental steps, with one new class added at each step.
At step $\#t$, the model is adapted using the available few-shot samples of the current new class, while the previously observed classes are retained through the memory buffer.
The evaluation set then includes all classes observed up to that step, allowing us to measure both the adaptation to the newly introduced attack and the preservation of previously acquired knowledge.
%
We keep the same experimental setup used in the main experiments;
values are averaged over $100$ tasks and reported as mean F1-score with standard deviation.

\begin{table}[t]
    \centering
    \caption{Multi-step incremental performance of \nafn on \CIC, \EDGEIIOT, and \IOTNID. Each step introduces one additional rare attack class. Evaluation is performed over all classes observed so far.}
    \label{tab:MultistepIncremental}
    \resizebox{\hsize}{!}{
        \begin{tabular}{lcc|ccc}
            \toprule
            \textbf{Dataset} & \textbf{Classifier} & \textbf{Step} & \textbf{F1-All} & \textbf{F1-New} & \textbf{F1-Old} \\ \midrule
            \multirow{6}{*}{\CIC}      & \multirow{3}{*}{\NN} & \#1 & 99.12\std{0.39} & 96.71\std{1.50} & 99.38\std{0.31} \\
                                       &                      & \#2 & 99.06\std{0.37} & 99.86\std{0.42} & 98.98\std{0.40} \\
                                       &                      & \#3 & 97.18\std{0.65} & 85.18\std{4.07} & 98.27\std{0.40} \\ 
                                       \cmidrule(lr){2-6}     
                                       & \multirow{3}{*}{\LR} & \#1 & 99.08\std{0.65} & 96.29\std{1.57} & 99.40\std{0.26} \\
                                       &                      & \#2 & 99.17\std{0.36} & 99.95\std{0.24} & 99.09\std{0.39} \\
                                       &                      & \#3 & 94.70\std{0.70} & 76.20\std{4.03} & 96.39\std{0.60} \\ \midrule
                                       
            \multirow{6}{*}{\EDGEIIOT} & \multirow{3}{*}{\NN} & \#1 & 82.23\std{1.07} & 92.84\std{3.12}  & 81.35\std{1.11} \\
                                       &                      & \#2 & 83.11\std{1.07} & 100.00\std{0.00} & 81.81\std{1.15} \\
                                       &                      & \#3 & 80.92\std{1.06} & 55.33\std{4.73}  & 82.75\std{1.01} \\ 
                                       \cmidrule(lr){2-6}     
                                       & \multirow{3}{*}{\LR} & \#1 & 82.70\std{1.04} & 92.88\std{3.28}  & 81.86\std{1.07} \\
                                       &                      & \#2 & 83.66\std{0.95} & 100.00\std{0.00} & 82.40\std{1.03} \\
                                       &                      & \#3 & 79.99\std{1.07} & 54.03\std{1.54}  & 81.85\std{1.08} \\ \midrule
                                       
            \multirow{6}{*}{\IOTNID}   & \multirow{3}{*}{\NN} & \#1 & 96.77\std{1.11} & 87.48\std{4.83} & 98.10\std{0.64} \\
                                       &                      & \#2 & 95.21\std{1.07} & 86.66\std{4.27} & 96.28\std{0.87} \\
                                       &                      & \#3 & 93.53\std{1.27} & 88.68\std{4.60} & 94.07\std{1.13} \\ 
                                       \cmidrule(lr){2-6}     
                                       & \multirow{3}{*}{\LR} & \#1 & 96.67\std{0.99} & 88.63\std{3.54} & 97.82\std{0.69} \\
                                       &                      & \#2 & 94.51\std{1.18} & 85.97\std{3.92} & 95.58\std{1.01} \\
                                       &                      & \#3 & 92.93\std{1.23} & 88.18\std{4.31} & 93.46\std{1.14} \\ \bottomrule

        \end{tabular}
    }
\end{table}

Table~\ref{tab:MultistepIncremental} shows that \nafn remains effective beyond a single incremental session.
Across datasets and classifiers, \textit{F1-Old} stays stable after successive updates, indicating that \nafn mitigates catastrophic forgetting even when new attacks are introduced one by one.
This behavior is particularly clear on \CIC, where \NN preserves \textit{F1-Old} above $98\%$ across all three steps, and on \IOTNID, where both classifiers retain more than $93\%$ \textit{F1-Old} at the final step.
The overall performance also remains competitive: after three incremental steps, \nafn achieves $97.18\%$, $80.92\%$, and $93.53\%$ \textit{F1-All} with \NN on \CIC, \EDGEIIOT, and \IOTNID, respectively.
The main limitation appears in \textit{F1-New} when the final introduced class is substantially harder or less separable.
For example, on \EDGEIIOT, \textit{F1-New} reaches $100.00\%$ at step $\#2$ but drops to $55.33\%$ for \NN and $54.03\%$ for \LR at step $\#3$.
A similar, though milder, pattern appears on \CIC with \LR, where the final-step \textit{F1-New} decreases to $76.20\%$.
These drops highlight that multi-step \fscil performance depends not only on the number of sessions but also on the intrinsic separability and biflow quality of the newly introduced attack class.
Nevertheless, the corresponding \textit{F1-Old} remains high, showing that the performance loss is concentrated on the current new class rather than caused by broad forgetting.

\bibliographystyle{IEEEtranN}
\begingroup
\footnotesize
\bibliography{IEEEabrv,bibliography}
\endgroup

\end{document}